\documentclass[aps,pre,twocolumn,letterpaper,floatfix,showpacs,superscriptaddress]{revtex4}
\usepackage{graphicx}
\usepackage{amsmath,amssymb,amsfonts} 
\usepackage{mathtools}
\usepackage[squaren]{SIunits}
\usepackage[hidelinks]{hyperref}
\usepackage{epstopdf}
\usepackage{wasysym}
\usepackage{import}
\usepackage{placeins}

\usepackage{tikz}
\usetikzlibrary{plotmarks}
\newcommand\marksymbol[4]{\tikz[#2,scale=#4,rotate=#3]\pgfuseplotmark{#1};}

\newcommand{\smax}{s_\mathrm{m}}

\newcommand{\mus}{\mu_s}

\newcommand{\e}[1]{\cdot10^{#1}}
\newcommand{\fthres}{f_\text{thres}}
\newcommand{\fslip}{f_\text{slip}}
\newcommand{\fnew}{f_\text{new}}

\newcommand{\taumax}{\tau_\text{max}}
\providecommand{\abs}[1]{\lvert#1\rvert}
\newcommand{\ThogersenAppMusEff}{F }

\usepackage{color}
\usepackage{subfig}
\usepackage{calc}
\makeatletter
\newcommand*{\getlength}[1]{\strip@pt#1}
\makeatother
\newlength\graphicheight
\newlength\graphicwidth

\newcommand{\fpf}[1]{({\MakeLowercase{#1}})}
\newcommand{\fpl}[1]{({\MakeLowercase{#1}})}
\newcommand{\fpt}[1]{\MakeLowercase{#1}}

\newcommand{\pnasreusefigure}{Figure adapted from \citep{Tromborg2014slow}.}
\newcommand{\pnasreusepanel}[1]{Panel {#1} adapted from \citep{Tromborg2014slow}.}
\newcommand{\pnasreusepanels}[1]{Panels {#1} adapted from \citep{Tromborg2014slow}.}

\newcommand{\finalfigurepath}[1]{./#1}

\newcommand{\coloronlineornothing}{Color) }

\newcommand{\JSa}[1]{{#1}}

\newcommand{\AMSa}[1]{{#1}}

\newcommand{\JKTa}[1]{{#1}}

\newcommand{\Fig}{Figure}
\newcommand{\fig}{Fig.}
\newcommand{\tab}{Table}

\newcommand{\ie}{i.e. }
\newcommand{\eg}{e.g. }


\begin{document}
\title{On the speed of fast and slow rupture fronts along frictional interfaces}
\author{J{\o}rgen Kjoshagen Tr{\o}mborg}
\email{j.k.tromborg@fys.uio.no}
\affiliation{Department of Physics\\University of Oslo\\Sem S{\ae}lands vei 24, \\ NO-0316, Oslo, Norway}
\affiliation{Laboratoire de Tribologie et Dynamique des Syst{\`e}mes, CNRS, Ecole Centrale de Lyon\\36, Avenue Guy de Collongue, \\ 69134 Ecully cedex, France}
\author{Henrik Andersen Sveinsson}
\author{Kjetil Th{\o}gersen}
\affiliation{Department of Physics\\University of Oslo\\Sem S{\ae}lands vei 24, \\ NO-0316, Oslo, Norway}
\author{Julien Scheibert}
\email{julien.scheibert@ec-lyon.fr}
\affiliation{Laboratoire de Tribologie et Dynamique des Syst{\`e}mes, CNRS, Ecole Centrale de Lyon\\36, Avenue Guy de Collongue, \\ 69134 Ecully cedex, France}
\author{Anders Malthe-S{\o}renssen}
\affiliation{Department of Physics\\University of Oslo\\Sem S{\ae}lands vei 24, \\ NO-0316, Oslo, Norway}

\date{\today} 

\begin{abstract}
The transition from stick to slip at a dry frictional interface occurs through the breaking of the junctions between the two contacting surfaces.
Typically, interactions between the junctions through the bulk lead to rupture fronts propagating from weak and/or highly stressed regions, whose junctions break first.
Experiments find rupture fronts ranging from quasi-static fronts with speeds proportional to external loading rates, via fronts much slower than the Rayleigh wave speed, and fronts that propagate near the Rayleigh wave speed, to fronts that travel faster than the shear wave speed.
The mechanisms behind and selection between these fronts are still imperfectly understood.
Here we perform simulations in an elastic 2D spring--block model where the frictional interaction between each interfacial block and the substrate arises from a set of junctions modeled explicitly.
We find that a proportionality between material slip speed and rupture front speed, previously reported for slow fronts, actually holds across the full range of front speeds we observe.
We revisit a mechanism for slow slip in the model and demonstrate that fast slip and fast fronts have a different, inertial origin.
We highlight the long transients in front speed even in homogeneous interfaces, and we study how both the local shear to normal stress ratio and the local strength are involved in the selection of front type and front speed.
Lastly, we introduce an experimentally accessible integrated measure of block slip history, the Gini coefficient, and demonstrate that in the model it is a good predictor of the history-dependent local static friction coefficient of the interface.
These results will contribute both to building a physically-based classification of the various types of fronts and to identifying the important mechanisms involved in the selection of their propagation speed.
\end{abstract}

\maketitle
\section{Introduction}
The onset of sliding at a frictional interface occurs through the breaking of the \JSa{many} contacts that \JSa{were preventing} the relative motion of the surfaces. When a \JSa{single} contact breaks, the stress it bore is redistributed to its neighbors, bringing them closer to or past their load-bearing capacity. \JSa{In extended frictional interfaces (\ie larger than a characteristic correlation length scale), this process can lead} to propagating ruptures -- rupture fronts. The recent direct observation of rupture fronts in laboratory friction experiments \JSa{(see \eg \citep{Rosakis1999cracks,Baumberger2002self-healing,Rubinstein2004detachment,Prevost2013probing,Romero2014probing})} have provided new insights and opened new questions. \JSa{It was found, for instance, that not all fronts span the entire interface \citep{Rubinstein2007dynamics,Maegawa2010precursors}. The selection of the propagation length \AMSa{of the fronts} has been intensely investigated \citep{Scheibert2010role,Tromborg2011transition,Amundsen20121D,Radiguet2013survival,Braun2014propagation,Ozaki2014finite,Kammer2014linear}. It was also found that} the fronts can propagate \JSa{at a variety of speeds, either quasi-statically \citep{Chateauminois2010friction,Prevost2013probing,Romero2014probing}}, at \JSa{speeds close to that} of surface waves \JSa{(sub-Rayleigh)} \citep{Rubinstein2004detachment,Ben-David2010dynamics,Schubnel2011photo-acoustic}, at speeds faster than the shear wave speed $c_s$ \JSa{(supershear)} \citep{Rosakis1999cracks,Rubinstein2004detachment,Ben-David2010dynamics,Schubnel2011photo-acoustic,Latour2011ultrafast}, \JSa{or} at speeds \JSa{one or two orders of magnitude slower} than the Rayleigh speed \JSa{(slow)} \citep{Rubinstein2004detachment,Ben-David2010dynamics,Audry2012slip,Broermann2013friction}. \JSa{In this context, the present paper is mainly devoted to the study of the mechanisms responsible for front speed selection.}

Experiments to shed light on the nature of the rupture fronts have been performed. The authors of \citep{Ben-David2010dynamics,McLaskey2012fault} placed arrays of sensors close to the interface and used continuum theory to infer the properties of the elastic fields at the interface. \JSa{\citet{Svetlizky2014classical} recently showed that the dynamic fields associated with sub-Rayleigh fronts are consistent with the ones predicted by linear elastic fracture mechanics \citep{Freund1990dynamic}}. The authors of \citep{Degrandi2012sliding,Broermann2013friction,Romero2014probing} used microtextured surfaces where each contact can be tracked individually to follow the rupture fronts \JSa{directly at the individual micro-contact level}. Common to all \JSa{the experiments mentioned till now} is that they seek to increase \JSa{both the spatial and the temporal resolution of the behavior of the very interface}. These experiments can be usefully complemented by computer simulations, which by their nature provide complete information of all quantities in the models they implement\JSa{, thus filling in} the information gaps that remain despite the experimental progress; for example, simulations \JSa{can} provide simultaneous access to shear stress, normal stress, local contact strength and front propagation, a combination which remains hard to access experimentally.

\AMSa{To represent the propagation of the front separating a stuck part of the interface from a slipping one, models} of the transition to sliding need to include at least one level of discretization of the macroscopic interface. Depending on the model, this so-called mesoscopic scale may itself host a population of smaller scaled micro-junctions, such as the micro-contacts forming the multi-contact between rough surfaces (see \eg \citep{Braun2009dynamics,Capozza2012static,Tromborg2014slow}). However, the emerging collective behavior of the micro-junctions can also be lumped into a local friction law, often inspired by laws developed empirically from macroscale experiments, acting directly at the meso-scale (see \eg \citep{Maegawa2010precursors,Scheibert2010role,Tromborg2011transition,Bouchbinder2011slow,BarSinai2012slow,Braun2012crack,DiBartolomeo2012wave,Kammer2012propagation,Otsuki2013systematic,BarSinai2013instabilities,Papangelo2014simple}). The existing models also differ in how they treat the stress transfer in the bulk. The options range from 1D systems \citep{Amundsen20121D,Braun2014propagation,Braun2009dynamics,Maegawa2010precursors,Scheibert2010role,Bouchbinder2011slow,BarSinai2012slow,Capozza2012static,Braun2012crack,BarSinai2013instabilities,Papangelo2014simple} via 2D models \citep{Tromborg2011transition,Radiguet2013survival,DiBartolomeo2012wave,Kammer2012propagation,Otsuki2013systematic,Tromborg2014slow} to a full 3D visco-elasto-plastic discretization. In principle, the most comprehensive friction models could be combined with the most comprehensive bulk models, but the dynamics that arise from this approach tend to be nearly as complicated and difficult to disentangle as in the experiments themselves, and instead, authors have focused on one or a few model properties.

\JSa{Continuum models of friction, from the Amontons--Coulomb description to more sophisticated rate-and-state friction laws \citep{Dieterich1979modeling1,Ruina1983slip,Marone1998laboratory-derived,Baumberger2006solid,Li2011ageing}, are successful when they reproduce a robust average behavior of the myriad micro-junctions that make up each mesoscopic region of the frictional interface. However, by their nature, these models do not explain how the individual micro-junctions evolve and interact to produce the overall friction behavior. To approach this question, numerical \citep{Braun2002transition,Farkas2005static,Persson1995theory} and analytical \citep{Persson2000sliding,Braun2008modeling,Braun2010master,Thogersen2014history-dependent} models have been made that explicitly include a set of junctions, each representing one or a few microscopic contacts. The main missing ingredient in these models is a solid foundation for the junction evolution laws. In principle, this could be addressed by molecular dynamics simulations \citep{Reguzzoni2010onset}, but these simulations enable too short time scales and too small length scales for a systematic study of the onset of macroscopic sliding. Even models in which junctions represent thermally activated single molecular bonds in a simplified way \citep{Filippov2004friction,Srinivasan2009binding} would require a very large number of junctions and prohibitive calculation times in order to resolve the spatiotemporal dynamics of the transition to sliding observed experimentally.}

The simplest bulk model that includes spatial and temporal structure in the transition to sliding is probably the 1D spring--block model. Friction has been of interest to the earthquake community at least since \citet{Brace1966stick-slip} suggested that shallow earthquakes are the slip events in a stick--slip cycle, and since the spring--block experiment and simulation by \citet{Burridge1967model} the 1D spring--block model has been popular in the earthquake literature, see \eg \citep{Carlson1991intrinsic,Carlson1994dynamics}. As spatiotemporal data for the onset of sliding became available in laboratory friction experiments, the 1D spring--block model was also applied in the friction literature \citep{Braun2009dynamics,Braun2012crack,Maegawa2010precursors,Amundsen20121D,Capozza2011stabilizing,Braun2014propagation,Capozza2012static}. The main limitations of the 1D spring--block model are its inability to accurately reproduce the stress fields that arise in the experiments and the lack of a physical length scale \citep{Myers1993rupture,Ben-Zion2001dynamic} unless such a length scale is introduced in the friction law (\eg \citep{Amundsen20121D}).

To better reproduce the experimental loading conditions and how they translate to heterogeneous shear and normal stress fields at the frictional interface, we used, in \citep{Tromborg2011transition}, a 2D spring--block model, which can be shown to reproduce 2D linear elasticity \citep{Yim2000numerical}. 
With Amontons--Coulomb friction this model agreed well with experiments for \JSa{static measures related to the} onset of sliding, such as for the length of precursors and the evolution of the normal stress along the interface, but the model did not capture the full dynamics of the rupture fronts. \citet{Radiguet2013survival} studied the memory of the stress state through the passage of multiple ruptures in a visco-elastic 2D finite element model with a slip-weakening friction law. They too focused on successive precursors rather than the dynamics of each rupture event. \citet{Kammer2012propagation} studied \JSa{the properties of fast rupture} front propagation in a 2D finite element model with a static+velocity weakening dynamic friction law. \citet{Otsuki2013systematic} \JSa{studied how the normal force and the size of the interface influence the effective macroscopic static friction coefficient, using a 2D finite element model with a local velocity-weakening Amontons--Coulomb friction law. While the above-mentioned} models were able to model various aspects of the properties of rupture \JSa{fronts, they} do not provide a framework that \AMSa{within the same model} is able to account for the full richness of front dynamics that was observed experimentally. \JSa{In particular}, the self-selection \AMSa{of front type} leading to \JSa{sub-Rayleigh, slow} and supershear fronts, as well as the transitions between them, was missing.

In \citep{Tromborg2014slow} we combined an asperity model of the friction at the interface with a 2D elastic solver \JSa{in order to accurately reproduce the experimental loading conditions used in \citep{Rubinstein2004detachment}}. In the asperity model we included a time scale \JSa{inspired by the time scale} identified in \JSa{the same experimental system} \citep{Ben-David2010slip-stick} that \JSa{in the model} \AMSa{controls} the healing of the interface back to a fully pinned state after slipping. This combined model produced spatio-temporal features of the rupture dynamics \JSa{very} similar to those observed in the experiment. In \JSa{particular, we reproduced} the abrupt transitions between fast and slow front propagation\JSa{, which can} be understood from the underlying fast and slow slip mechanisms. Here, we use this \JSa{very same} model to \JSa{gain significant additional insights into} the relation between the micro-scale junction dynamics, the meso-scale slow and fast slip dynamics, and the macro-scale friction dynamics \JSa{and front propagation}. Our \JSa{results} show that while fast slip and fast fronts are inertially controlled, slow slip and slow fronts are non-inertial and instead depend on properties of the friction model\JSa{. Nevertheless,} the same proportionality relates front speed and slip speed in both velocity regimes. This suggests that \JSa{the key to understanding complicated experimental front dynamics lies in understanding the underlying slip dynamics.} Further, we \JSa{show} that the spatial extent of transients in front speed \JSa{is} comparable to the size of our system (we take the system size from experiments). \JSa{This suggests that, although front speed is influenced strongly by local quantities like the stress state and the local strength \citep{Tromborg2014slow}, front speed selection is intrinsically a non-local quantity which depends on the rupture history of the whole interface.} Finally, we \JSa{show} that by applying the Gini coefficient, which is an integrated measure of inequality commonly used in demography \citep[][p.~186]{Giorgi1999income,Giorgi1990bibliographic}, we can characterize the complicated slip dynamics \JSa{of a rupture event} in the \JSa{simulation} and predict the \JSa{subsequent (slip-history-dependent)} local frictional strength of the interface.

The paper is structured as follows. In Section~\ref{sec:model_description} we introduce the model. We first describe the motivating physical picture, and then make the model explicit by defining its equations and parameters. Then, we highlight some similarities and differences with other popular friction models. Next follow four results sections. In Section~\ref{sec:loading_curves_and_slip_dynamics} we show that the friction law introduced on the junction-level leads to stick--slip on the system level. We also revisit the slow slip mechanism identified in \citep{Tromborg2014slow}. In Section~\ref{sec:front_type_results} we present simulations with both fast--only and fast--slow--fast rupture events. We identify a possible signature of slow fronts in the macroscopic loading curve, and demonstrate the inertial nature of fast slip and fast fronts. We show how changes to parameters in the model or the state of the interface before rupture can turn fast--slow--fast fronts into fast--only fronts and vice versa. We map out the initial conditions that lead to fast--only and to slow fronts. In Section~\ref{sec:front_speed_results} we study the range of front speeds within each event and within each type of front (fast or slow). We show that in the model, the spatial extent of front speed transients is comparable to the current system size. We show that the influence of local stress and strength on front speed within each type of front is \JSa{similar} to their role in selecting the front type. We show that fast front speed is proportional to fast slip speed and that the constant of proportionality is the same as for slow fronts vs slow slip, providing a universal relationship across the velocity scales in the model. In Section~\ref{sec:history_dependence_with_gini} we come full circle by resolving how a rupture event sets up the state of the interface, which in turn determines the properties of the next rupture event. This links back to our results in \citep{Thogersen2014history-dependent}. Section~\ref{sec:discussion} is a brief discussion. The appendices provide additional details we deemed important to the understanding of the model, but that would hamper the flow of the arguments were they to appear within the main text. These include the stencil used for finding the front speed numerically, a data collapse demonstrating the utility of some of our dimensionless parameters, and a detailed account of how we apply initial and boundary conditions.

\section{Model description\label{sec:model_description}}
The frictional stability of a system made of two solids in contact locally depends on the level of normal and shear stresses at the contact interface. These interfacial stresses result from the external forces applied at the boundaries of the solids, transmitted through the bulk. Slip motion will in general be triggered when the local interfacial shear stress reaches a threshold, the level of which crucially depends on the interface behavior law at the microscale. The model we employ here is the same as in \citep{Tromborg2014slow}. In this section we begin by describing the physical aspects that underlie the model assumptions; we then describe the model and its parameters in detail; we end by discussing briefly the relationship to other models.

\subsection{Physical aspects}
The net contact between two solids generically consists of a large number of stress bearing micro-junctions. The \JKTa{properties} of these junctions depend on the type of interface. For rough solids, each micro-junction corresponds to a micro-contact between asperities on the opposing surfaces, whereas for smoother surfaces the junctions can be solidified patches of an adsorbate layer \citep{Persson1995theory}. We include in our model the following three physical aspects of the junction behavior. 1) A micro-junction in its pinned state \AMSa{behaves elastically and} can bear a shear force $f_T$, provided $f_T$ remains smaller than a threshold $\fthres$. When $\fthres$ is reached, a local fracture-like event occurs, and the junction enters a slipping state. 2) In the slipping state, the micro-junction moves with the slider's surface. The physical picture can be \eg the micro-slipping of micro-asperities in contact or the fluidization of an adsorbate layer. During slip, the micro-junction sustains some residual force $f_T = \fslip$, with $\fslip$ smaller than $\fthres$. 3) Slipping micro-junctions have a certain probability to disappear or relax. For example, a micro-contact disappears when an asperity moves away from its antagonist asperity by a typical distance equal to the mean size of micro-contacts, as classically considered for slow frictional sliding, e.g. in rate-and-state friction laws. However, another picture may arise from the sudden release of energy when pinned junctions break. This energy will dissipate as heat in the region around the micro-junction \citep{Ben-David2010slip-stick}. The rise in temperature will significantly increase the rate of a thermally activated relaxation of the slipping micro-junction during the time it takes for the interface to cool down \citep{Persson1995theory}. The effects of such temperature rises on friction have recently received renewed attention (see e.g. \citep{DiToro2011fault}), but remain poorly understood. In an attempt to include such thermal processes in our model, we recognize that they will lead to time- rather than distance-controlled relaxations, which will distribute the shear force drop in time. In order for the interface to continue bearing the normal forces applied to it, the micro-junctions that relax are replaced by new, pinned junctions that bear a small tangential force $\fnew$.

\subsection{Technical aspects}
The physical aspects described above have been modeled in a simple way using the following assumptions. We consider the rough frictional interface between a horizontal track and a thin linear elastic slider (\fig~\ref{fig:sketch_and_loading_curves}\fpt{A}). The slider has mass $M$ and sizes $L$ and $H$ in the horizontal ($x$) and vertical ($z$) directions, respectively. We present the values of all parameters in \tab~\ref{tab:parameters} in Appendix~\ref{appsec:simulation_setup}. The bulk elastodynamics of the slider are solved using a square lattice of blocks connected by internal springs (\fig~\ref{fig:sketch_and_loading_curves}\fpt{B}) \citep{Tromborg2011transition,Yim2000numerical}. The slider is divided into a square lattice of $N = N_x N_z$ blocks of mass $m=M/N$. Blocks are coupled to their four nearest neighbours and their four next-nearest neighbours by springs of equilibrium lengths $l=L/(N_x-1)=H/(N_z-1)$ and $\sqrt{2}l$ and stiffnesses $k$ and $k/2$, respectively, giving an isotropic elastic model with Poisson's ratio $1/3$. The force exerted on block $i$ by block $j$ is thus $k_{ij} (r_{ij}-l_{ij}) \frac{\Delta \textbf{x}_{ij}}{r_{ij}}$ when blocks are connected, 0 otherwise, where $\textbf{x}=(x,z)$, $\Delta \textbf{x}_{ij} = \textbf{x}_j-\textbf{x}_i$, $r_{ij} = \left|\Delta \textbf{x}_{ij}\right|$ and $k_{ij}$ and $l_{ij}$ are the stiffness and equilibrium length of the spring connecting blocks $i$ and $j$. Block oscillations are damped \AMSa{by introducing} a viscous force $\eta(\dot{\textbf{x}}_{j}-\dot{\textbf{x}}_{i})$ on the relative motion of connected blocks. We chose the coefficient $\eta=\sqrt{0.1 k m}$ so that blocks are underdamped and event-triggered oscillations die out well before the next event. The non-frictional boundary conditions are the same as setup~2 in \citep{Tromborg2011transition}, which was also used in \citep{Tromborg2014slow}. The top blocks are submitted to uniformly distributed, time-independent vertical forces $\frac{F_N}{N_x}$. The bottom blocks lie on an elastic foundation of modulus $k_f=k/2$, i.e. each block is submitted to a vertical force of amplitude $p_i=k_f \left|z_i\right|$ if $z_i<0$ or 0 otherwise, where $z_i$ is the vertical displacement of block $i$. Both vertical boundaries are free, except for a horizontal driving force $F_T=K(Vt-x_h)$ applied on the left-side block situated at height $h$ above the interface, where $x_h$ is the $x$-displacement of this block. This models a pushing device of stiffness $K$ driven at a small constant velocity $V$.

The multi-contact nature of the interface is modeled through an array of $N_s$ tangential springs representing individual micro-junctions, attached in parallel to each interfacial block (\fig~\ref{fig:sketch_and_loading_curves}\fpt{C}) \citep{Braun2009dynamics,Capozza2012static}. The individual spring behavior is as follows (\fig~\ref{fig:sketch_and_loading_curves}\fpt{F}, \citep{Tromborg2014slow,Thogersen2014history-dependent}). A spring pinned to the track stretches linearly elastically as the block moves, acting with a tangential force $f_T$ on the block. When the force reaches the static friction threshold $\fthres$ (we neglect aging, so that $\fthres$ is time independent), the micro-junction ruptures and the spring becomes a slipping spring acting with a dynamic friction force $f_T=\fslip$. After a random time $t_R$ drawn from a distribution $T(t_R)$, the slipping spring relaxes. It is replaced immediately by a pinned, unloaded spring ($\fnew=0$) and a new cycle starts. Here we use $T(t_R)$ as a simplified way of modeling the distribution of times after which micro-junctions relax. Due to the variety and the complexity of the underlying thermal processes, we did not try to derive $T(t_R)$ for a specific situation. Rather, we chose to model $T(t_R)$ in the simplest way, as a Gaussian with average time $\langle t_R\rangle$ and width $\delta t_R$. The shape of $T(t_R)$ is not crucial: we obtain qualitatively similar results with an exponential distribution. The width of $T(t_R)$ is the only source of randomness in our model and causes the interface springs of a block to evolve differently from each other.

The $2N$ equations of motion are solved simultaneously using a leapfrog / velocity Verlet integrator \footnote{The leapfrog method and the velocity Verlet method are different names for the same algorithm. When the force terms in the equations of motion depend on velocities (here through the viscous damping) the method looses second order accuracy. However, its computational cost is the same as that of other explicit first order methods, and in our experience the discontinuities in the junction force law negate the benefits of computationally more expensive higher-order methods.} on a uniform temporal grid of resolution $\Delta t$.

\subsection{Relationship to other models}
\subsubsection{Bulk modeling}
A spring--block discretisation of the bulk elasticity is particularly convenient for models where the friction is described as an ensemble of micro-junctions rather than a continuum law, because the blocks provide natural units on which to couple the frictional and the bulk elastic behaviour.

Let us note that like finite element (FEM) and finite difference (FDM) methods, the spring--block discretization satisfies the equations of linear elasticity. In particular, longitudinal (P) and shear (S) waves in the bulk propagate with the correct speeds and the right reflection and refraction properties \citep{Yim2000numerical}. To verify our implementation we checked that the code reproduces the expected bulk wave speeds.

The choice of the spatial resolution, the size of a single block, is made according to the following physical arguments. First, as discussed for example by \citet{Persson2000sliding}, by \citet{Caroli1998hysteresis} and by \citet{Braun2012collective}, below a characteristic length scale $\lambda$, called the elastic screening length, the interface behaves rigidly. $\lambda$ is thus the maximum block size allowing for a correct representation of the elasticity of the interface. For a linear elastic rough interface, $\lambda \sim d^2 /a$, with $a$ the typical lateral size of microcontacts and $d$ the typical distance between them. For micrometer-ranged roughnesses, we expect $a \sim \unit{1}{\micro\meter}$ and $d \sim\unit{10-100}{\micro\meter}$, yielding $\lambda \sim\unit{0.1-10}{\milli\meter}$. Second, the frictional behaviour of each block is then a statistical average over the many micro-junctions connected to it. This statistical approach is increasingly relevant for larger blocks involving more junctions. One thus looks for the largest possible block size. Combining both requirements, $\lambda$ appears as the natural block size for such spring-block models.

\subsubsection{Interface modeling}
The rate-and-state picture of friction, which includes displacement-controlled disappearance of micro-contacts, has proved to be adequate for slow (typically up to $\unit{100}{\micro\meter\per\second}$ range) sliding in a variety of materials. Such slow velocities imply a negligible temperature rise of the interface and thus a slipping state the duration of which is of purely geometrical origin, i.e. it is controlled by a length scale of the order of the mean micro-contact size.

Here, we focus on a drastically different situation, in which the transition from static to kinetic friction is extremely short (millisecond range) and is accompanied by fast slip ($\unit{100}{\milli\meter\per\second}$ range). As recognized in recent experiments by \citet{Ben-David2010slip-stick, Ben-David2010short-time}, the sudden rupture of the interface and its subsequent slip will generate a significant heating of the interface, sufficient to melt the broken micro-asperities. In these severe conditions, precise knowledge about the micro-contact behavior at the millisecond time scale is currently lacking.

In our description, we acknowledge the fact that the onset of sliding is far from the slow steady sliding situation. During the time in which the interface is significantly heated, thermal activations for the transitions from slipping to pinned states are highly probable. Such activation is classically described by time-rates, rather than by displacement-related quantities. It is therefore natural to propose an alternative picture that incorporates the possibility that transitions between the slipping and the pinned states can be controlled by a time.

We emphasize that experimental data did show that the dynamics at the onset of sliding involves a transition from fast to slow slip which occurs after a constant time, rather than a constant displacement \citep[see][]{Ben-David2010slip-stick}. This timescale controls the fast dynamics with which the interface comes back to a fully pinned state after slipping. It thus drastically differs from the classical time scale for aging, also found in \citep{Ben-David2010slip-stick}, and which controls the slow strength recovery of the interface when it is at rest.

Note, however, that we would not be surprised if a length scale would also be relevant to the dynamics of the onset of sliding in the same experimental system. However, the study of a complete model involving both a time scale and a length scale is far beyond the scope of the present work.

In terms of modeling approach, let us also stress the fact that a number of reference models from the literature considered, before us, time-controlled transitions between micro-junction states \citep[see e.g.][]{Braun2009dynamics,Capozza2012static,Persson1995theory,Braun2011dependence,Thogersen2014history-dependent}. \JSa{As shown in \citet{Thogersen2014history-dependent}, our friction law is actually one particular case of a more general family of models.}

\section{Loading curves and slip dynamics\label{sec:loading_curves_and_slip_dynamics}}
\subsection{Loading curves\label{sec:loading_curves}}

\begin{figure}
\centering
\includegraphics[width=\columnwidth]{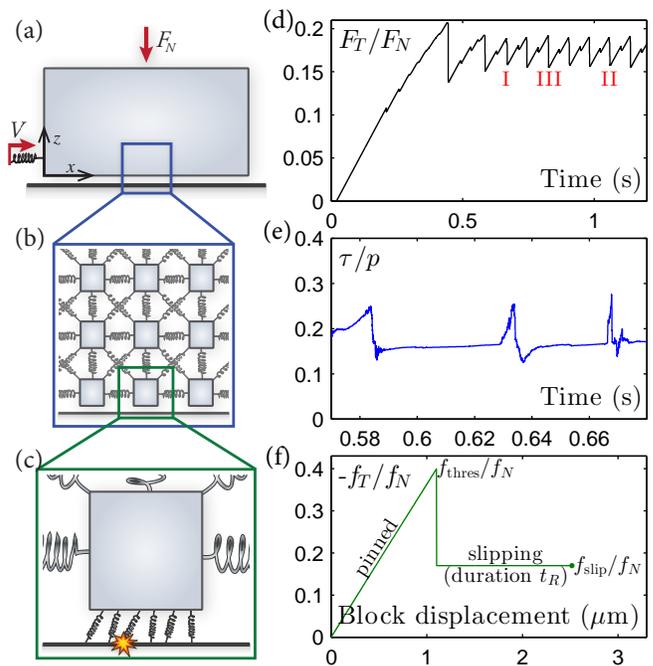}%
\caption{\coloronlineornothing Sketch and behaviour of the multiscale model. \fpf{a} Slider and external loading conditions. \fpf{b} Spring--block network modeling elastodynamics. \fpf{c} Surface springs modeling friction on a block. \fpf{d} Macroscopic loading curve, the ratio $F_T/F_N$ of driving shear force to total normal force. \fpf{e} Mesoscopic loading curve, the ratio $\tau/p$ of shear to normal stress on a block. \fpf{f} Microscopic friction model for the spring loading curve, the ratio $f_T/f_N$ of friction to normal force for one spring ($f_N=p/N_s$). \pnasreusefigure\label{fig:sketch_and_loading_curves}}
\end{figure}

The evolution in time of the driving force $F_T$ is called the loading curve. It is readily measured in experiments, and is used to characterize the motion as smooth sliding, or regular or chaotic stick--slip. The loading curve for our chosen set of parameters is shown in \fig~\ref{fig:sketch_and_loading_curves}\fpt{d}. It starts with an initial buildup from zero load. This initial buildup is linear, as we apply the load through a linearly elastic spring, one end of which is pushing the sample while its other end is being driven with a constant speed. From about $\unit{0.2}{\second}$ the linear increase is interrupted by small drops in the driving force. These are associated with precursors: rupture events confined to only part of the interface \JSa{\citep[see e.g.][]{Rubinstein2007dynamics, Braun2009dynamics, Maegawa2010precursors, Scheibert2010role, Tromborg2011transition, Amundsen20121D, Katano2014novel, Kammer2014linear,Braun2014propagation}}. The first event where the entire interface breaks and the slider moves macroscopically occurs at about $\unit{0.45}{\second}$ and is seen in the loading curve as a larger drop in $F_T$ whose duration is not resolved on this figure. Then the system enters regular stick--slip, with alternating events showing partial and full breaking of the interface (small and large drops in $F_T$).
In the model, as in experiments, the behaviour on larger scales emerges from the interactions on the scales below. \Fig~\ref{fig:sketch_and_loading_curves}\fpt{f} illustrates the single junction law described in Section~\ref{sec:model_description}.

Each mesoblock is coupled to the track with many junctions (here we use $N_s=100$). An example of the evolution of the net force in the junctions, which apart from a small inertial term is equivalent to the sum of forces on the block from its neighbours, is shown in \fig~\ref{fig:sketch_and_loading_curves}\fpt{e}. The data is from the stick--slip phase. The temporal structure is rich in detail, but can overall be described as periods of very slow increase (the driving is slow compared to the internal dynamics) separated by faster increase and decrease when rupture fronts pass.

The force maxima attained on the mesoblock level are smaller than the sum of the individual junction thresholds, which would give $\tau/p=0.4$. The explanation is simple: due to the disorder in the individual stretching states, some junctions will break before others, and so they will not all contribute their maximum force simultaneously \JSa{(see detailed discussion in \citep{Tromborg2014slow, Thogersen2014history-dependent})}. Similarly, the maxima in the macroscopic loading curve of \fig~\ref{fig:sketch_and_loading_curves}\fpt{d} are smaller than those on the mesoblock level, as the mesoblocks do not reach their individual maxima at the same time. \JSa{This difference between macroscopic and local friction coefficients has been discussed in e.g. \citep{Maegawa2010precursors, Scheibert2010role, Tromborg2011transition, Ben-David2011static, Capozza2012static, Otsuki2013systematic}.} That the overall strength in a system is usually smaller than the sum of the individual strengths of the constituents is familiar from other fields, for example fracture mechanics \citep[see e.g.][]{Bonamy2011failure} and fiber bundle theory \citep[see e.g.][]{Pradhan2010failure}.

\subsection{Block slip dynamics and a slow slip mechanism\label{sec:slip_dynamics_and_a_slow_slip_mechanism}}
\begin{figure}
\centering
\includegraphics[width=\columnwidth]{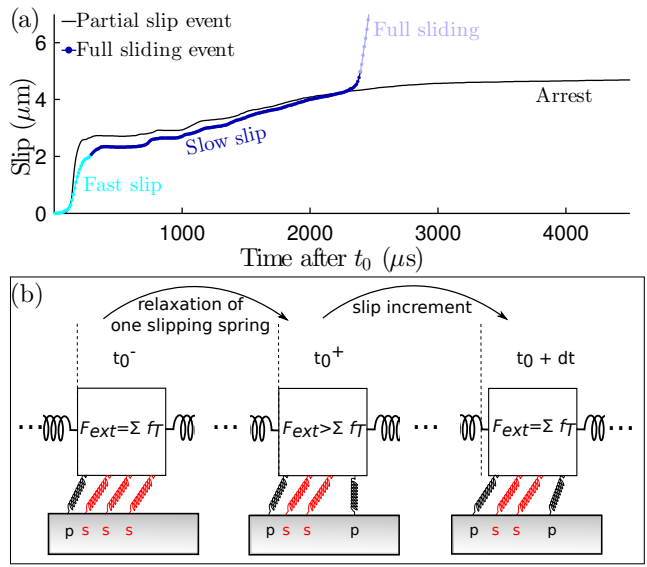}%
\caption{\coloronlineornothing \fpf{a} Example slip profiles from a partial slip and a full sliding event. Line without markers (black): slip profile for block at $x=0.16L$ with $t_0=\unit{0.7872}{\second}$, \ie the partial slip event preceding event III. Line with markers (grey/multicolored): slip profile for block at $x=0.34L$ with $t_0=\unit{1.0571}{\second}$, \ie event II (more details of this event in \fig~\ref{fig:slow_slip_to_slow_front}). In both cases we chose a block located near the middle of the region where the rupture front speed was fast. For blocks closer to the fast--slow transition or to the front arrest point, the amplitude of fast slip is smaller. \fpf{b} Sketch showing how the relaxation of force associated with the junctions' relaxation from the slipping (s) to the pinned state (p) can lead to a slow slip motion of the block.\label{fig:slow_slip_mechanism}}
\end{figure}

\Fig~\ref{fig:slow_slip_mechanism}\fpt{a} shows the slip dynamics of a block in a partial slip event and a block in a full sliding event. \JSa{We find both fast and slow slip regimes of the motion, in excellent agreement with the experiments reported in \citep{Ben-David2010slip-stick}.} The initial fast slip regime begins after the passage of a fast rupture front. In both these cases the fast slip is followed by the block coming nearly to rest (no visible increase in net slip between $450$ and $\unit{550}{\micro\second}$) and then by a slow slip regime with roughly linear increase of slip vs time, i.e. constant slip speed. The slow slip regime can end in two ways. In full sliding events, slow slip changes back to fast slip when the slider enters the full sliding regime. For arresting events, the slow slip regime ends when the block comes to rest.

\JSa{The initial fast slip regime corresponds to an inertial motion of the block when a large number of junctions break in a short time interval as the rupture front passes by. The net friction force is rapidly reduced, bringing the block out of mechanical equilibrium. The result is a large positive acceleration (in the direction of the net force due to the neighboring blocks). The inertial nature of this motion is demonstrated in \fig~\ref{fig:inertial}.}

The subsequent slow slip observed in the model has a different physical origin that was explained in \citep{Tromborg2014slow, Thogersen2014history-dependent}. This slow slip mechanism is illustrated in \fig~\ref{fig:slow_slip_mechanism}\fpt{b}. When $\fnew<\fslip$ the net friction force on a block is reduced slightly whenever a spring leaves the slipping state, yielding a small positive acceleration as the net friction drops below the net external force from neighboring blocks. \JSa{The friction reduction is soon balanced by the changes in the pinned junctions and the external forces on the block as it slowly moves. This slow slip mechanism is present as long as some junctions are going from the slipping to the pinned state and $\fnew<\fslip$, but it is masked by the fast slip while the fast slip lasts. The dependence of slow slip speed on model parameters was discussed in detail in \citep{Tromborg2014slow}.}

\section{Front type results\label{sec:front_type_results}}
In this section we present our results on the qualitative features of rupture fronts. That is, we discuss the conditions under which we observe fast rupture, slow rupture, and the transitions between these regimes.

\subsection{Rupture front characteristics\label{sec:rupture_fronts}}
\begin{figure}
\centering
\includegraphics[width=\columnwidth]{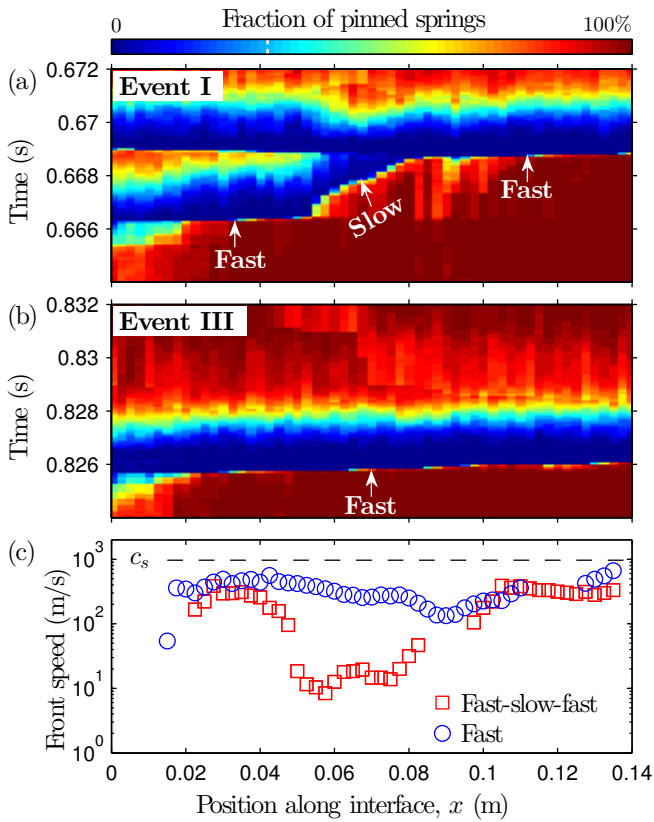}%
\caption{\coloronlineornothing Two interface-sized events. \fpf{a} A fast--slow--fast event (I in \fig~\ref{fig:sketch_and_loading_curves}\fpt{d}). Spatiotemporal plot of the fraction of pinned springs. \fpf{b} A fast--only event (III in \fig~\ref{fig:sketch_and_loading_curves}\fpt{d}) shown as in \fpl{a}. \fpf{c} Rupture front speed $v_c$ vs. front location for both events. Block rupture is defined to occur when 70\% of interface junctions have broken (white dashed line in the colorbar). Front speed is measured as the inverse slope of the rupture line (indicated by arrows in \fpl{a} and \fpl{b}) using the endpoints in a five-point-wide moving stencil. \pnasreusefigure\label{fig:events_with_rupture_speeds}}
\end{figure}

With the driving force applied on the trailing edge of the slider, the blocks near the trailing edge are the first to reach their effective static friction thresholds. A block that slips increases the load on its neighbors, which can start to slip in turn. The rupture front tip, i.e. the boundary between a region of stuck blocks and a region of slipping blocks, then propagates away from the nucleation point. If the rupture arrests before reaching the leading edge of the slider the event is called a partial slip event; if the leading edge is reached, we use the name full/global sliding event; precursors are partial slip events that occur before the first global sliding event.

In the interaction law the distinction between pinned and slipping states is made on the junction rather than the block level. We therefore define slipping on the block level to mean that a certain fraction of the block's junctions are in the slipping state. This criterion is robust to the choice of threshold fraction, because as is seen in \fig~\ref{fig:events_with_rupture_speeds} and other figures (\ref{fig:slow_slip_to_slow_front}\fpt{ab}, \ref{fig:fast_slow_fast_into/from_fast_only}, \ref{fig:135x}), it is typical for a block to go from having nearly all its junctions pinned to having nearly all of them broken in a time short compared to the other time scales in the simulation. For the events in \fig~\ref{fig:events_with_rupture_speeds} the time to go from 80\% of springs pinned to 20\% of springs pinned is approximately $\unit{0.03}{\milli\second}$ in the fast part of the fronts and $\unit{0.3}{\milli\second}$ in the slow part of the fronts. We have used a threshold value of 30\%, that is, the start time of block slipping is taken as the instant when the fraction of pinned junctions dropped below 30\%.

\Fig~\ref{fig:events_with_rupture_speeds}\fpt{a} shows a rupture front whose speed $v_c$ changes from fast ($v_c\sim c_s/3$) to slow ($v_c\sim c_s/100$) and back to fast again. The left-travelling front that starts when the primary front is reflected from the leading edge re-breaks the junctions that had healed behind the front tip.
 \Fig~\ref{fig:events_with_rupture_speeds}\fpt{b} shows a rupture front that is fast across the entire interface. By defining the transition to block sliding as above, the location of the front tip in time can be measured. The local front speed is then the ratio of the distance travelled by the front to the time for that propagation. Because of disorder in the junction state along the interface remaining from earlier events, the propagation time from one block to the next can vary significantly. We have found that using only the end-points in a 5 blocks wide moving stencil (Appendix~\ref{appsec:5_point_stencil}) gives the best balance between robustness and spatial resolution in the calculated front speed. The results are shown in \fig~\ref{fig:events_with_rupture_speeds}\fpt{c}.

\subsection{The influence of front type on the loading curve\label{sec:influence_of_front_type_on_the_loading_curve}}
\begin{figure}
\centering
\includegraphics[width=\columnwidth]{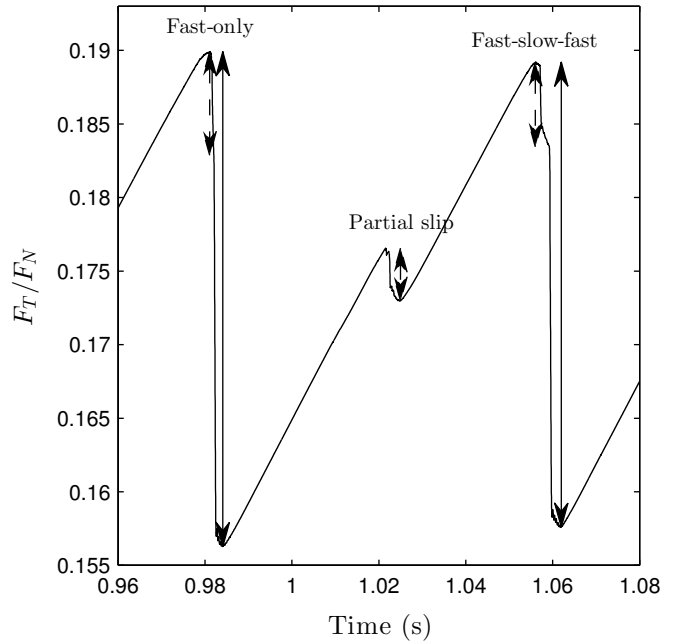}%
\caption{A zoom on the macroscopic loading curve $F_T/F_N$. The drop in a full sliding event is comprised of two parts: first, the drop associated with deformation of the slider during rupture front passage (dashed arrow); second, the drop associated with motion of the entire slider (full arrow minus dashed arrow). The amplitude of the first drop is approximately the same as the amplitude of the drop associated with a partial slip event (see \fig~\ref{fig:drop_amplitudes}).\label{fig:loading_curve_zoom}}
\end{figure}

The two events in \fig~\ref{fig:events_with_rupture_speeds} are different. One is a fast--slow--fast front, the other is fast across the entire interface. Both events are marked in \fig~\ref{fig:sketch_and_loading_curves}\fpt{d}, and the associated drops in $F_T$ are seen to have approximately the same amplitude. This indicates that the details of the front propagation do not influence the loading curve strongly, at least not for these events. Nevertheless, the slow front propagation does have a signature in the loading curve that appears when we zoom in on a few events as in \fig~\ref{fig:loading_curve_zoom}. Namely, the drop in the loading force has a significant change in slope while the slow front lasts, which distinguishes it visually from the force drop associated with a fast--only event.

Let us consider the evolution of $F_T$ in more detail. As long as the slider remains pinned, $F_T$ increases with the motion of the driving stage (the driving stage moves the end of the driving spring that is not attached to the slider). $F_T$ decreases only when the point on the slider where the driving spring attaches, the trailing edge, moves away from the driving stage. This occurs in two distinct ways. First, the trailing edge moves away from the driving stage when the slider deforms in compression during the passage of a rupture front, which happens both for partial slip and full sliding events. Second, the trailing edge moves with the rest of the slider when the entire interface is slipping in a full sliding event. For full sliding events we define the boundary between rupture front passage and full sliding as the moment the rupture front reaches the leading edge, seen \eg in \fig~\ref{fig:events_with_rupture_speeds}. The relative amplitude of the $F_T$ reductions associated with each of these two motions of the trailing edge depends on the relative stiffnesses of the slider and the driving spring. When the slider is stiff compared to the driving spring, so that little motion of the trailing edge can occur unless the entire slider moves, the drop in $F_T$ is only appreciable during the sliding part of full sliding events. With a softer slider, the deformation occurring in partial slip events and during the rupture front passage in a full sliding event accounts for a larger fraction of the net reduction in $F_T$. As seen by the relative amplitudes of the force drops in \fig~\ref{fig:loading_curve_zoom}, with the present parameters the (trailing edge) slip associated with the slider deformation accounts for about a fifth of the net slip of full sliding events.

\begin{figure}
\centering
\includegraphics[width=\columnwidth]{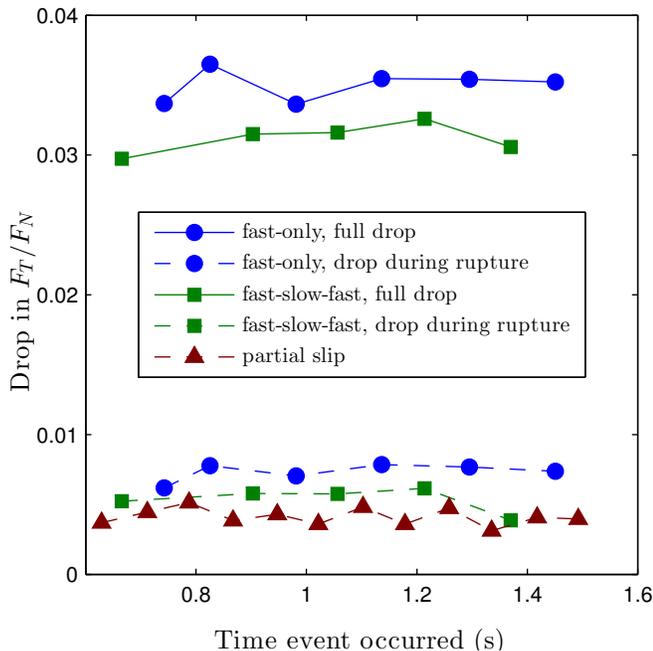}%
\caption{\coloronlineornothing The force drops in the loading curve (\fig~\ref{fig:sketch_and_loading_curves}\fpt{d}) grouped according to event type. We include partial slip events and full sliding events occurring between $t=\unit{0.63}{\second}$ and $t=\unit{1.50}{\second}$ (the developed stick--slip regime). The drops occurring during the passage of the rupture front have comparable amplitude between the partial slip events, fast--slow--fast events and fast--only events.  Also, the net force drop in full sliding events have comparable amplitudes for fast--slow--fast and fast--only events.\label{fig:drop_amplitudes}}
\end{figure}

Further quantification of this feature is presented in \fig~\ref{fig:drop_amplitudes}, where the force drops associated with either the entire event or with the rupture front passage only are presented for all events during the developed stick--slip regime. We find that the loading force drop occurring during the rupture front passage, regardless of the type of event, is always much smaller than the one associated with sliding. The net force drop in full sliding events has comparable amplitudes for fast--slow--fast and fast--only events. This is because the block motion during full sliding accounts for the largest part of the net block motion in an event, and after the rupture front reaches the leading edge, the distinction between fast--slow--fast events and fast--only events is unimportant.


\subsection{Fast slip and fast front speeds are inertial\label{sec:fast_slip_and_fast_front_speeds_are_inertial}}
\begin{figure}
\centering
\includegraphics[width=\columnwidth]{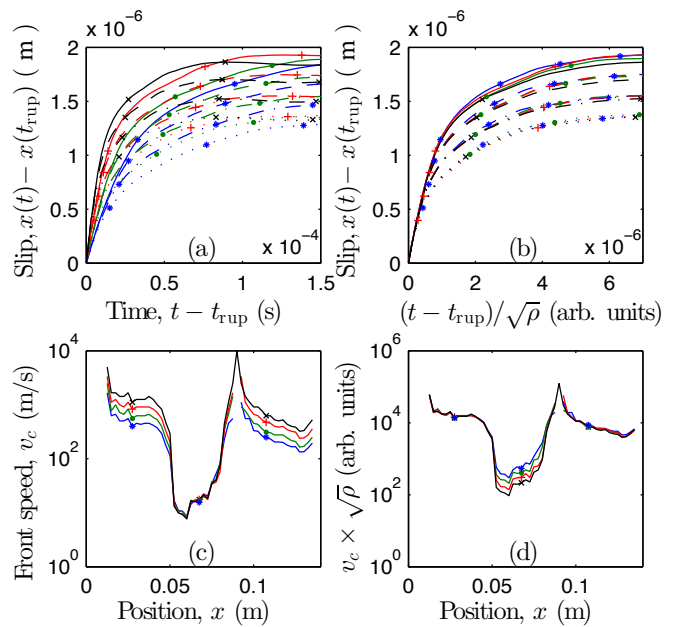}%
\caption{\coloronlineornothing The fast slip and the fast front speeds scale with inertia. \fpf{a} Block slip motion for four neighboring blocks within the fast front region, for four simulations of the same fast--slow--fast event. In each simulation the initial state is the same, but the mass density $\rho$ is different between the simulations. The block slip is measured from the rupture time $t_\text{rup}$ of each block as defined in \fig~\ref{fig:events_with_rupture_speeds}. Lines with the same color and marker are from the same simulation. Lines with the same line style represent the same block in different simulations. \fpf{b} Rescaling the time of slip with $\rho^{-1/2}$ collapses the data in \fpl{a}. \fpf{c} The rupture front speed as a function of position for the same simulations as the data in \fpl{a} (corresponding colors and markers). The change of density modified the fast front speed, while the slow front speed remained nearly unchanged. \fpf{d} Rescaling the front speed data in \fpl{c} by $\rho^{-1/2}$ collapses the fast front speed measurements, but splits the slow front measurements.\label{fig:inertial}}
\end{figure}

In \fig~\ref{fig:slow_slip_mechanism}\fpt{a} we showed example block slip profiles and the regimes of fast slip, slow slip and full sliding. In section~\ref{sec:slip_dynamics_and_a_slow_slip_mechanism} we focused on the slow slip part, and discussed how the junction evolution law leads to a slow slip mechanism on the block level. In \fig~\ref{fig:events_with_rupture_speeds} we showed that the model exhibits both fast--only and fast--slow--fast fronts, and explained how we measure the front speed. In this section we demonstrate that both the fast slip part of the block slip evolution and the fast front speed are of inertial origin. That is, like the bulk wave speeds, these speeds scale as $\rho^{-1/2}$, where $\rho=M/(LBH)$ is the mass density of the system.

To isolate the effect of inertia from the stress and frictional state at the interface we have performed four simulations starting from the same state (Appendix~\ref{appsec:simulation_setup}), but with $\rho$ decreased to $1$, $1/2$, $1/4$ and $1/8$ of its value in \tab~\ref{tab:parameters}; we changed the mass and kept the system size constant. \Fig~\ref{fig:inertial}\fpt{a} shows the fast slip dynamics for four neighboring blocks located within the part of the interface where the front speed was high as the front passed, \AMSa{for each of the four simulations}. The figure \AMSa{clearly} shows that the fast slip speed was modified by the change of density. \Fig~\ref{fig:inertial}\fpt{b} demonstrates that a $\rho^{-1/2}$ scaling collapses the data. The reference time $t_\text{rup}$ for each block is the time of block rupture as defined in \fig~\ref{fig:events_with_rupture_speeds}, and the block slip is measured with respect to the block position at $t_\text{rup}$.

\Fig~\ref{fig:inertial}\fpt{c} shows the front speed as a function of position along the interface for the four simulations, while \fig~\ref{fig:inertial}\fpt{d} shows the same data with the front speed rescaled by $\rho^{-1/2}$. The fast front speeds are collapsed onto each other by this rescaling. The slow front speeds, in contrast, remain the same in all four simulations (\fig~\ref{fig:inertial}\fpt{c}), and are split from each other by the scaling. This indicates a non-inertial origin of the slow front speed, which is the topic of the next section.

\subsection{From slow slip to slow fronts}
\begin{figure}
\centering
\includegraphics[width=\columnwidth]{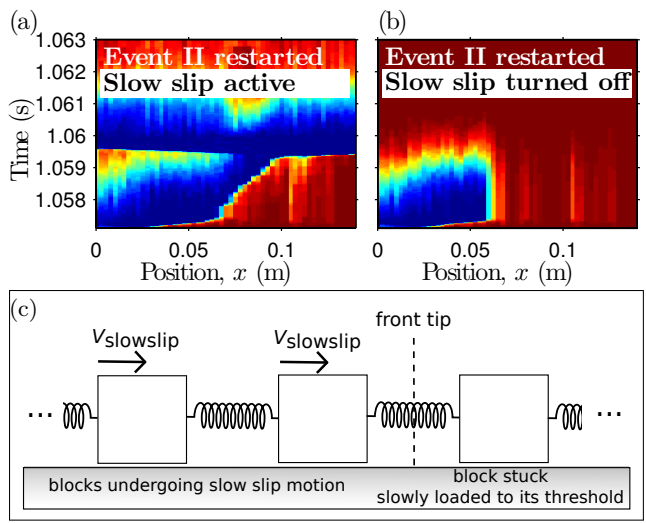}%
\caption{\coloronlineornothing \fpf{a} and \fpf{b} Event II restarted at $\unit{1.0571}{\second}$ with driving speed $V=0$, shown as in \fig~\ref{fig:events_with_rupture_speeds}. Slow slip is either \fpf{a} active or \fpf{b} turned off. \fpf{c} A sketch showing how slip behind the front tip leads to front tip propagation. \pnasreusepanels{\fpl{a} and \fpl{b}}\label{fig:slow_slip_to_slow_front}}
\end{figure}
In the model the propagation of fast rupture and slow rupture is governed by different mechanisms. To demonstrate this we start from the fast--slow--fast event in \fig~\ref{fig:slow_slip_to_slow_front}\fpt{a}. We then modify the junction law by increasing $\fnew$ to $\fslip$, so that there is no change in the friction force when a slipping junction relaxes and is replaced by a pinned junction. This does not affect the initial force relaxation of pinned junctions reaching $\fthres$, but it does affect the second relaxation of slipping junctions relaxing back to the pinned state. In practice, this change turns off the slow slip mechanism that was discussed in Section~\ref{sec:slip_dynamics_and_a_slow_slip_mechanism}. The result of restarting from the same state as in \fig~\ref{fig:slow_slip_to_slow_front}\fpt{a} and with only this modification to the junction law is seen in \fig~\ref{fig:slow_slip_to_slow_front}\fpt{b}. The first, fast part of the rupture is unaffected, but the slow rupture is suppressed, and the front stops where the fast-slow transition used to occur.

\Fig~\ref{fig:slow_slip_to_slow_front}\fpt{c} illustrates how slow slip in the region behind the front tip can result in additional front propagation: as the blocks at and behind the front tip move towards the stuck region, the external forces on the block just ahead of the front tip increase. Two outcomes are possible. Either the block remains stuck and the front arrests, or the forces on this block eventually overcome its effective static friction threshold, and the block starts to slip, moving the front tip one block ahead. This last part of the explanation would be the same for fast slip and fast fronts. The difference between fast and slow fronts is thus simply a matter of the origin of the underlying slip motion, which is inertial for fast fronts (see \fig~\ref{fig:inertial}), and is related to the intrinsic arrest dynamics of the interface for slow fronts (see \citep{Tromborg2014slow, Thogersen2014history-dependent}).

Our understanding of the transition from fast to slow fronts is that the fast propagation stops at the same point in both \fig~\ref{fig:slow_slip_to_slow_front}\fpt{a} and \fpt{b}. In \fig~\ref{fig:slow_slip_to_slow_front}\fpt{a}, slow slip becomes important when the fast slip ends, and results in a slow front propagating. In \fig~\ref{fig:slow_slip_to_slow_front}\fpt{b}, the slow slip mechanism is turned off, and the event is over once the initial fast slip ends.

We emphasize that the physical origin of the slip on the block level is unimportant to the way slip behind the front tip leads to rupture front propagation. We showed in \citep{Tromborg2014slow} that a mechanism for slow slip completely different from the one in the present model would also lead to slow front propagation.

\subsection{Junction force distribution affects front type selection\label{sec:modifyingFrontTypeFromWidth}}
\begin{figure}
\centering
\includegraphics[width=\columnwidth]{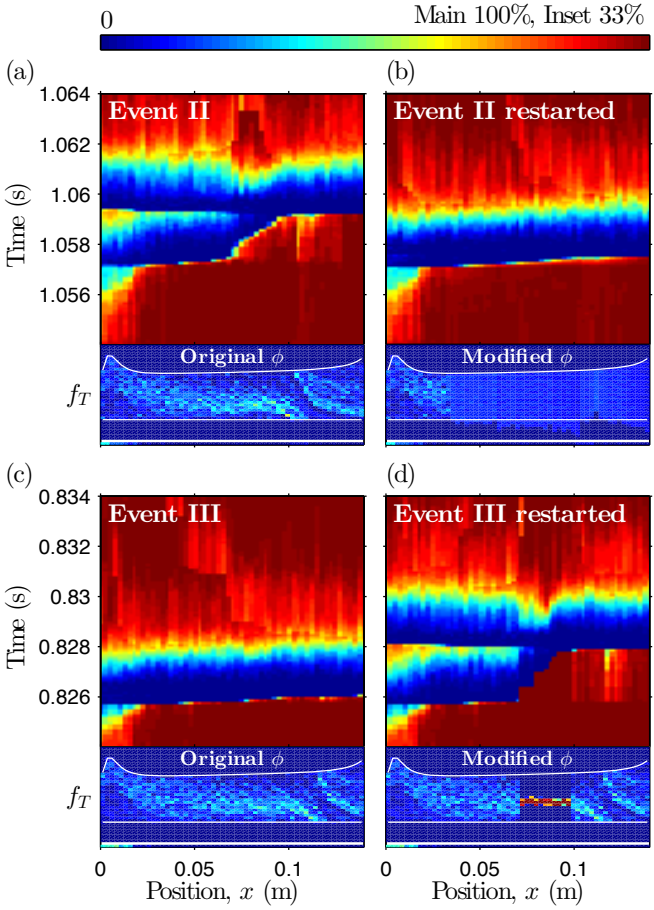}%
\caption{\coloronlineornothing Junction force distributions affect front propagation. Main panels: events shown as in \fig~\ref{fig:events_with_rupture_speeds}. Insets: histograms of junction force distributions. (a, b): A fast--slow--fast event turned into a fast--only event by increasing the width of the stretching distribution while maintaining the shear stress (the modified distributions are uniform). The insets are taken at $t=\unit{1.054}{\second}$. (c, d): A fast--only event turned into a fast--slow--fast event by decreasing the width of the stretching distribution along part of the interface. The shear stress was unchanged and the new distribution was uniform with the width shown in the inset of panel \fpl{d}. The insets are taken at $t=\unit{0.824}{\second}$. All insets: for each block along the interface, a color coded histogram of $\phi(f_T)$. The vertical axis shows the force level in individual springs, which extends up to $f_\mathrm{thres}$. Upper white line: the level $f_\mathrm{thres}$, which is different for each block because it varies with normal force. Lower white line: $f_T=\unit{0}{\newton}$. The color scale denotes the fraction of springs found at each value of $f_T$ using an arbitrary bin width. Offset data: fraction of slipping springs at $\unit{1.054}{\second}$ (a, b), at $\unit{0.824}{\second}$ (c, d). \pnasreusepanels{\fpl{a} and \fpl{b}}\label{fig:fast_slow_fast_into/from_fast_only}}
\end{figure}

We argued in the previous section that even when the slow slip mechanism is active, its effect on the rupture front speed can be masked by fast slip and the associated fast front propagation. To illustrate this (\fig~\ref{fig:fast_slow_fast_into/from_fast_only}), we will here re-simulate a fast--slow--fast event with the initial state modified so that the event becomes fast--only. Similarly, we will re-simulate a fast--only event with the initial state modified so that the event becomes fast--slow--fast.

To understand these results, recall the connection between the distribution $\phi(f_T)$ of forces among the junctions attached to one block and the corresponding effective static friction coefficient of this same block. As shown numerically in e.g. \citep{Tromborg2014slow} and theoretically in e.g. \citep{Farkas2005static, Thogersen2014history-dependent} and Appendix~\ref{appsec:mus_from_width}, for friction, and more generally for the rupture of heterogeneous systems in which a number of junctions are loaded in parallel \citep{Pradhan2010failure}, the maximum load that an interface can bear is related to the width of the load distribution and/or threshold distribution of the various junctions. Homogeneous systems (vanishing distribution width) have the maximum possible macroscopic rupture threshold because all junctions will contribute with their maximum force when collective rupture is reached. In contrast, in heterogeneous systems (finite width) the weaker and/or initially more highly loaded junctions will break first, so that when macroscopic rupture occurs, only a fraction of the initial population of junctions will contribute to the total force. With this in mind, the strategy for turning fast--slow--fast events into fast--only fronts and the other way around was to modify the width of the initial force distribution of selected blocks along the interface.

Slow slip becomes important to front propagation only after a fast front stops. We can keep the initial fast part of the event in the re-simulation equal to the original event by leaving the stress state unchanged. By increasing the width of the junction force distribution we make the interface weaker, which may enable fast propagation across the whole interface. An example of this is shown in \fig~\ref{fig:fast_slow_fast_into/from_fast_only}\fpt{b}.

Conversely, in \fig~\ref{fig:fast_slow_fast_into/from_fast_only}\fpt{d}, by significantly decreasing the width of the initial junction force distribution, we made the interface stronger along a part of the interface, so that the original fast--only propagation was stopped. This allowed the slow slip mechanism to become important, and the front was turned into a fast--slow--fast one.

We note that in both cases, the stress state in the original and modified simulations are the same, and the only change is in the width of $\phi(f_T)$ in the region where it is modified. To keep the initiation of the events the same in the modified and original simulations, we did not modify $\phi$ near the trailing/left edge of the system.

\subsection{Front type phase diagram and its predictive power\label{sec:front_type_phase_diagram_and_its_predictive_power}}
\begin{figure}
\centering
\includegraphics[width=\columnwidth]{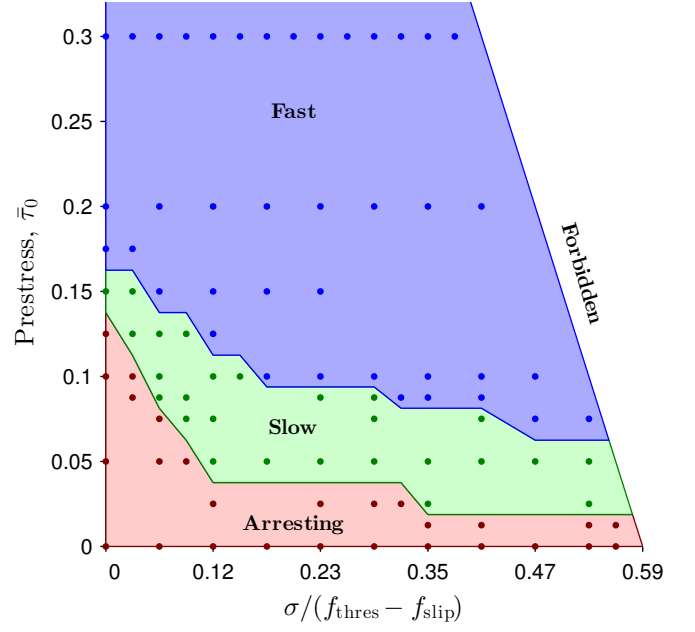}%
\caption{\coloronlineornothing A front type "phase diagram" based on simulation of homogeneous prepared interfaces. Front type observed vs initial width $\sigma$ of junction force distribution and prestress $\bar\tau_0=(\tau_0/p-\fslip/f_N)/(\fthres/f_N-\fslip/f_N)$. Initial states in the forbidden region would have had some junctions stretched beyond their breaking threshold and are therefore excluded. \pnasreusefigure\label{fig:phase_diagram}}
\end{figure}

We have seen that increasing (decreasing) the width of the junction force distribution makes the interface weaker (stronger) and that this favours fast (slow) front propagation. It also makes intuitive sense that higher prestress, hence smaller distance to the breaking threshold, would favor fast front propagation (we return to this in more detail in Section~\ref{sec:front_speed_results}). We have performed simulations where these two parameters are varied systematically (see Section~\ref{sec:front_speed_depends_on_local_stress_state} and Appendix~\ref{appsec:simulation_setup}), and the observed front types are presented in \fig~\ref{fig:phase_diagram}. In the arresting region, the fronts are partial slip events, that is, they stop before reaching the leading edge (some of them stop early, some almost reach the leading edge). The region labelled slow includes all those events that have a slow front part, even if the event is fast along most of the interface. They share the characteristic that the events would arrest in the absence of the slow slip mechanism. In the fast region, events are fast across the entire interface.

In Section~\ref{sec:front_speed_results} we discuss the transient behavior of the rupture fronts. For now, let us stress that while the arresting, slow and fast regions of the rupture fronts are robust in their relative positions (the fast front region is found at higher values of normalized prestress $\bar\tau_0$ and junction distribution width $\sigma$ than the slow front region, which is itself found at higher values than the arrest region), the precise locations of the boundaries between them depend also on how the events are triggered. For example, it is possible to prepare two simulations with the same stresses and distributions in the propagation region, but with different stresses in the triggering region (these regions are defined in \fig~\ref{fig:fast_front_speed_vs_prestress_width_0}\fpt{c} and in Appendix~\ref{appsec:simulation_setup}) such that the simulation with the more highly stressed triggering region produces a fast--only front and the other a fast--slow--fast front. For this reason, all the simulations in \fig~\ref{fig:phase_diagram} have the same settings in the triggering region.

\begin{figure}
\centering
\includegraphics[width=\columnwidth]{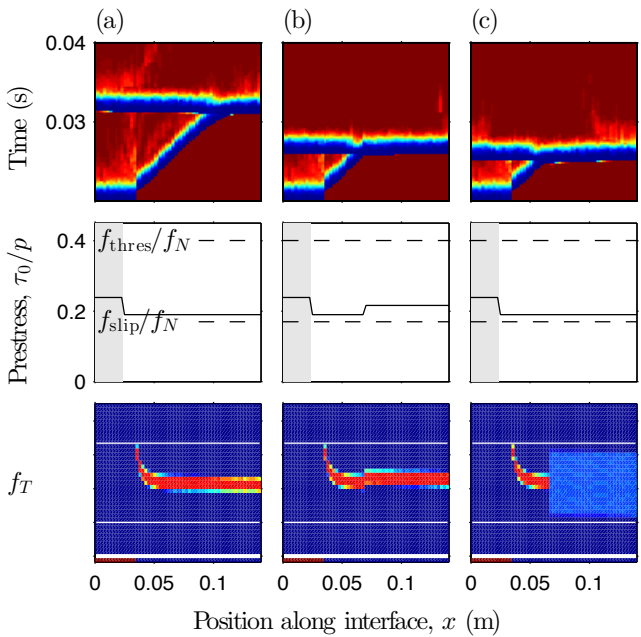}%
\caption{\coloronlineornothing Simulations showing that even though the phase diagram in \fig~\ref{fig:phase_diagram} is an interface-wide and not a local measure, the change in front characteristics upon local changes to prestress and spring stretching width are consistent with the predictions of the phase diagram. \fpf{a} The original simulation with homogeneous stress and junction distribution states in the propagation region. \fpf{b} A step increase in the initial shear stress within the propagation region leads to a slow--fast transition. \fpf{c} A step increase in the width of the junction force distribution within the propagation region also leads to a slow--fast transition. Top: events shown as in \fig~\ref{fig:events_with_rupture_speeds}. Middle: Spatial distributions of initial prestress $\tau_0/p$ (before event triggering) is shown with drawn line. The grey region is the triggering region (see Appendix~\ref{appsec:simulation_setup}). Bottom: histograms of junction force distributions shown as in \fig~\ref{fig:fast_slow_fast_into/from_fast_only}, taken after the triggering of the event, at $t=\unit{0.0207}{\second}$.\label{fig:phase_diagram_local_changes}}
\end{figure}

We also wish to note that the front type phase diagram is not a local measure. For example, as mentioned above, many of the events that populate the slow front region are of the fast--slow--fast type, meaning that even though the prestress and junction force distributions are the same for all the blocks in the propagation region, the rupture passes some of them as a slow front and others as a fast front. However, if these limitations are kept in mind, the diagram is still a powerful tool for guiding our intuition. For example, in an interface where there is a sudden change of initial conditions within the propagation region, the phase diagram tells us what change to expect in the rupture front. In \fig~\ref{fig:phase_diagram_local_changes} we show three simulations. The reference simulation (\fig~\ref{fig:phase_diagram_local_changes}\fpt{a}) for this figure is the simulation behind the point at $(0.088,\,0.0875)$ in \fig~\ref{fig:phase_diagram}. The two other simulations show that when there is a sudden increase of prestress or junction force distribution width in the propagation region, this can lead to a transition from slow to fast front propagation right at the location where the change occurs. Conversely, some (but not all) simulations with a change of opposite sign (from a high to a low prestress, for example), lead to a fast--slow transition.

\section{Front speed results\label{sec:front_speed_results}}
The rupture front speed is the speed at which the slipping region grows into (invades) the region which is still stuck. In the previous section we distinguished slow rupture and fast rupture and found that they are governed by different mechanisms. In this section we go on to consider the difference in speed between two events that are either both fast, or both slow.

In the language of fracture mechanics, rupture occurs when the energy available at the front tip reaches the energy required to break the contacts there. Thus, the speed at which rupture propagates depends on the energy that is already present (related to the stress state), the energy level needed to break the contacts (related to the local strength) and on how quickly the missing energy can be supplied (which depends on the slip motion behind the tip and is therefore transient). Although it has been shown that fracture mechanics satisfactorily describes fast shear rupture events \citep{Svetlizky2014classical,Kammer2014linear}, in the model, it is more straightforward to argue in terms of forces rather than energies; however, we will see that considerations similar to those in fracture mechanics apply.

\subsection{Front speed is transient\label{sec:front_speed_is_transient}}
If the front speed did not have transients, or if the transients were short compared to the length scale at which we study the front propagation, the rupture front speed would be uniquely determined by the local state of the interface at the rupture tip, i.e. local stresses, strength, stiffnesses etc. Indeed, various local interfacial parameters have been shown to be correlated with the local front speed (see \eg \citep{Ben-David2010dynamics, Tromborg2011transition, Kammer2012propagation}). However, our simulations show that the speed at any point depends not only on the state at that point, but also on the region the front has just passed through, that is, the front speed has transients. Near the trailing edge, within the length required for the speed of the newly nucleated front to converge, the region behind the front that influences the propagation extends back to the region where the front was triggered, and so in addition to the state of the interface, the precise way rupture starts also is part of what determines front speed.

\begin{figure}
\centering
\includegraphics[width=\columnwidth]{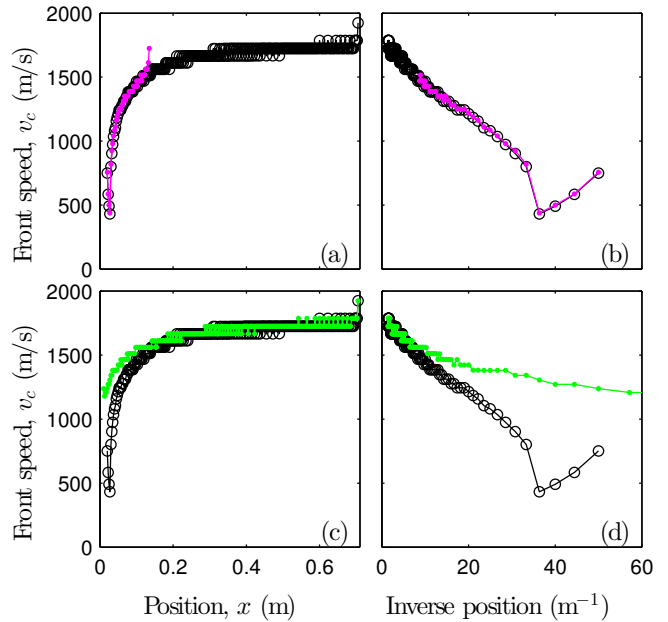}
\caption{\coloronlineornothing Results for transient behavior of fast rupture fronts. Front speed vs position \fpf{a} and inverse position \fpf{b} for simulations where the local state is homogeneous along the propagation region. The magenta/grey line with dotted markers is for a system of our reference length, the black line with circular markers for a system five times longer. \fpf{c} and \fpf{d} Difference in front speed for two systems of the same length (five times reference length) and the same initial state. Black line with circular markers: the same data as the black lines in \fpl{a} and \fpl{b}, which was triggered by simultaneously breaking all junctions for all blocks in the triggering region as explained in Appendix~\ref{appsec:simulation_setup}. Green/grey line with dotted markers: front triggered by driving from the trailing edge, which modifies the stress in the loading region prior to rupture. In all panels, rupture velocity is defined as in \fig~\ref{fig:events_with_rupture_speeds}. In \fpl{b} and \fpl{d}, the speed for the ten blocks closest to the leading edge are excluded from the plot, to avoid the region where the edge effects dominate.\label{fig:front_speed_transients}}
\end{figure}

To illustrate the transient nature of the front speed, \fig~\ref{fig:front_speed_transients}\fpt{a} shows the front speed as a function of position for two simulations in which the local stress state and local frictional strength are homogeneous along the propagation region (see details about \AMSa{how} these single event simulations \AMSa{were set up} in Appendix~\ref{appsec:simulation_setup}). We first consider the shorter of the two systems (magenta dotted line). We observe that the front speed changes throughout, that is, even though the local state is the same along the slider, the front speed is not. The length of this system is $N_x=57$, $L_x = \unit{140}{\milli\meter}$, and the front is in the transient part of propagation throughout. Note that this is a fast--only front; even though the front speed starts low, its propagation does not depend on the slow front mechanism.

To investigate the convergence length of the transients and the spatial extent over which edge effects dominate we performed another simulation. \AMSa{This simulation} differs only in the total length of the sample, which we increased by a factor of five. This gives the black line with circular markers in \fig~\ref{fig:front_speed_transients}\fpt{a}. We observe that the two rupture events have very similar front speeds. When the front tip approaches the leading edge, there is a change in curvature that we interpret as an edge effect. This interpretation is supported by the fact that the effect is present near the leading edge of the sample \AMSa{for both the short and the long system}, but \AMSa{the transient effect observed at the end of the short sample, around blocks 50--57,} is \AMSa{not observed} at blocks 50--57 in the longer sample, where these blocks are far from the edge.

\Fig~\ref{fig:front_speed_transients}\fpt{b}, where front speed is plotted against inverse position along the interface, shows that the front speed does converge to a finite value. Note that we do not expect a $1/x$ behaviour to hold in general; rather, we plot front speed versus inverse position because it is the simplest function that brings very large $x$ values close to the ordinate axis, and thus facilitates a crude extrapolation to the expected front speed reached in very long samples.

\Fig~\ref{fig:front_speed_transients}\fpt{c} and \fpt{d} also indicate convergence. They shows two events. The local stresses and strengths are the same in both, but the initialization of the events is different. This leads to front speeds that are initially also different, but which converge to the same value.

\subsection{Front speed depends on local stress state\label{sec:front_speed_depends_on_local_stress_state}}

\begin{figure}
\includegraphics[width=\columnwidth]{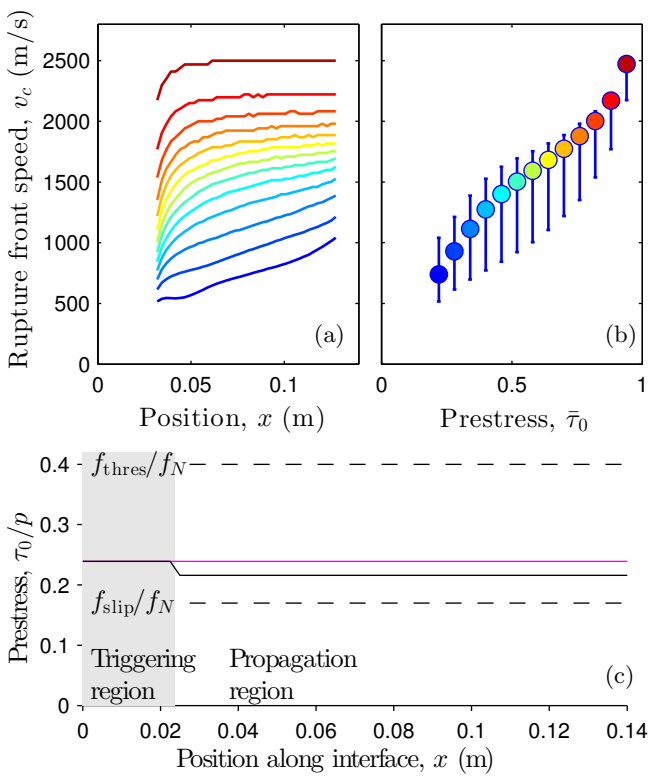}%
\caption{\coloronlineornothing Rupture front speed depends on local stress state. \fpf{a} Rupture front speed vs position along the interface for a series of simulations with various homogeneous initial shear stresses in the propagation region. The junction distribution width was zero, so these simulations would plot on top of the vertical axis in \fig~\ref{fig:phase_diagram}. They are all of the fast--only type. The transient nature of the fronts discussed in Section~\ref{sec:front_speed_is_transient} can be seen here as well, in the increase in front speed towards the right of the sample. \fpf{b} The front velocity data in \fpl{a} vs the prestress value for each simulation. Dots: mean values. Bars: min/max values. Each dot+bar corresponds to the line in \fpl{a} with the same color / grey scale + vertical extent. The prestress values are 0.22 to 0.94 in steps of 0.06. \fpf{c} Spatial distributions of prestress $\tau_0/p$. More details in the text and in Appendix~\ref{appsec:simulation_setup}. For these simulations, a time step of $\Delta t=\unit{5\e{-8}}{\second}$ was used to obtain smooth measurements at high front speeds. \pnasreusepanel{\fpl{c}}\label{fig:fast_front_speed_vs_prestress_width_0}}
\end{figure}

While we and others \citep{Kammer2012propagation} argue that the ratio $\tau_0/p$ of shear to normal stress at the interface as the rupture front begins \AMSa{provides} insufficient information for predicting the front speed \AMSa{on its own}, it remains true \AMSa{with}in the model, as is also seen in experiments \citep{Ben-David2010dynamics}, that the front speed has a strong dependence on $\tau_0/p$. Higher prestress $\tau_0$ results in higher front propagation speed. This result agrees with our intuition: for a given strength, a region that is more highly prestressed requires less stress change to start slipping, and it releases more energy when doing so. Both favor higher front speed. This is also consistent with \fig~\ref{fig:phase_diagram}, which showed that increasing the initial shear stress in the propagation region will bring a system into a regime where fronts are fast across the entire interface.

\Fig~\ref{fig:fast_front_speed_vs_prestress_width_0}\fpt{a} shows the front speed for events in which the initial stress state was varied systematically. A similar variation in the junction force distributions' width $\sigma$ is the topic of Section~\ref{sec:front_speed_depends_on_local_strength}. To isolate the effect of varying prestress from other variations that arise in the full simulations we used the protocol illustrated in \fig~\ref{fig:fast_front_speed_vs_prestress_width_0}\fpt{c}, which shows the initial shear and normal stress states. The normal stress $p$ is homogeneous, \ie taking the same value everywhere along the interface. The shear stress is also homogeneous within each of two regions. The region on the left is called the triggering region, because we start each event by simultaneously and artificially detaching all junctions for the blocks within this region. This ensures that the starting conditions are the same for all the fronts. We always use the same shear stress level in this region, $\bar\tau_0=(\tau_0/p-\fslip/f_N)/(\fthres/f_N-\fslip/f_N)=0.3$. In the rest of the interface, which we call the propagation region, the initial shear stress is also uniform, but not necessarily the same as in the triggering region. For clarity only two prestress values are shown in the propagation region of \fig~\ref{fig:fast_front_speed_vs_prestress_width_0}\fpt{c}.

In agreement with our argument above we observe in \fig~\ref{fig:fast_front_speed_vs_prestress_width_0}\fpt{a} that higher prestress (dark red line) corresponds to higher mean front velocities.

Note that in the model the longitudinal wave speed $c_L=\unit{1677}{\meter\per\second}$ does not set an upper limit for the front propagation speed. This is consistent with the results of other experimental (see \eg \citep{Coker2003dynamic}) and numerical (see \eg \citep{Kammer2012propagation}) studies of bi-material interfaces, in which the maximum front speed is expected to be the one of the stiffest material. Here the track is perfectly rigid in the front propagation direction, so that we do not expect any conceptual limitation to the front speed.

\subsection{Front speed depends on local strength\label{sec:front_speed_depends_on_local_strength}}
In the same way as for the influence of stress state on front speed discussed above, it makes intuitive sense that for a given prestress, the front propagates faster if the frictional strength, \ie the threshold stress for slip inception, is lower. In the model, the frictional strength is controlled to first order by the values of $\fthres$ and $\fslip$ on the junction level, but even when these are the same everywhere, the local strength can vary. We show in \fig~\ref{fig:mu_s^eff_analytical_and_numerical} (Appendix~\ref{appsec:mus_from_width}) how increasing the width, $\sigma$, of the initial junction force distribution, $\phi(f_T)$, results in a lower effective static friction threshold, and we expect that if conditions are otherwise the same, this will result in faster front propagation. In the case (not studied here) of a more complete model, variations in local strength would also be due to variations in individual strength of the junctions attached to each block.

\begin{figure}
\centering
\includegraphics[width=\columnwidth]{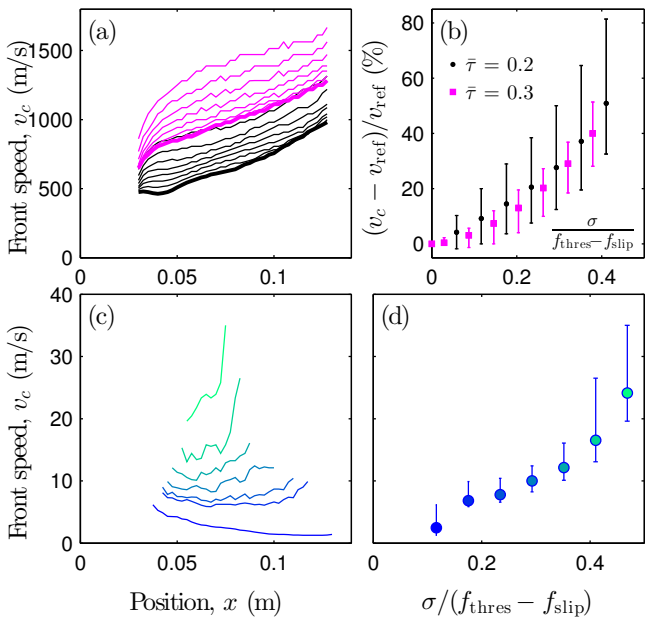}%
\caption{\coloronlineornothing Rupture front speed depends on local strength. \fpf{a} Rupture front speed vs position along the interface for a series of simulations in which the distribution of junction forces was varied systematically and two values of shear prestress were used. All of these events are of the fast--only type. The magenta (grey) and black lines have the shear stress values listed in the legend of \fpl{b}, these also correspond to the magenta (grey) and black lines in \fig~\ref{fig:fast_front_speed_vs_prestress_width_0}\fpt{c}. The width $\sigma$ of the junction force distribution is increasing from the bottom to top line of each color; the values can be found in \fpl{b}. For each value of junction force distribution width, the front propagation speed is higher when the prestress is higher, that is, each magenta (grey) line is higher than the corresponding black line. \fpf{b} Fast front speed data from \fpl{a} vs junction force distribution width. For this dataset we define, for each prestress series, the front speed at $\sigma=0$ as the reference speed $v_\text{ref}$. Then, for each simulation we calculate, for every position, $(v_c(x)-v_\text{ref}(x))/v_\text{ref}(x)$. The dots are the mean of these values, the bars show the min/max values. The values on the $\sigma/(\fthres-\fslip)$ axis range from $0$ to $0.41$ in 14 equal steps of $0.029$. \fpf{c} Slow front speed vs junction force distribution width. The region over which the slow front extends was determined from visual inspection of the underlying data, which is shown in \fig~\ref{fig:135x}. The prestress is $\bar\tau_0=0.3$ in the triggering region and $\bar\tau_0=0.05$ in the propagation region; the junction distribution widths can be found in \fpl{d}. \fpf{d} The slow front speed data in \fpl{c} vs junction force distribution width. The same colors/grey levels are used in \fpl{c} and \fpl{d}. Dots: mean values. Bars: min/max values. The data values on the $\sigma/(\fthres-\fslip)$ axis range from $0.12$ to $0.47$ in 6 equal steps of $0.059$.\label{fig:fast_and_slow_front_variation_vs_width}}
\end{figure}

We observe in \fig~\ref{fig:fast_and_slow_front_variation_vs_width}\fpt{a} that for the same prestress, higher $\sigma$ result\AMSa{s} in higher front speed. Within our sample size, which is the same as in previous sections and was selected from experimental parameters, the front speed is changing throughout the front propagation. In order to investigate the effect of $\sigma$ quantitatively and to isolate it from the effect of prestress, we compare each event to a reference event having the same prestress state and with width $\sigma=0$. We measure along the entire interface and find average and min/max values. The results are presented in \fig~\ref{fig:fast_and_slow_front_variation_vs_width}\fpt{b}. By rescaling the front speeds with the reference speed for $\sigma=0$ the similarity between the two series for different prestress becomes apparent. The range of results for each value of $\sigma$ due to the transient nature of the front speed is large, but there is a clear trend \AMSa{that} higher $\sigma$ (weaker interfaces) \AMSa{correspond} to higher front speeds. The effect of $\sigma$ on speed can be very significant, \AMSa{giving} up to a 50\% relative increase for the data shown in \fig~\ref{fig:fast_and_slow_front_variation_vs_width}\fpt{b}.

\Fig~\ref{fig:fast_and_slow_front_variation_vs_width}\fpt{a} and \fpt{b} show data for fast--only fronts. The influence of $\sigma$ on front speed is qualitatively similar for slow fronts. This is shown in \fig~\ref{fig:fast_and_slow_front_variation_vs_width}\fpt{c} and \fpt{d}. For the fronts, which have both fast and slow propagation parts, we used visual inspection of the underlying data in Appendix~\ref{appsec:slow_front_variation_with_sigma_underlying_data} to determine which part of the fronts to include. Our selection can be read off of the horizontal extent of the lines in \fig~\ref{fig:fast_and_slow_front_variation_vs_width}\fpt{c}. Again, the effect of $\sigma$ on the front speed is significant, with an increase of about one order of magnitude when $\sigma/(\fthres-\fslip)$ is increased from 0.12 to 0.47.

\subsection{Front speed is proportional to slip speed\label{sec:front_speed_vs_slip_speed}}
The motion of the rupture front tip and the motion of the material of the slider are interrelated: the slider cannot move while the interface is in the pinned state (the slip depends on the front), and the rupture propagates as deformation of the slider transfers stress and energy to the front tip (the front depends on the slip). In \citep{Tromborg2014slow} we demonstrated that the slow front speed in the model is proportional to the slow slip speed and worked out the constant of proportionality. In this section we show that the fast front speed in the model depends on the fast slip speed through the exact same relationship.

\begin{figure}
\centering
\includegraphics[width=\columnwidth]{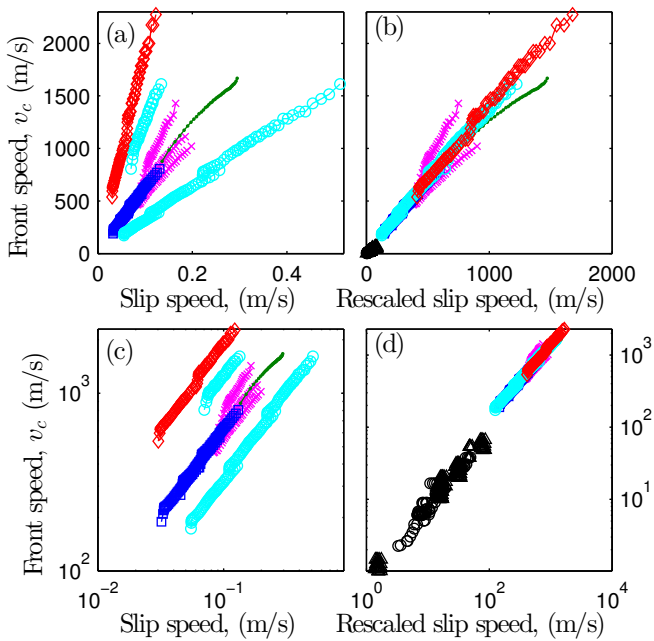}%
\caption{\coloronlineornothing Front speed is proportional to slip speed. \fpf{a} Front speed vs slip speed from simulations with controlled initial states in the regime that produces fast fronts. Taking the simulation that plots at $(0.23,\,0.2)$ in \fig~\ref{fig:phase_diagram} as an arbitrary reference, we have varied the junction force distribution $\phi(f_T)$ (magenta crosses), the prestress $\bar\tau_0$ (green dots), the mass density $\rho$ (blue squares), the bulk and interfacial stiffnesses (keeping $k/k_i$ constant) and $\rho$ (cyan circles), and the normal force $F_N$ and $\rho$ (red diamonds). \fpf{b} The data in \fpl{a} with slip speed rescaled as $v_\text{slip, rescaled} = v_\text{slip}\frac{k_il_0}{\tau_\text{thres}-\tau_0}$. The black markers near the origin show the data for slow slip and slow fronts from \citep[Fig. 3D]{Tromborg2014slow}. \fpf{c} and \fpf{d} The data in \fpl{a} and \fpl{b}, respectively, on logarithmic axes.\label{fig:fast_front_vs_fast_slip}}
\end{figure}

\Fig~\ref{fig:fast_front_vs_fast_slip}\fpt{a} shows the fast front speed vs fast slip speed for the blocks between $x=\unit{3}{\centi\meter}$ and $x=\unit{13}{\centi\meter}$ in a series of simulations with controlled initial states and variation in mass density, normal force, junction distribution width, prestress and bulk and interface stiffnesses (\fig~\ref{fig:fast_front_vs_fast_slip}\fpt{c} shows the same data on logarithmic axes). These blocks are all in the propagation region and away from the triggering zone and the leading edge. The front speed at each point along the interface is measured in the same way as in \fig~\ref{fig:events_with_rupture_speeds}. The corresponding slip speed is the average block slip speed measured from the time the block, $i$, starts slipping to the time when the block $\unit{5}{\milli\meter}$ away (block $i+2$) starts slipping. We choose this time interval because it is during this period that the motion of block $i$ most directly affects the rupture front propagation.

\Fig~\ref{fig:fast_front_vs_fast_slip}\fpt{b} shows the data in \fig~\ref{fig:fast_front_vs_fast_slip}\fpt{a} with the fast slip speed in each simulation rescaled according to the parameters used (\fig~\ref{fig:fast_front_vs_fast_slip}\fpt{d} shows the same data on logarithmic axes). We also include (in black) the slow front and slow slip data from \citep[Fig. 3D]{Tromborg2014slow}. Arguments for the scaling were presented in \citep[SI Equations]{Tromborg2014slow}; we repeat only the conclusions here. The rescaled slip speed is $v_\text{slip, rescaled} = v_\text{slip}\frac{k_il_0}{\tau_\text{thres}-\tau_0}$, where $k_i=\sum_j k_{ij}$ is the stiffness of the connection between a block and the interface, $l_0=\unit{7}{\milli\meter}$ is the characteristic decay length of the shear stress field, which depends on the bulk to interfacial stiffness ratio $k/k_i$, $\tau_\text{thres} = N_s\fthres$, and $\tau_0$ is the initial shear stress before the event. The normal force enters in the scaling indirectly because it modifies $\fthres$. We expect deviations from the rescaling due to two sources that have not been included in the scaling equation. First, the effective force threshold $\tau_\text{thres}^\text{eff}$ on a block with a given junction force distribution is always less than $\tau_\text{thres}$, as discussed in Appendix~\ref{appsec:mus_from_width}. The data where the junction force distribution was varied collapses better when $\tau_\text{thres}^\text{eff}$ replaces $\tau_\text{thres}$ in the scaling relation. However, because $\tau_\text{thres}^\text{eff}$ depends on the microscopic state of the interface, we choose to omit this correction in order to keep the factors in the scaling equation at the mesoscopic level. Second, the shape of the shear stress field also appears in the argument for the scaling, but not in our final scaling equation, which only includes the characteristic stress decay length $l_0$; we have kept its variation between simulations to a minimum by keeping the ratio $k/k_i$ constant.

\Fig~\ref{fig:fast_front_vs_fast_slip}\fpt{b} shows that the above rescaling allows all data from \fig~\ref{fig:fast_front_vs_fast_slip}\fpt{a} to nicely collapse on a single straight line, going through the origin and having a slope very close to 1. This demonstrates that the front speed is directly proportional to the concurrent slip speed, with the proportionality coefficient being $\frac{k_il_0}{\tau_\text{thres}-\tau_0}$. Strikingly, the data for slow slip and slow fronts collapse on the very same line, as clearly seen in \fig~\ref{fig:fast_front_vs_fast_slip}\fpt{d}. We return to this result in the discussion, Section~\ref{sec:discussion}. 

\subsection{Front speed compares well to experiments\label{sec:front_speed_compares_well_to_experiments}}
So far in Section~\ref{sec:front_speed_results} we have sought to isolate the influence on front speed of transients, prestress and interface strength from each other and from the event triggering. We now combine these results in a form suitable for comparison with experiments.

\begin{figure}
\includegraphics[width=\columnwidth]{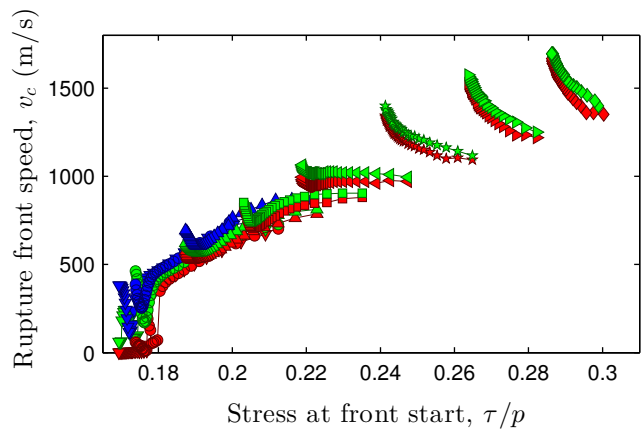}%
\caption{\coloronlineornothing Rupture front speed vs prestress for several events where the prestress and the junction force distribution have been varied systematically. Different colors correspond to different widths of the junction force distribution: $\sigma/(\fthres-\fslip) = 0.014$ (red), $0.143$ (green), $0.286$ (blue); the bell-shaped distributions defined in \fig~\ref{fig:mu_s^eff_analytical_and_numerical} were used. Different markers correspond to different initial levels of prestress (before additional loading was applied): $\bar\tau=-0.02$ (\protect\marksymbol{triangle*}{black}{180}{1.4}), $0.0$ (\protect\marksymbol{*}{black}{0}{1}), $0.06$ (\protect\marksymbol{triangle*}{black}{0}{1.4}), $0.13$ (\protect\marksymbol{square*}{black}{0}{1}), $0.2$ (\protect\marksymbol{triangle*}{black}{90}{1.4}), $0.3$ ($\bigstar$), $0.4$ (\protect\marksymbol{triangle*}{black}{270}{1.4}), $0.5$ (\protect\marksymbol{diamond*}{black}{0}{1.4}). In terms of $\tau/p=\fslip/f_N+\bar\tau(\fthres/f_N-\fslip/f_N)$, these prestresses are 0.165, 0.170, 0.184, 0.200, 0.216, 0.239, 0.262 and 0.285, respectively. Edge effects, which can dominate visually, have been excluded by showing only the region between $x=\unit{2.5}{\centi\meter}$ and $x=\unit{11}{\centi\metre}$. Prestress is measured for all blocks when the first block starts slipping (as defined from the number of pinned junctions). To get a sense for the range of values on the $\tau/p$ axis, which differs from the ones in \citep{Ben-David2010dynamics}, consider that values much below $\fslip/f_N=0.17$ lead to arresting fronts and that the maximum possible value of $\fthres/f_N=0.4$ is reached only for narrow junction force distributions and near the loading point, so it is excluded by excluding the edges of the system. Recall from \fig~\ref{fig:phase_diagram} that combinations of high prestress and high $\sigma$ are forbidden: this is the reason why there are fewer blue entries. For these simulations, a time step of $\Delta t=\unit{5\e{-8}}{\second}$ was used to obtain smooth measurements at high front speeds.\label{fig:v_front_vs_tau_over_p}}
\end{figure}

\Fig~\ref{fig:v_front_vs_tau_over_p} shows the rupture front speed vs prestress ratio $\tau/p$ for several events where the shear prestress and the junction force distributions were varied systematically. In order to both enable the tuning of prestress and force distributions and to mimic the experimental driving conditions more closely, as in the full simulations, initial conditions were prepared in two steps. First, homogeneous distributions and prestresses were assigned. Then, the trailing edge driving spring was allowed to move, eventually triggering the front. This raises the prestress most prominently near the trailing edge, but for the present parameters (\AMSa{which} are \AMSa{found from} comparison with experiments) the ratio of shear stress profile decay length, $l_0$, to system length is such that the shear stress is non-negligibly changed everywhere. For each event, rupture speed is measured at all points between $x=\unit{2.5}{\centi\metre}$ and $x=\unit{11}{\centi\metre}$, \ie avoiding the immediate vicinity of the system edges.

\Fig~\ref{fig:v_front_vs_tau_over_p} is analogous to Fig.~3 in \citet{Ben-David2010dynamics}, which shows the rupture front speed in selected points away from the sample edge vs prestress. Both figures show the same trend, namely that the front speed increases with increasing prestress, and that there exists a continuum of front speeds ranging from close to zero (slow) up to super-shear wave speeds. In addition, the numerical data indicates a systematic effect of varying the width of the junction force distribution, in a way which is completely consistent with the observations reported in section \ref{sec:front_speed_depends_on_local_strength}: larger widths yield lower local strength and thus promote faster fronts.

\section{History dependence of mesoscopic static friction coefficient\label{sec:history_dependence_with_gini}}
In the previous section we argued that the rupture front speed depends on the local strength, i.e. the effective static friction threshold on the block level. Here we investigate the extent to which variations in the mesoscale strength occur even when the individual \JSa{micro-junctions attached to the block all have} the same strength. This effect comes in addition to \JSa{spatial heterogeneities} in stresses, due for instance to previous ruptures of the interface, which also cause \JSa{spatial} variations in the mesoscale strength.

We showed in \citep{Thogersen2014history-dependent} (and \JSa{recall} in Appendix~\ref{appsec:mus_from_width}) how the local strength depends on the distribution of forces among the microjunctions of a block. We also showed how, for simple velocity profiles, the distribution of forces at \JSa{block} arrest depends on the deceleration \JSa{that led to this arrest}. More fundamentally (in the model), the distribution of forces at arrest depends on the slip profile $x(t)$ of the block during the re-pinning of \JSa{its} junctions. As simplified velocity (and thus slip) profiles were studied in \citep{Thogersen2014history-dependent}, we focus here on the more complicated spontaneously occurring events in the full simulations. We are primarily interested in the effective static friction threshold \JSa{$\mu_s^\text{eff}=\taumax/p$} after an event, where $\taumax$ is the largest tangential force that the block will bear before it starts to slip in the next event. \JSa{$\mu_s^\text{eff}$} can be calculated directly from the junction force distribution if we assume that the rupture is fast compared to the mean re-pinning time. As we will show, however, $\mu_s^\text{eff}$ can also be estimated from the preceding slip dynamics.

This section has four figures where the first three serve to introduce new concepts and definitions and the fourth is the main result of the section. \Fig~\ref{fig:gini_sketch} defines the Gini coefficient, which we will use as a scalar, integrated quantifier of complicated block slip dynamics. \Fig~\ref{fig:gini_theory} provides a link to \citep[Section~IV]{Thogersen2014history-dependent} in using the same constant deceleration slip profiles as were used there. These will also serve as the basis for our prediction of effective static friction as a function of Gini coefficient. \Fig~\ref{fig:gini_simulation_slip_profiles} shows example slip data from full simulations and how we measure the Gini coefficient from these data. Finally, \fig~\ref{fig:mu_s_eff_vs_gini} shows that the Gini coefficient is a good predictor of effective static friction.


\subsection{Gini coefficient: definition and measurement\label{sec:introduce_gini}}
A feature of the microscopic junction law that we also studied in \citep{Thogersen2014history-dependent} is how the junction force distribution $\phi(f_T)$ of a block evolves with the slip history of the block. As the block starts moving, \JSa{slips} and comes to rest, junctions are continually breaking and reforming, and the complete distribution $\phi(f_T)$ depends on the full slip velocity history $v(t)$. In practice, however, it is the very end of the preceding event that has the largest influence on $\phi$, and this allows us to predict the strength of $\phi$ after an event (which is the effective static friction threshold for the next event) from a characterization of the block slip motion.

First, we pick a time $t'$ between two full sliding events at which we wish to predict the strength of $\phi$. The blocks should be at rest at this point in time. Second, we identify the segment \JSa{$[t_0,t_1]$} of the slip history that needs to be considered to predict the strength of $\phi$ at the chosen time. Following each block's motion $x(t)$ backwards from $t'$, we define $t_0$ as the time when the block was one junction breaking length \JSa{$\smax=\fthres/k_{ij}$} away from where it is at $t'$; $x(t')-x(t_0)=\smax$. The slip history before $t_0$ can be neglected because all the junction attachment points that existed before $t_0$ have been broken and renewed during $t\in[t_0,t']$. Further, we define $t_1=t_0+\bar t_R+\delta t_R$, so that $[t_0,t_1]$ includes most of the junction renewal occurring after $t_0$. We require $t_1<t'$, but as long as the blocks remain at rest, any $t_1$ that satisfies this constraint can be chosen. It is possible to create pathological cases where $[t_0,t_1]$ fails to include the bulk of re-pinning events that set up $\phi(f_T)$ at $t'$, but in practice, in all our simulations this is the most important segment of the slip history. Third, we characterize $v(t)$ for $t\in[t_0,t_1]$ and compare it to \JSa{velocity} profiles that can be treated analytically or with simple numerical integration.

The feature of $v(t)$ that has the most prominent impact on $\phi$ is how evenly the slip motion $x(t)$ is distributed in time. If the block comes fully or nearly fully to rest during a time interval much shorter than $\bar t_R$, most of the junctions will be renewed while the block is at rest; they will therefore have approximately the same stretching and $\phi(f_T)$ will be narrow. If the block moves more uniformly as it comes to rest, the stretching of the junctions will distribute more uniformly. This simple argument assumes that the rate of junction renewal is constant, a simplification that is a reasonable approximation when the block comes to rest after \JSa{significant slip motion} \citep{Thogersen2014history-dependent}.

\newcommand{\Gini}{\mathcal{G}}
\newcommand{\GA}{\mathcal{A}}
\newcommand{\GB}{\mathcal{B}}
\begin{figure}
\centering
\includegraphics{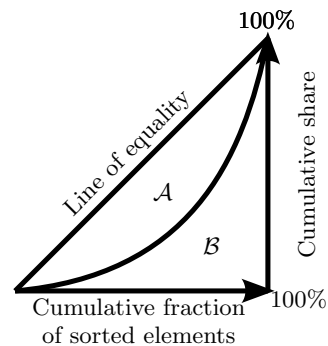}%
\caption{The areas that define the Gini coefficient, a measure of inequality.\label{fig:gini_sketch}}
\end{figure}

A robust measure of inequality is the Gini coefficient, which we introduce in \fig~\ref{fig:gini_sketch}. It is commonly used to characterize the inequality of the income or wealth distribution in a population \citep{Giorgi1990bibliographic,Giorgi1999income}, but can be applied to our case without modification. The Gini coefficient of a set is defined from the areas in \fig~\ref{fig:gini_sketch} as $\Gini=\GA/(\GA+\GB)$ (which equals $2\GA$, since $\GA+\GB=1/2$). \JSa{For any set of elements, each having a share of a some quantity (\eg wealth), the line separating $\GA$ and $\GB$ is constructed as follows. Elements are first ordered by the size of their share and placed along the abscissa. The ordinate for a given share then corresponds to the sum of all shares that are smaller than the chosen share. The line thus defines the cumulative distribution of shares.} $\GB$ is the area under the cumulative and $\GA$ is the area between the cumulative and the line of equality.
The line of equality is the cumulative for a set where each element has the same share. Thus, if each element has the same share, $\Gini=0$, while in the other extreme where one element has everything and the other elements have zero share, $\Gini=1$. To measure the Gini coefficient of the block slip during $t\in[t_0,t_1]$ we create a set consisting of $N_e=10$ elements. We define $\Delta t_\text{elem}=(t_1-t_0)/N_e$ and assign $x_\text{share}$ to each element so that the elements tile the time interval. That is, the share of the first element is the block slip that occurred between $t_0$ and $t_0+\Delta t_\text{elem}$, the share of the second element is the block slip that occurred during $[t_0+\Delta t_\text{elem},t_0+2\Delta t_\text{elem}]$, and so on. Equivalently, each element's share is the average slip velocity during the element's time interval (net slip equals average slip velocity times the length of the time interval, but \JSa{the} common normalization factor \JSa{$\Delta t_\text{elem}$} can be ignored).

\begin{figure}
\centering
\includegraphics[width=\columnwidth]{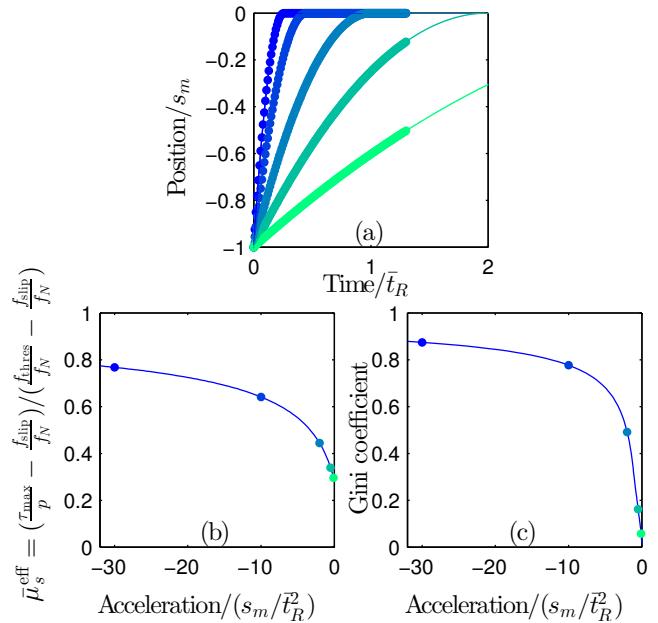}%
\caption{\coloronlineornothing \fpf{a} Slip profiles $x(t)$ for constant deceleration slip. The markers extend over the interval $[t_0,t_1]$. \fpf{b} \JSa{Rescaled} effective static friction \JSa{$\bar\mu_s^\text{eff}$} for the slip profiles in \fpl{a} and the junction parameters in \tab~\ref{tab:parameters}. \fpf{c} Gini coefficients of the slip profiles in \fpl{a}. The line extends to the origin. The markers in \fpl{b} and \fpl{c} show the data from the slip profiles in \fpl{a} of the corresponding color/grayscale; the drawn lines are based on a denser set of slip profiles that were omitted from panel \fpl{a} for clarity. Combining \fpl{b} and \fpl{c} gives the drawn blue line with circular markers in \fig~\ref{fig:mu_s_eff_vs_gini}.\label{fig:gini_theory}}
\end{figure}

\Fig~\ref{fig:gini_theory}\fpt{a} shows block slip profiles $x(t)$ for constant deceleration slip. These slip profiles have the advantage that they \JSa{are completely} characterized by the deceleration amplitude; they help build intuition before we consider the full simulation data, where none of the basic kinematic quantities (slip, velocity, acceleration) are even approximately constant. In \fig~\ref{fig:gini_theory}\fpt{a} the time interval has been shifted so that $t_0=0$. With the parameters in \tab~\ref{tab:parameters}, $(t_1-t_0)/\bar t_R=1.3$. \Fig~\ref{fig:gini_theory}\fpt{b} shows the effective static friction, i.e. the strength of $\phi(f_T)$, resulting from the slip profiles in \fig~\ref{fig:gini_theory}\fpt{a}. We showed in \citep[Fig.~9]{Thogersen2014history-dependent} that for \JSa{(i)} initial velocities large enough that the blocks move longer than the junction breaking length $\smax$ before coming to rest, and \JSa{(ii)} reasonable initial junction distributions, the effective static friction for constant deceleration slip depends only on the deceleration value and the junction law parameters; here we use the same initial conditions as in \citep[Fig.~9]{Thogersen2014history-dependent} and the junction law parameters in \tab~\ref{tab:parameters}. \Fig~\ref{fig:gini_theory}\fpt{c} shows the Gini coefficient of the slip profiles in \fig~\ref{fig:gini_theory}\fpt{a} as a function of their constant deceleration parameter (Gini coefficients were calculated using \citep{Komarov2010gini}). The shape of the curve can be intuitively understood. For deceleration close to zero, the slip motion from $t_0$ to $t_1$ occurs with low and nearly constant speed, so the \JSa{cumulative} is close to the line of equality, and $\Gini\approx0$. For large deceleration values, the block comes to rest sharply, $[t_0,t_1]$ includes a long period where the block is at rest\JSa{, and the cumulative is more peaked}. In the limit of very large deceleration values, $\Gini\rightarrow1$. We can \JSa{combine \fig~\ref{fig:gini_theory}\fpt{b}} and \fig~\ref{fig:gini_theory}\fpt{c} to obtain a prediction for \JSa{$\mu_s^\text{eff}$} vs $\Gini$. This gives the full drawn lines in \fig~\ref{fig:mu_s_eff_vs_gini}.

\begin{figure}
\centering
\includegraphics[width=\columnwidth]{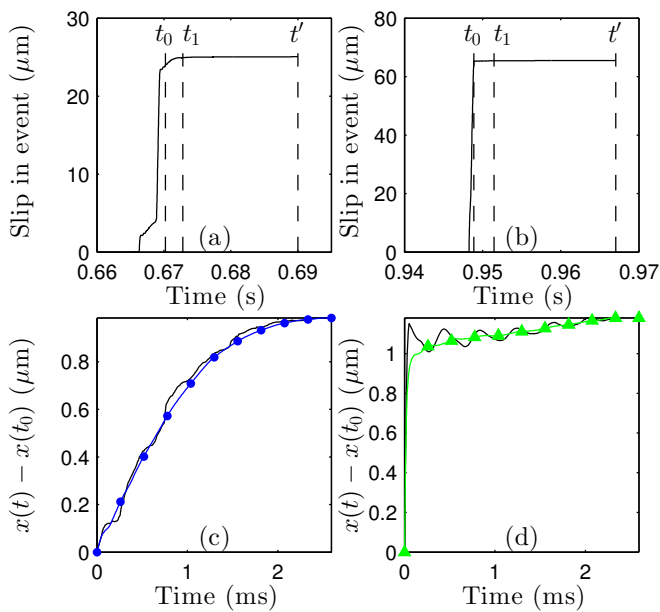}%
\caption{\coloronlineornothing Example slip profiles from full simulations. \fpf{a} and \fpf{b} Slip profiles for full sliding events. Only the slip occurring between $t_0$ and $t_1$ (defined in the text) are used to calculate the Gini coefficient. Effective static friction is measured from the junction force distribution $\phi(f_T)$ at $t'$. The data in \fpl{a} are \JSa{for the block at} $x=\unit{4}{\centi\meter}$ in Event~I, while the data in \fpl{b} are from a simulation with $\fslip/f_N=0.05$. \fpf{c} and \fpf{d} Detailed views of the slip profiles in \fpl{a} and \fpl{b}, respectively\JSa{, for the time interval between $t_0$ and $t_1$}. Black line without markers: $x(t)$. Colored line with markers: a moving average filter of width $\unit{1}{\milli\second}$ was applied to $x(t)$ and then $N_e=10$ equally sized intervals were defined \JSa{(markers indicate intervals' boundaries)}. The smoothing helps avoid negative increments that while possible to include are not part of the basic formulation of the Gini coefficient.\label{fig:gini_simulation_slip_profiles}}
\end{figure}

\Fig~\ref{fig:gini_simulation_slip_profiles} shows example slip profiles from full sliding events in full simulations. Two events are shown, with the data in \fpt{c} being a detailed view of the data in \fpt{a} and the data in \fpt{d} being a detailed view of the data in \fpt{b}.

\subsection{Gini coefficient predicts effective static friction\label{sec:gini_results}}
We are now ready to test how well the Gini coefficient, an integrated property of the block slip history, predicts the effective static friction threshold. The link between the two is the distribution of junction forces that exist after a sliding event: the Gini coefficient is an averaged measure of how uniform or non-uniform the distribution becomes; the effective static friction threshold is essentially a property of the distribution (it also depends on the junction evolution law, but this remains fixed within a simulation).

For the limiting cases of a block coming to rest very abruptly or very gradually, the Gini coefficient and the effective static friction threshold can both be predicted exactly from simple arguments. The abrupt stopping case is the limit where the block comes to rest from a speed high compared to $\smax/\bar t_R$ within a distance short compared to $\smax$ and a time short compared to $\bar t_R$. In this case the entire motion between $t_0$ and $t_1$ will be assigned to a single element and the Gini coefficient will be $\Gini=1$. The distribution of junction forces will be a $\delta$-function as all junctions reform after the block has stopped moving, and so $\taumax/p=\fthres/f_N$ and the \JSa{rescaled} effective static friction threshold is $\bar\mu_s^\text{eff}=(\taumax/p-\fslip/f_N)/(\fthres/f_N-\fslip/f_N)=1$. The gradual stopping case is the limit where the block spends a long time at velocities small compared to $\smax/\bar t_R$. In this case the velocity is nearly constant from $t_0$ to $t_1$ and the Gini coefficient will be $\Gini=0$. A uniform junction force distribution is set up (see \citep[Section~III~A]{Thogersen2014history-dependent} for a detailed account). From equation~\eqref{eq:nu_s^eff_scaled} we find $\bar\mu_s^\text{eff}=(\fthres-\fslip)/(2\fthres)$.

\begin{figure}
\centering
\includegraphics[width=\columnwidth]{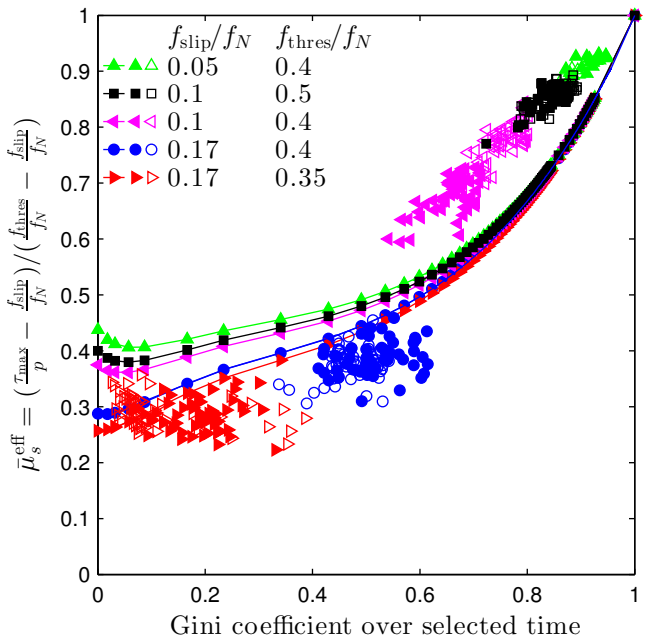}%
\caption{\coloronlineornothing \JSa{Rescaled effective} static friction after full sliding events vs Gini coefficient, an integrated quantifier of block slip motion. All data from full simulations, with variation in junction law parameters between simulations as shown in the legend. See \fig~\ref{fig:gini_theory} for an explanation of the Gini coefficient and the definition of the theoretical lines, and \fig~\ref{fig:gini_simulation_slip_profiles} for sample block slip histories. For each simulation we show data from two arbitrary full sliding events \JSa{(solid and open symbols)}.\label{fig:mu_s_eff_vs_gini}}
\end{figure}

As a slipping block slows down the slow slip mechanism becomes important. If the amount of slow slip is large compared to the junction breaking length $\smax$, the motion is always far from the abrupt stopping limit. Consequently, to span out the range of Gini coefficients from $0$ to $1$ we have performed simulations where we indirectly varied the relative amplitude of slow slip and $\smax$ by varying $\fthres$ and $\fslip$. \Fig~\ref{fig:mu_s_eff_vs_gini} shows results from this series of full simulations. As expected, the variation between events within one simulation is small compared to the variation between simulations. In all cases, the Gini coefficient is a good predictor of effective static friction. \JSa{In particular, we find} near equality between $\Gini$ and \JSa{$\bar\mu_s^\text{eff}$} for $\Gini>0.5$.

The agreement between the line corresponding to constant deceleration slip and the simulation data is \JSa{rather} good, \JSa{but far from perfect, as expected given }the approximations inherent in describing the complicated block slip motion arising in the full simulations with a single number. Important differences in the underlying motion are: (i) In the full simulations the duration of sliding events is on the order of $\bar t_R$, see \fig~\ref{fig:gini_simulation_slip_profiles}\fpt{a} and \fpt{b}. To reach steady state junction distributions, \JSa{as was assumed in \citep{Thogersen2014history-dependent},} a longer sliding period is necessary. (ii) In the full simulations, the oscillations in the block velocity set up junction force distributions that are non-smooth. This is why the simulation data is not bounded by the same lower limit as the constant deceleration data.

\section{Discussion\label{sec:discussion}}
We begin this section with an overview of the robustness of the observed front dynamics to variations in model parameters. We then revisit the results and provide interpretations and comparisons to earlier work in the literature.

\subsection{Robustness\label{sec:robustness}}
To elucidate the robustness of the system dynamics to variations in the model parameters we have performed simulations where one or two parameters at a time were varied from their values in \tab~\ref{tab:parameters}. We find that most of the parameters are in a range where the results remain qualitatively the same under halving and doubling of the parameter value. Others, notably the bulk and interface compliances, are in a range where these changes lead to significant changes in the system dynamics.

As mentioned in the model description in Section~\ref{sec:model_description}, $T(t_R)$ is a simplified way of modeling the distribution of times after which \JSa{slipping} micro-junctions relax. It is therefore reassuring that the front dynamics are not sensitive to the details of $T(t_R)$: 
(i) We observe fast--slow--fast fronts in a simulation with an exponential $T(t_R)$. 
(ii) In \citep{Tromborg2014slow} we varied the average, $\langle t_R\rangle$, systematically and observed that this changes the speed, but not the nature of slow slip and slow fronts.
\JSa{(iii)} We have performed full simulations where we halved and doubled the width, $\delta t_R$, from its value in \tab~\ref{tab:parameters}. Although the details of the fronts change, we still observe precursors, fast--only events, fast--slow--fast events and intermediate partial slip events. 

The number of junctions per block, $N_s$, could be adjusted to match the area density of junctions in a given experiment. However, for $N_s\gg1$ we expect to recover the results from \citep{Thogersen2014history-dependent}, where instead of tracking individual junctions we described the state evolution in terms of \JSa{continuum} distributions of pinned and slipping junctions. \JSa{With $N_s=100$, the present simulations and those in \citep{Tromborg2014slow} appear to be within this continuum limit: when we compare to full simulations with $N_s=50$, $200$} and $500$ we again observe precursors, fast--only events, fast--slow--fast events and intermediate partial slip events, and although due to the randomness in the junction dynamics the events are not exactly the same between these simulations, they are very similar. 

There are four stiffness parameters in the model: the driving spring modulus $K$; the elastic foundation modulus $k_f$; Young's modulus $E$, from which the bulk spring stiffness $k$ follows; and the interface spring stiffness $k_{ij}$.
We have performed simulations where the driving spring modulus $K$ was increased twofold, tenfold and hundredfold from its value in \tab~\ref{tab:parameters}. This changes the relatively regular stick--slip seen in the loading curve (\fig~\ref{fig:sketch_and_loading_curves}\fpt{d}) to a more chaotic stick--slip pattern, and as predicted in Section~\ref{sec:influence_of_front_type_on_the_loading_curve} \AMSa{it} causes the amplitude of the drops in $F_T$ during partial slip events to increase. We still observe both fast--only and fast--slow--fast fronts. 
For the elastic foundation modulus, $k_f$, doubling and halving \JSa{its} value in full simulations still allows the full range of system dynamics. 
For $k_f\ll k$, the normal stress profile $p(x)$ is essentially flat, as significant variations in $p$ would then require large vertical deformations of the slider, and the experimental stress profiles of \citep{Rubinstein2007dynamics, Ben-David2010dynamics} will no longer be reproduced.

For the bulk and interface stiffnesses $k$ and $k_{ij}$ we have shown in \fig~\ref{fig:fast_front_vs_fast_slip} and in \citep{Tromborg2014slow} that when their ratio remains constant, changes to the value of these parameters affect the slip and front speed, but dynamics remain qualitatively the same. When their ratio changes, however, qualitative changes to the system dynamics occur. Doubling $k$ or halving $k_{ij}$ increases the loading region where the interface stress is significantly influenced by the increasing force in the driving spring between events. This leads to multiple front tips at different locations propagating simultaneously instead of one front propagating from the trailing to the leading edge, so that a single front speed cannot readily be defined. Halving $k$ or doubling $k_{ij}$, \JSa{on the other hand}, reduces the extent of the loading region, so fronts can still be defined as before, but these settings favor fast front propagation, so very few fast--slow--fast fronts were observed. Of course, the slow slip mechanism is still active, but as discussed before it is masked by the fast slip and fast front propagation. 

\JSa{Variations} in the force levels in the junction evolution law, $\fthres$ and $\fslip$, are included in the scaling results that are explained in Appendix~\ref{appsec:mus_from_width}. We also fully expect that parameter values can be found where full simulations produce precursors, fast--only events, fast--slow--fast events and intermediate partial slip events with larger values of $\fthres$ and $\fslip$ than we use. Because this change modifies both the triggering and propagation of events, other parameters would have to be changed simultaneously.

\subsection{Interpretations and comparisons}
In Section~\ref{sec:influence_of_front_type_on_the_loading_curve} we identified a signature of slow front propagation in the macroscopic loading curve. If this signature turns out to also be present in the experiments where slow fronts are known to occur, it could be a useful first indicator of slow fronts in experiments where the nature of rupture front propagation is unknown. The underlying insight is that the reduction in $F_T$ during an event has two distinct parts, both of which are due to the motion of the trailing edge of the slider: first, $F_T$ drops during front propagation, as the slider deforms when some parts of the interface are slipping and others are stuck; then, $F_T$ continues to drop during full sliding. If the reduction in $F_T$ during front propagation is sufficiently large to be reliably measured, a fast--slow transition will show up in $F_T(t)$ as a reduction in slope. If the front transitions back to fast propagation or reaches the leading edge, a change back to large negative slope will occur in $F_T(t)$.

In Section~\ref{sec:fast_slip_and_fast_front_speeds_are_inertial} we demonstrated by changing the slider mass density $\rho$ and hence the bulk wave speeds that the fast slip and fast fronts in the model are of inertial origin. This result, which was to be expected, further supports our claim from \citep{Tromborg2014slow} that fast slip and slow slip are due to distinct mechanisms. It follows that the ratio of their speeds can vary \JSa{if system parameters are changed. For instance, in the model:} fast slip speed is independent of and slow slip speed is inversely proportional to the time scale $\bar t_R$ \AMSa{which determines} \JSa{interface healing after rupture} \citep[equation~S1]{Tromborg2014slow}; fast slip speed is inversely proportional the square root of slider mass density and slow slip speed is independent of slider mass density. In contrast, reducing (increasing) the interface stiffness increases (reduces) both slow and fast slip speeds \JSa{and thus does not affect their ratio}. Because front speed is proportional to slip speed across the velocity scales \JSa{that }we have observed, see \fig~\ref{fig:fast_front_vs_fast_slip}, it follows that large variations in the ratio of slow to fast front speeds is \JSa{also} possible between systems. This \JSa{paradoxically suggests that the usual definition of slow fronts in terms of their velocity (one to several orders of magnitude slower than the Rayleigh wave speed), although natural in the system in which they were discovered \citep{Rubinstein2004detachment}, is not satisfying. This definition does not capture the physical characteristics of slow fronts -- dynamic fronts propagating due to the intrinsic relaxation dynamics of the frictional interface after arrest -- or \AMSa{differentiate them from} other front types, and may lead to misinterpretations in systems where the velocity range of slow fronts significantly overlap the velocity range of fast fronts (of the order of sound speed) or quasi-static fronts (speed proportional to the rate of external loading).}

In Section~\ref{sec:front_type_phase_diagram_and_its_predictive_power} we \JSa{first} repeated our finding from \citep{Tromborg2014slow} that the type of front observed (arresting, fast--slow--fast, or fast--only) depends systematically on both the strength and the prestress of the interface, with weaker interfaces and higher prestresses favoring faster front propagation. We then showed that although the front type phase diagram in \fig~\ref{fig:phase_diagram} is not a local measure, because front speed is transient and also depends on the interface state behind the front tip, the diagram can \AMSa{still} give useful predictions of the way front type changes with sudden changes in interface conditions. This could help in the interpretation of experiments on interfaces with heterogeneous friction properties like \citep{Latour2013effect}, where material heterogeneities in the form of rock pebbles embedded at a gel--glass interface were observed to cause the rupture front speed to sometimes increase, sometimes decrease.


In Section~\ref{sec:front_speed_depends_on_local_stress_state} we showed that, as expected from experiments and other numerical work, front speed increases with increasing shear to normal prestress ratio $\tau/p$. However, knowing \JSa{the local} $\tau/p$ alone is insufficient information to predict the front speed. \JSa{Indeed, in} Section~\ref{sec:front_speed_is_transient} we showed that in the model \JSa{and for experimentally realistic system sizes, front speed is transient at all points and as such the selection of its value is both history-dependent and non-local. In} Section~\ref{sec:front_speed_depends_on_local_strength} we showed that both fast and slow front speeds also depend on the local strength of the interface\JSa{, which is generally out of experimental reach, especially when the interfacial junctions bear some randomness, \eg in their size or stiffness.}

In \citep{Tromborg2014slow} we demonstrated that the slow front speed in the model is proportional to the slow slip speed and worked out the constant of proportionality. \citet[equation~(9)]{BarSinai2012slow} (see also \citep[equation~(9)]{Bouchbinder2011slow}) found \JSa{a similar} result in an analytical model with a rate-and-state-type friction law. In Section~\ref{sec:front_speed_vs_slip_speed} we found that the same constant of proportionality relates fast slip to fast front speed. The proportionality follows from relating how far the blocks behind the front tip have to move to advance the rupture, to how quickly they move (the slip speed)\JSa{. For instance, less} motion is required, leading to faster fronts, when the interface is \JSa{stiffer or more highly} prestressed, \JSa{or} when the stress field in the bulk has a \JSa{longer} decay length. Taken together, these results demonstrate that the origin of the front speed, be it slow or fast, lies in the relevant slip speed during front propagation. The selection of a fast or a slow front regime is thus the direct consequence of the selection of a fast (inertial) or a slow (driven by the internal relaxation of an arrested interface) slip regime. We therefore claim that understanding front type selection and the transition between front types essentially reduces to understanding the slip motion and its time evolution during an event. Note, however, that even when a mechanism for slow slip is present in a system, slow fronts will not appear if the fast fronts always traverse the entire interface: being fast, they will mask any slow propagation.

\JSa{In \citet{BarSinai2012slow}, the slow slip speed was selected as the speed for which the steady-state friction law had a minimum. However, the existence of such a minimum is not a necessary condition for slow fronts to appear in a system. As a matter of fact, in our model, and for the parameters used, the steady state friction law is purely decreasing. The apparent contradiction disappears as soon as one remembers that in our model, the slow slip speed arises from a completely different physical mechanism, namely the time distributed relaxation of the interface after its rupture and arrest.}

For the scaling of the shear to normal stress ratio we have used $\bar\tau_0=(\tau_0/p-\fslip/f_N)/(\fthres/f_N-\fslip/f_N)$ throughout. It \JSa{would have been} consistent with the observed dependence of front speed on both local stress state and local strength to replace $\fthres/f_N$ with $\taumax/p=\mu_s^\text{eff}$ to obtain $\bar\tau_0^\text{eff}=(\tau_0/p-\fslip/f_N)/(\taumax/p-\fslip/f_N)$, and use this in \fig~\ref{fig:fast_front_speed_vs_prestress_width_0}\fpt{b}. \JSa{Nevertheless, we} have chosen to stick with a scaling relation that depends only on input \JSa{model} parameters and not on the \JSa{emerging} effective strength of the interface, because quantitative validation of the more advanced scaling is not directly possible due to the \AMSa{transient nature} of the fronts. In an application where fronts \JSa{were} known to have converged, it would be interesting to try the $\bar\tau_0^\text{eff}$ scaling when plotting the fronts' stationary speed against prestress. Similarly, replacing $\tau_\text{thres}$ with $\taumax$ in the scaling relation between front speed and slip speed in \fig~\ref{fig:fast_front_vs_fast_slip} would improve the scaling of the data where $\phi(f_T)$ was varied, at the cost of introducing emerging properties into the scaling.

\citet[Fig.~3]{Ben-David2010dynamics} reported a relationship between front speed and shear to normal prestress ratio. In Section~\ref{sec:front_speed_compares_well_to_experiments} we showed that although our results are quantitatively different (in particular, the range of $\tau_0/p$ values that are stable in the model is \JSa{about half of that} in the experiments), qualitatively we capture the experimental observation well. Further, our results for front speed transients (Section~\ref{sec:front_speed_is_transient}) and dependence on local strength (Section~\ref{sec:front_speed_depends_on_local_strength}) are possible explanations for the relatively large range of front speed values that were observed for each value of $\tau/p$ in \citep{Ben-David2010dynamics}. Namely, we expect variation in local strength to be present in the experiments, probably to a larger degree than in the simulations, as we have explicitly avoided spatial \JSa{heterogeneities in the friction parameters. It} is also possible that the experimental fronts share with our simulations the property that over the sample length studied, the front speed is in the transient regime\JSa{: the fronts in \eg Fig.~2 of \citep{Ben-David2010dynamics} actually} change their speed throughout the interface, but whether this is due purely to changing interface conditions or also to a ``true'' transient behavior is difficult to assess.

We argued in the model description (Section~\ref{sec:model_description}) that a time-dependent rule for the reformation of junctions has experimental justification. A consequence of this microscopic friction law is that the local microscopic state, specifically the distribution of forces in the junctions, depends on the block slip dynamics of the preceding event \citep{Thogersen2014history-dependent}. The microscopic junction state in turn determines the effective static friction, which affects the front propagation. In Section~\ref{sec:history_dependence_with_gini} we showed that the Gini coefficient, an integrated estimator of the non-uniformity of the block slip history, predicts the effective local static friction even in full simulations were the block slip dynamics is highly complicated. We were thus able to link two very different aspects of the system behavior. Further, while the distribution of junction forces and the resulting effective friction are difficult to measure experimentally, the local slip history can more readily be measured, at least on the side of the slider \citep[see e.g.][]{Svetlizky2014classical,Ben-David2010slip-stick} or in setups which track the motion of patterns or markers at the interface \citep[see e.g.][]{Prevost2013probing,Romero2014probing}. We \JSa{recognize} that there are challenges that need to be overcome to adjust our protocol for measuring the Gini coefficient and apply it to experiments \JSa{or real systems like seismic faults}. However, if the length scale for junction breaking and the time scale for junction reformation can be determined, we believe that the underlying idea, namely to characterize the crucial part of the slip motion with a robust measure of inequality \JSa{and to use it to predict the (still unknown) strength of the interface, is applicable.}

We have studied the onset of sliding friction in a model that couples a microscopic friction law involving the state of a large number of individual junctions to a 2D elastic solver. In the range of possible combinations of friction and bulk models outlined in the introduction this is, we believe, an example of bringing together a moderately complicated friction law with a moderately complicated bulk law. For example, the friction law excludes both aging and a distribution of junction strengths, and the bulk law, while 2-dimensional, excludes among other things plasticity and melting, and is solving for the deformation of the slider only. Nevertheless, the simulations reproduce many of the features of the spatio-temporal dynamics in experiments to which the model was attuned, and we were able to use the more complete information of the system and the fine control of initial conditions available in simulations to better understand how some of the central material properties and system state characteristics modify the dynamics. We believe important avenues for future research to be the inclusion of the bulk deformations of the substrate, which brings with it the complication of defining frictional properties at a non-planar interface, and the parametrization from more fundamental models of the individual junction evolution laws.

\acknowledgments{%
This work was supported by the bilateral researcher exchange programme Aurora (Hubert Curien Partnership), financed by the Norwegian Research Council and {the French Ministry of Foreign Affairs and International Development} (Grant No. 27436PM, 213213). K.T. acknowledges support from VISTA, a basic research programme funded by Statoil, conducted in close collaboration with The Norwegian Academy of Science and Letters. J.S. acknowledges support from the People Programme (Marie Curie Actions) of the European Union's 7th Framework Programme (FP7/2007-2013) under REA grant agreement 303871.}

\appendix
\section{Stencil for determining rupture front speeds\label{appsec:5_point_stencil}}
In this appendix we give the details of how we measure the front propagation speed $v_c$. We define it as
\begin{align}
v_c = \frac{\mathrm{d}x_\mathrm{tip}}{\mathrm{d}t},
\end{align}
where $x_\mathrm{tip}(t)$ is the location of the front tip. To use this expression in practice, the position of the front tip in time must be found and a suitable stencil used to take the derivative numerically. We use the rupture criterion on the block level, that less than $30 \%$ of a block's junctions remain in the pinned state, to find rupture times $t_{\mathrm{rup},i}$ for every block that partakes in an event. We also define the front tip at any time to be located at the last block to reach this sliding criterion. The simplest formula for the local rupture speed is then to take the average speed of the front on its way between two blocks, that is
\begin{align}
v_{c,i} = \frac{x_{i+1}-x_{i}}{t_{\mathrm{rup},i+1}-t_{\mathrm{rup},i}}.
\end{align}
However, this is very sensitive to the discreteness and non-smoothness of the front and gives very large fluctuations along the interface. Smoother results are obtained by averaging over a larger area. We have used
\begin{align}
v_{c,i} = \frac{x_{i+\mathrm{vwidth}}-x_{i-\mathrm{vwidth}}}{t_{\mathrm{rup},i+\mathrm{vwidth}}-t_{\mathrm{rup},i-\mathrm{vwidth}}}\label{eq:vwidth},
\end{align}
with $\mathrm{vwidth}$ a number of blocks. We have found that $\mathrm{vwidth}=2$, which gives a 5-point wide stencil, strikes a good balance between smoothness and spatial resolution for the parameters we employ. Of course, for higher $N_x$, $N_z$, $\mathrm{vwidth}$ can be increased proportionally.

In principle, another option for finding a larger kernel is to use standard stencils for numerical derivatives of first order. Since the raw data are the progressive $t_{\mathrm{rup},i}$ over an approximately constantly spaced spatial grid, the inverse rupture speed can be found from the standard 5-point stencil as
\begin{align}
\frac{1}{v_{c,i}} &= \frac{-t_{\mathrm{rup},i+2}+8t_{\mathrm{rup},i+1}-8t_{\mathrm{rup},i-1}+t_{\mathrm{rup},i-2}}{12\Delta x}\label{eq:5-point}.
\end{align}
Although this is a kernel with five nodes, the weights are such that the nodes near the centre dominate, and its fluctuations are comparable to those for $\mathrm{vwidth}=1$ in Equation~\eqref{eq:vwidth} and larger than for $\mathrm{vwidth}=2$. The standard 7- and 9-point stencils are also dominated by the centre nodes, giving poor results.

When a front nucleates, either as a first front nucleating in a fully pinned interface, or as a secondary front nucleating ahead of the main front due to inhomogeneities in the stress state, it travels both left and right. This leads to points on both sides of the nucleation site having nearly the same rupture time $t_\mathrm{rup}$, and to the spurious conclusion by a naive measurement scheme that the front speed was suddenly very high. To simplify automatic front speed measurement we have chosen to ignore any points for which the stencil in use extends across a region where the front is left-travelling. This is the reason for the front speed being undefined for some positions in \fig~\ref{fig:events_with_rupture_speeds}c.


\section{The dependence of effective local static friction on the width of \texorpdfstring{$\phi(f_T)$}{phi(f\_T)}, and the analytical result that underlies the scaling of \texorpdfstring{$\mu_s^\text{eff}$ and $\sigma$}{mu\_s\^eff and sigma} axes\label{appsec:mus_from_width}}
\newcommand{\sthres}{s_\text{thres}}
\newcommand{\swidth}{{s_\text{supp}}}
\newcommand{\sdyn}{{s_\text{slip}}}
\newcommand{\kjunc}{k_\text{jun}}
\newcommand{\fsupp}{f_\text{supp}}
In this appendix we show that the scaling that we use here and in \citep{Tromborg2014slow} for the effective static friction threshold, $\bar\mu_s^\text{eff}=(\taumax/p-\fslip/f_N)/(\fthres/f_N-\fslip/f_N)$, and for the width of the junction force distribution, $\tilde\sigma = \sigma/(\fthres-\fslip)$, can be derived analytically for uniform junction force distributions $\phi$ and holds also for other shapes of $\phi$. The arguments were omitted from \cite{Tromborg2014slow} for brevity. We include a figure that demonstrates the scaling and repeats the relationship between the shape and width of $\phi$ and the effective static friction threshold.

The main result in this appendix is equation~\eqref{eq:nu_s^eff_scaled}. We will derive it using results from \citet{Thogersen2014history-dependent}, and some simple bookkeeping is required to set up the correspondence between our choice of parameters in that work and here. Namely, in the present paper and in \citep{Tromborg2014slow} we express the state of the pinned junctions by the distribution $\phi$ of forces $f_T$ in the junctions. In \citep{Thogersen2014history-dependent} we expressed this state in terms of a length scale, the distribution $S$ of junction stretching lengths $s$. In the linearly elastic regime that we consider in all three works, these are related by $\phi = \kjunc S$, where $\kjunc\equiv k_{ij}$ is the single junction stiffness.

In Appendix~\ThogersenAppMusEff of \cite{Thogersen2014history-dependent} we calculated the effective static friction for $S$ uniform and with support of extent $\swidth$. The assumptions were (i) linear springs with force $\kjunc s$ at stretching $s$, and (ii) breaking of the complete set of junctions attached to the block in a time interval small compared to the average time spent by a junction in the slipping state, so that the first contacts that break stay broken until $\mus^\text{eff}$ is reached. Defining notation consistent with the present paper, we write equations~(F1) and (F2) in \citep{Thogersen2014history-dependent} as
\begin{widetext}
\begin{align}
\frac{\taumax}{N_s} =
  \left\{
    \begin{array}{ll}
      \frac{\kjunc}{2}\left(2\sthres-\swidth\right) &,\sthres-\swidth>\sdyn\\
      \frac{\sdyn-\sthres+\swidth}{\swidth}\fslip + \frac{\kjunc}{2\swidth}\left(\sthres^2-\sdyn^2\right) &,\sthres-\swidth\leq\sdyn,
    \end{array}
  \right.\label{eq:museff_unscaled}
\end{align}
where $\sthres\equiv\smax=\fthres/\kjunc$ is the stretching at the junction breaking threshold and $\sdyn=\fslip/\kjunc$ is the stretching corresponding to the sliding state friction.

To rewrite equation~\eqref{eq:museff_unscaled} in terms of forces we define $\fsupp=\kjunc\swidth$. Subtracting $\fslip$ and dividing by $(\fthres-\fslip)$ on both sides, and recalling that $N_sf_N=p$, we arrive at
\begin{align}
\bar\mu_s^\text{eff}
=\frac{\taumax/p-\fslip/f_N}{\fthres/f_N-\fslip/f_N}
= 
  \left\{
    \begin{array}{ll}
      1 - \frac{1}{2}\frac{\fsupp}{\fthres-\fslip} &,\fthres-\fsupp>\fslip \\
      \frac{1}{2}\frac{\fthres-\fslip}{\fsupp} &,\fthres-\fsupp\leq\fslip.
    \end{array}
  \right.\label{eq:nu_s^eff_scaled}
\end{align}
\end{widetext}
On this form it is apparent that scaling $\mu_s^\text{eff}$ by using the expression on the left hand side and scaling the width by using $\frac{\sigma}{\fthres-\fslip}$ (a uniform distribution with support of extent $\fsupp$ has standard deviation $\sigma=\fsupp/(2\sqrt{3})$) provides a data collapse under changes to $\kjunc$, $\fslip$ and $\fthres$.

\Fig~\ref{fig:mu_s^eff_analytical_and_numerical}\fpt{a} shows unscaled data for uniform $\phi$. To show that the scaling in equation~\eqref{eq:nu_s^eff_scaled} is not limited to this particular shape of $\phi$, we include data for bell-shaped $\phi$ in \fig~\ref{fig:mu_s^eff_analytical_and_numerical}\fpt{b}. When we apply the scaling, in \fig~\ref{fig:mu_s^eff_analytical_and_numerical}\fpt{c}, each distribution shape gets a data collapse. The master curves for the two distribution shapes are slightly different.

A point which we have discussed in both of \citep{Tromborg2014slow,Thogersen2014history-dependent} and which we use in Sections~\ref{sec:modifyingFrontTypeFromWidth} and \ref{sec:front_speed_depends_on_local_strength} is that wider $\phi$ correspond to lower effective local static friction. This can be seen directly in \fig~\ref{fig:mu_s^eff_analytical_and_numerical}.
Finally, we note that the scaling of $\mu_s^\text{eff}$ is the same as the scaling of prestress $\bar\tau$, a scaling that was also used in \citep{Tromborg2011transition,Amundsen20121D,Muratov1999traveling,Scholz2002mechanics}.

\begin{figure}
\centering
\includegraphics[width=\columnwidth]{\finalfigurepath{Tromborg_Fig20}}%
\caption{\coloronlineornothing Scaling of effective static friction vs junction force distribution width. \fpf{a} Unscaled analytical results for uniform junction distributions. Changes made to $\kjunc$, $\fslip$ and $\fthres$ independently. Markers indicate $\fsupp=\fthres-\fslip$. \fpf{b} Unscaled numerical results for junction distributions that are bell-shaped polynomials with roots at $\pm a$ and the functional form $\phi(\xi) = 5/(4a)(1+3\abs{\xi/a})(1-\abs{\xi/a})^3, \xi\in[-a,a]$. Changes made to $\fslip$ and $\fthres$ independently. \fpf{c} Data collapse using exact scaling.\label{fig:mu_s^eff_analytical_and_numerical}}
\end{figure}

\section{Initialisation, boundary and driving conditions\label{appsec:simulation_setup}}
This appendix supplements the model description found in the main text with detailed information on how we initialize the system and apply the boundary conditions. The simulations we have performed can be grouped in three categories. (i) Full simulations. In these the sample is initially unstressed; then the normal force is applied; then the shear force is applied through the driving spring, and the spring drives the slider through tens of events: precursors, full sliding events, and possibly partial slip events between the full sliding events. (ii) Simulations that restart at a point within a full simulation. In these we make well-defined changes to identify or highlight a particular mechanism, for example the slow slip mechanism. (iii) Smaller simulations where the shear and normal prestresses are controlled and a single event is studied. In these we have done triggering both through (iii)a breaking the junctions in a predefined region, and (iii)b motion of the driving spring as in the full simulations. In this appendix we explain how each type of simulation was set up, and give parameter values in \tab~\ref{tab:parameters}.

\subsection{Full simulations}
In full simulations, the slider is initialized with full normal load $F_N$ and no tangential load $F_T$ by gradually applying $F_N$ without allowing springs to break, a technicality required because the normal forces on the springs, $f_{Nij}$, start at zero and therefore springs, if allowed to, would break under any stretching. We distribute the load $F_N$ uniformly on the top blocks; apart from this we use the same non-frictional boundary conditions as in \citep{Tromborg2011transition}. The bottom blocks interact with a vertically elastic foundation that \JSa{produces} a normal force $-k_fz_i$ when $z_i<0$, zero when $z_i\geq0$; \JSa{$z_i$ is the vertical displacement of interface block $i$; }the value of $k_f$ is given in \tab~\ref{tab:parameters}. Zero force boundary conditions are applied to the right and left edges (the top and bottom forces still apply to the corner blocks). The unique equilibrium is found through damped relaxation of typical duration $\unit{10}{\milli\second}$. After relaxation, we check that no spring is stretched beyond its strength and introduce the driving spring starting from zero applied driving force $F_T$. Then $F_T$, which acts on the block on the left side of the slider situated at height $h$ above the interface, through the driving spring, increases as the driving point moves to the right with speed $V$.
Full simulations were used for \fig~\ref{fig:sketch_and_loading_curves}, \ref{fig:slow_slip_mechanism}, \ref{fig:events_with_rupture_speeds}, \ref{fig:loading_curve_zoom}, \ref{fig:drop_amplitudes}, \ref{fig:gini_simulation_slip_profiles} and \ref{fig:mu_s_eff_vs_gini}.

\subsection{Restarted simulations}
It can be difficult to analyze events from the full simulations, because the initial conditions for each event arise dynamically from the preceding events. To make direct comparisons between events possible, restarted simulations begin from an interesting state arising in a full simulation. The state consists of the instantaneous positions and velocities of all the blocks and the driving stage, and the pinned/slipping and lifetime information of all the junctions. We modify one or a few model parameters or state properties, rerun the simulation, and compare the output of the restarted simulation with the original, or with another restarted simulation. Restarted simulations were used for \fig~\ref{fig:inertial} (the mass density was varied), \ref{fig:slow_slip_to_slow_front} (junction evolution law was varied) and \ref{fig:fast_slow_fast_into/from_fast_only} (junction force distribution, a part of the state, was varied).

\subsection{Single event simulations}
Single event simulations, like restarted simulations, were performed to elucidate the qualitative and quantitative importance of individual parameters or interface states by varying them individually, keeping the rest of the simulation setup unchanged. This allows us to go beyond the insights from the full simulations. To simplify analysis, these systematic studies were done with different normal forces and different initialization from the full and restarted simulations. The normal force boundary conditions on the top and bottom were exchanged: this simplifies the analysis by setting a constant normal force $p_i=F_N/N_x$ on all blocks $i$ at the interface. To maintain stability against global rotation, the top blocks interacted with an elastic ceiling with the same properties as the elastic foundation used in the full simulations.

To obtain an initial state with a prescribed interfacial shear stress profile we turned the interface springs off during the initialisation. In their place we added to each bottom block the force corresponding to the shear stress to be prescribed. We also introduced the driving spring, but let $V=0$. During relaxation, the sample moved along the $x$-axis until the force in the driving spring balanced the net force from the interfacial shear stress. To get rid of oscillations more efficiently we added damping forces $-\alpha(\vec v_i)$ on the blocks' motion. After relaxation, the extra forces and the extra damping were turned off and the interfacial springs were introduced, with their attachment points $x_{ij}$ chosen such that the net force on each block was unchanged and the desired distribution of spring forces, $\phi(f_T)$, appeared. We then waited a few timesteps to ensure that the transition from pre- to post-relaxation involved no force discontinuities.

For type (iii)a single event simulations, instead of driving the system with $V\neq­0$ until rupture was triggered, we started fronts by simultaneously de-pinning all junctions for all blocks to the left of $x_\text{trigger}$. The shear stress in the triggering region has a strong influence on the rupture fronts, and in order to simplify the comparison of results between simulations we used a constant value $\bar\tau_\text{trigger}=0.3$. The triggering and propagation regions and the initial stress configuration for this type of simulations are shown in \fig~\ref{fig:fast_and_slow_front_variation_vs_width}\fpt{c}.
Type (iii)a simulations were used for \fig~\ref{fig:phase_diagram}, \ref{fig:phase_diagram_local_changes}, \ref{fig:front_speed_transients}, \ref{fig:fast_front_speed_vs_prestress_width_0}, \ref{fig:fast_and_slow_front_variation_vs_width}, \ref{fig:fast_front_vs_fast_slip} and \ref{fig:135x}.

For type (iii)b single event simulations we matched the experimental event triggering conditions more closely. Again, homogeneous stress and junction force distribution states were prepared as described above. Then, an event was started by moving the driving spring as in the full simulations. This modifies the shear stress state before the event starts, particularly near the loading point. As in our other simulations, the width of the junction force distribution $\phi$ affects the effective static friction threshold $\mus^{\text{eff}}$ and by that the energy released as the event starts. To isolate the effect of varying parameters in the propagation region, we kept the width of the junction force distribution constant at $\sigma_\text{load}/(\fthres-\fslip)=0.143$ to the left of $x_\text{load}=\unit{22.5}{\milli\meter}$.
Type (iii)b simulations were used for \fig~\ref{fig:front_speed_transients} and \fig~\ref{fig:v_front_vs_tau_over_p}. 

For \fig~\ref{fig:front_speed_transients}, which includes both type (iii)a and (iii)b simulations, we used the procedures described above, except we used $\bar\tau=0.4$, $\sigma=0$ along the entire interface.

\begin{table*}
\centering
\caption{Model parameters. We used the same parameters in \citep{Tromborg2014slow}. Parameters above the horizontal line were also used, in the same way, in \citep{Tromborg2011transition}.\label{tab:parameters}}
\begin{tabular}{lll}\hline
Name				& Symbol	& Value				\\\hline
Slider length ($x$)		& $L$		& $\unit{140}{\milli\metre}$	\\
Slider height ($z$)		& $H$		& $\unit{75}{\milli\metre}$	\\
Slider width ($y$)		& $B$		& $\unit{6}{\milli\metre}$	\\
Number of blocks		& $N_x$		& $57$				\\
				& $N_z$		& $31$				\\
Slider mass			& $M$		& $\unit{75.6}{\gram}$		\\
Block mass			& $m$		& $M/(N_xN_z)$			\\
Young's modulus			& $E$		& $\unit{3}{\giga\pascal}$	\\
Bulk spring modulus		& $k$		& $3BE/4$		\\
Bulk spring length		& $l$		& $L/(N_x-1) = H/(N_z-1)$	\\
Damping {coefficient}		& $\eta$	& $\sqrt{0.1km}$		\\
Normal load			& $F_N$		& $\unit{1920}{\newton}$	\\
Elastic foundation modulus	& $k_f$		& $k/2$				\\
Driving spring modulus		& $K$		& $\unit{4}{\mega\newton\per\meter}$	\\
Driving height			& $h$		& $\unit{5}{\milli\metre}$	\\
Driving speed			& $V$		& $\unit{0.4}{\milli\metre\per\second}$	\\
\hline
Threshold force coefficient	& $\mu_s = f_\mathrm{thres}/f_N$	& $0.4$				\\
Slipping force coefficient	& $\mu_d = f_\mathrm{slip}/f_N$	& $0.17$			\\
Number of interface springs per block	& $N_s$		& $100$				\\
Interface spring stiffness	& {$k_{ij}$}	& $\sqrt{\unit{39.2}{\giga\newton\per\meter^2}f_{N,ij}}$ \\
Slipping time mean		& $\bar t_R$	& $\unit{2}{\milli\second}$	\\
Slipping time standard deviation		& $\delta t_R$	& $\unit{0.6}{\milli\second}$	\\
Triggering region width		& $x_\mathrm{trigger}$ & $\unit{22.5}{\milli\metre}$ \\
Triggering region prestress	& {$\bar\tau_\mathrm{trigger}$}& 0.3\\
Time step duration		& $\Delta t$ & $\unit{2\e{-7}}{\second}$	\\
Extra damping coefficient	& $\alpha$	& $\eta/40$\\
\hline
\end{tabular}
\end{table*}
\null
\vfill

\section{Slow fronts for systematical variation in junction force distributions\label{appsec:slow_front_variation_with_sigma_underlying_data}}
In this appendix we present the data underlying \fig~\ref{fig:fast_and_slow_front_variation_vs_width}\fpt{c} in the main text.

\Fig~\ref{fig:135x} shows the full evolution of the events corresponding to a range of junction force distribution widths $\sigma$ and with $\bar\tau_0=0.05$. For the lowest $\sigma$ (strong interfaces) the front arrests, which is also reflected in \fig~\ref{fig:phase_diagram}. For the fronts that do reach the leading edge, the fronts are gradually changing from slow to fast propagation. These events plot in the ``Slow'' region of the front type phase diagram in \fig~\ref{fig:phase_diagram} because they would arrest if the slow front mechanism was turned off. For each simulation we define the extent of the slow front region by visual inspection of the panels in \fig~\ref{fig:135x}. We then measure the front speed in the same way as usual, described in Appendix~\ref{appsec:5_point_stencil}. From these data we determine the average, minimum and maximum slow front speed values for these events.

\begin{figure}
\centering
\includegraphics[width=\columnwidth]{\finalfigurepath{Tromborg_Fig21}}
\caption{\coloronlineornothing Underlying data for \fig~\ref{fig:fast_and_slow_front_variation_vs_width}\fpt{c}: the events that lie at $\bar\tau = 0.05$ in \fig~\ref{fig:phase_diagram}, shown as in \fig~\ref{fig:events_with_rupture_speeds}. From top left to bottom right, the width of $\phi(f_T)$ is $\sigma/(\fthres-\fslip)=0.059$, $0.088$, and then $0.117$ to $0.528$ in steps of $0.059$; the event for $\sigma=0$ is not shown, but it too arrests early on. For each panel, the extent of the slow front was determined by visual inspection. On the left, the slow front starts after the step (ca. $x=\unit{0.04}{\metre}$). On the right, because the transition to a fast front is more gradual, the boundary was set where the trend of near constant front velocity is broken. The exact limits chosen can be seen from the horizontal extent of the data in \fig~\ref{fig:fast_and_slow_front_variation_vs_width}\fpt{c}. Panels \fpl{a--c} show a larger time interval than the rest of the panels. No data for \fig~\ref{fig:fast_and_slow_front_variation_vs_width}\fpt{c} were extracted from the bottom right panel: even though the front has a slow part and plots in the slow part of the front type phase diagram, the slow part is too intermixed with the transition to fast rupture for the analysis to be carried out.\label{fig:135x}}
\end{figure}

\FloatBarrier
\bibliography{./Tromborg}

\begin{thebibliography}{75}
\expandafter\ifx\csname natexlab\endcsname\relax\def\natexlab#1{#1}\fi
\expandafter\ifx\csname bibnamefont\endcsname\relax
  \def\bibnamefont#1{#1}\fi
\expandafter\ifx\csname bibfnamefont\endcsname\relax
  \def\bibfnamefont#1{#1}\fi
\expandafter\ifx\csname citenamefont\endcsname\relax
  \def\citenamefont#1{#1}\fi
\expandafter\ifx\csname url\endcsname\relax
  \def\url#1{\texttt{#1}}\fi
\expandafter\ifx\csname urlprefix\endcsname\relax\def\urlprefix{URL }\fi
\providecommand{\bibinfo}[2]{#2}
\providecommand{\eprint}[2][]{\url{#2}}

\bibitem[{\citenamefont{Rosakis et~al.}(1999)\citenamefont{Rosakis, Samudrala,
  and Coker}}]{Rosakis1999cracks}
\bibinfo{author}{\bibfnamefont{A.~J.} \bibnamefont{Rosakis}},
  \bibinfo{author}{\bibfnamefont{O.}~\bibnamefont{Samudrala}},
  \bibnamefont{and} \bibinfo{author}{\bibfnamefont{D.}~\bibnamefont{Coker}},
  \bibinfo{journal}{Science} \textbf{\bibinfo{volume}{284}},
  \bibinfo{pages}{1337} (\bibinfo{year}{1999}),
  \urlprefix\url{http://www.sciencemag.org/content/284/5418/1337.abstract}.

\bibitem[{\citenamefont{Baumberger et~al.}(2002)\citenamefont{Baumberger,
  Caroli, and Ronsin}}]{Baumberger2002self-healing}
\bibinfo{author}{\bibfnamefont{T.}~\bibnamefont{Baumberger}},
  \bibinfo{author}{\bibfnamefont{C.}~\bibnamefont{Caroli}}, \bibnamefont{and}
  \bibinfo{author}{\bibfnamefont{O.}~\bibnamefont{Ronsin}},
  \bibinfo{journal}{Phys. Rev. Lett.} \textbf{\bibinfo{volume}{88}},
  \bibinfo{pages}{075509} (\bibinfo{year}{2002}), ISSN
  \bibinfo{issn}{0031-9007},
  \urlprefix\url{http://link.aps.org/doi/10.1103/PhysRevLett.88.075509}.

\bibitem[{\citenamefont{Rubinstein et~al.}(2004)\citenamefont{Rubinstein,
  Cohen, and Fineberg}}]{Rubinstein2004detachment}
\bibinfo{author}{\bibfnamefont{S.}~\bibnamefont{Rubinstein}},
  \bibinfo{author}{\bibfnamefont{G.}~\bibnamefont{Cohen}}, \bibnamefont{and}
  \bibinfo{author}{\bibfnamefont{J.}~\bibnamefont{Fineberg}},
  \bibinfo{journal}{Nature} \textbf{\bibinfo{volume}{430}},
  \bibinfo{pages}{1005} (\bibinfo{year}{2004}),
  \urlprefix\url{http://dx.doi.org/10.1038/nature02830}.

\bibitem[{\citenamefont{Prevost et~al.}(2013)\citenamefont{Prevost, Scheibert,
  and Debr\'{e}geas}}]{Prevost2013probing}
\bibinfo{author}{\bibfnamefont{A.}~\bibnamefont{Prevost}},
  \bibinfo{author}{\bibfnamefont{J.}~\bibnamefont{Scheibert}},
  \bibnamefont{and}
  \bibinfo{author}{\bibfnamefont{G.}~\bibnamefont{Debr\'{e}geas}},
  \bibinfo{journal}{Eur. Phys. J. E} \textbf{\bibinfo{volume}{36}}
  (\bibinfo{year}{2013}), ISSN \bibinfo{issn}{1292-8941},
  \urlprefix\url{http://dx.doi.org/10.1140/epje/i2013-13017-0}.

\bibitem[{\citenamefont{Romero et~al.}(2014)\citenamefont{Romero, Wandersman,
  Debr\'{e}geas, and Prevost}}]{Romero2014probing}
\bibinfo{author}{\bibfnamefont{V.}~\bibnamefont{Romero}},
  \bibinfo{author}{\bibfnamefont{E.}~\bibnamefont{Wandersman}},
  \bibinfo{author}{\bibfnamefont{G.}~\bibnamefont{Debr\'{e}geas}},
  \bibnamefont{and} \bibinfo{author}{\bibfnamefont{A.}~\bibnamefont{Prevost}},
  \bibinfo{journal}{Phys. Rev. Lett.} \textbf{\bibinfo{volume}{112}},
  \bibinfo{pages}{094301} (\bibinfo{year}{2014}), ISSN
  \bibinfo{issn}{0031-9007},
  \urlprefix\url{http://link.aps.org/doi/10.1103/PhysRevLett.112.094301}.

\bibitem[{\citenamefont{Rubinstein et~al.}(2007)\citenamefont{Rubinstein,
  Cohen, and Fineberg}}]{Rubinstein2007dynamics}
\bibinfo{author}{\bibfnamefont{S.~M.} \bibnamefont{Rubinstein}},
  \bibinfo{author}{\bibfnamefont{G.}~\bibnamefont{Cohen}}, \bibnamefont{and}
  \bibinfo{author}{\bibfnamefont{J.}~\bibnamefont{Fineberg}},
  \bibinfo{journal}{Phys. Rev. Lett.} \textbf{\bibinfo{volume}{98}},
  \bibinfo{pages}{226103} (\bibinfo{year}{2007}),
  \urlprefix\url{http://link.aps.org/doi/10.1103/PhysRevLett.98.226103}.

\bibitem[{\citenamefont{Maegawa et~al.}(2010)\citenamefont{Maegawa, Suzuki, and
  Nakano}}]{Maegawa2010precursors}
\bibinfo{author}{\bibfnamefont{S.}~\bibnamefont{Maegawa}},
  \bibinfo{author}{\bibfnamefont{A.}~\bibnamefont{Suzuki}}, \bibnamefont{and}
  \bibinfo{author}{\bibfnamefont{K.}~\bibnamefont{Nakano}},
  \bibinfo{journal}{Tribol. Lett.} \textbf{\bibinfo{volume}{38}},
  \bibinfo{pages}{313} (\bibinfo{year}{2010}), ISSN \bibinfo{issn}{1023-8883},
  \urlprefix\url{http://dx.doi.org/10.1007/s11249-010-9611-7}.

\bibitem[{\citenamefont{Scheibert and Dysthe}(2010)}]{Scheibert2010role}
\bibinfo{author}{\bibfnamefont{J.}~\bibnamefont{Scheibert}} \bibnamefont{and}
  \bibinfo{author}{\bibfnamefont{D.~K.} \bibnamefont{Dysthe}},
  \bibinfo{journal}{Europhys. Lett.} \textbf{\bibinfo{volume}{92}}
  (\bibinfo{year}{2010}), ISSN \bibinfo{issn}{0295-5075},
  \urlprefix\url{http://stacks.iop.org/0295-5075/92/i=5/a=54001}.

\bibitem[{\citenamefont{Tr{\o}mborg et~al.}(2011)\citenamefont{Tr{\o}mborg,
  Scheibert, Amundsen, Th{\o}gersen, and
  Malthe-S{\o}renssen}}]{Tromborg2011transition}
\bibinfo{author}{\bibfnamefont{J.}~\bibnamefont{Tr{\o}mborg}},
  \bibinfo{author}{\bibfnamefont{J.}~\bibnamefont{Scheibert}},
  \bibinfo{author}{\bibfnamefont{D.~S.} \bibnamefont{Amundsen}},
  \bibinfo{author}{\bibfnamefont{K.}~\bibnamefont{Th{\o}gersen}},
  \bibnamefont{and}
  \bibinfo{author}{\bibfnamefont{A.}~\bibnamefont{Malthe-S{\o}renssen}},
  \bibinfo{journal}{Phys. Rev. Lett.} \textbf{\bibinfo{volume}{107}},
  \bibinfo{pages}{074301} (\bibinfo{year}{2011}), ISSN
  \bibinfo{issn}{0031-9007},
  \urlprefix\url{http://link.aps.org/doi/10.1103/PhysRevLett.107.074301}.

\bibitem[{\citenamefont{Amundsen et~al.}(2012)\citenamefont{Amundsen,
  Scheibert, Th{\o}gersen, Tr{\o}mborg, and
  Malthe-S{\o}renssen}}]{Amundsen20121D}
\bibinfo{author}{\bibfnamefont{D.~S.} \bibnamefont{Amundsen}},
  \bibinfo{author}{\bibfnamefont{J.}~\bibnamefont{Scheibert}},
  \bibinfo{author}{\bibfnamefont{K.}~\bibnamefont{Th{\o}gersen}},
  \bibinfo{author}{\bibfnamefont{J.}~\bibnamefont{Tr{\o}mborg}},
  \bibnamefont{and}
  \bibinfo{author}{\bibfnamefont{A.}~\bibnamefont{Malthe-S{\o}renssen}},
  \bibinfo{journal}{Tribol. Lett.} \textbf{\bibinfo{volume}{45}},
  \bibinfo{pages}{357} (\bibinfo{year}{2012}), ISSN \bibinfo{issn}{1023-8883},
  \urlprefix\url{http://dx.doi.org/10.1007/s11249-011-9894-3}.

\bibitem[{\citenamefont{Radiguet et~al.}(2013)\citenamefont{Radiguet, Kammer,
  Gillet, and Molinari}}]{Radiguet2013survival}
\bibinfo{author}{\bibfnamefont{M.}~\bibnamefont{Radiguet}},
  \bibinfo{author}{\bibfnamefont{D.~S.} \bibnamefont{Kammer}},
  \bibinfo{author}{\bibfnamefont{P.}~\bibnamefont{Gillet}}, \bibnamefont{and}
  \bibinfo{author}{\bibfnamefont{J.-F.} \bibnamefont{Molinari}},
  \bibinfo{journal}{Phys. Rev. Lett.} \textbf{\bibinfo{volume}{111}},
  \bibinfo{pages}{164302} (\bibinfo{year}{2013}), ISSN
  \bibinfo{issn}{0031-9007},
  \urlprefix\url{http://dx.doi.org/10.1103/PhysRevLett.111.164302}.

\bibitem[{\citenamefont{Braun and Scheibert}(2014)}]{Braun2014propagation}
\bibinfo{author}{\bibfnamefont{O.~M.} \bibnamefont{Braun}} \bibnamefont{and}
  \bibinfo{author}{\bibfnamefont{J.}~\bibnamefont{Scheibert}},
  \bibinfo{journal}{Tribol. Lett.} \textbf{\bibinfo{volume}{56}},
  \bibinfo{pages}{553} (\bibinfo{year}{2014}), ISSN \bibinfo{issn}{1023-8883},
  \urlprefix\url{http://dx.doi.org/10.1007/s11249-014-0432-y}.

\bibitem[{\citenamefont{Ozaki et~al.}(2014)\citenamefont{Ozaki, Inanobe, and
  Nakano}}]{Ozaki2014finite}
\bibinfo{author}{\bibfnamefont{S.}~\bibnamefont{Ozaki}},
  \bibinfo{author}{\bibfnamefont{C.}~\bibnamefont{Inanobe}}, \bibnamefont{and}
  \bibinfo{author}{\bibfnamefont{K.}~\bibnamefont{Nakano}},
  \bibinfo{journal}{Tribology Letters} \textbf{\bibinfo{volume}{55}},
  \bibinfo{pages}{151} (\bibinfo{year}{2014}), ISSN \bibinfo{issn}{1023-8883},
  \urlprefix\url{http://dx.doi.org/10.1007/s11249-014-0343-y}.

\bibitem[{\citenamefont{Kammer et~al.}(2014)\citenamefont{Kammer, Radiguet,
  Ampuero, and Molinari}}]{Kammer2014linear}
\bibinfo{author}{\bibfnamefont{D.~S.} \bibnamefont{Kammer}},
  \bibinfo{author}{\bibfnamefont{M.}~\bibnamefont{Radiguet}},
  \bibinfo{author}{\bibfnamefont{J.-P.} \bibnamefont{Ampuero}},
  \bibnamefont{and} \bibinfo{author}{\bibfnamefont{J.-F.}
  \bibnamefont{Molinari}}, \bibinfo{journal}{arXiv} pp. \bibinfo{pages}{1--10}
  (\bibinfo{year}{2014}), \eprint{1408.4413},
  \urlprefix\url{http://arxiv.org/abs/1408.4413v1}.

\bibitem[{\citenamefont{Chateauminois
  et~al.}({2010})\citenamefont{Chateauminois, Fretigny, and
  Olanier}}]{Chateauminois2010friction}
\bibinfo{author}{\bibfnamefont{A.}~\bibnamefont{Chateauminois}},
  \bibinfo{author}{\bibfnamefont{C.}~\bibnamefont{Fretigny}}, \bibnamefont{and}
  \bibinfo{author}{\bibfnamefont{L.}~\bibnamefont{Olanier}},
  \bibinfo{journal}{Phys. Rev. E} \textbf{\bibinfo{volume}{{81}}},
  \bibinfo{pages}{026106} (\bibinfo{year}{{2010}}), ISSN
  \bibinfo{issn}{{1539-3755}},
  \urlprefix\url{http://link.aps.org/doi/10.1103/PhysRevE.81.026106}.

\bibitem[{\citenamefont{Ben-David
  et~al.}(2010{\natexlab{a}})\citenamefont{Ben-David, Cohen, and
  Fineberg}}]{Ben-David2010dynamics}
\bibinfo{author}{\bibfnamefont{O.}~\bibnamefont{Ben-David}},
  \bibinfo{author}{\bibfnamefont{G.}~\bibnamefont{Cohen}}, \bibnamefont{and}
  \bibinfo{author}{\bibfnamefont{J.}~\bibnamefont{Fineberg}},
  \bibinfo{journal}{Science} \textbf{\bibinfo{volume}{330}},
  \bibinfo{pages}{211} (\bibinfo{year}{2010}{\natexlab{a}}), ISSN
  \bibinfo{issn}{0036-8075},
  \urlprefix\url{http://www.sciencemag.org/content/330/6001/211.abstract}.

\bibitem[{\citenamefont{Schubnel et~al.}(2011)\citenamefont{Schubnel, Nielsen,
  Taddeucci, Vinciguerra, and Rao}}]{Schubnel2011photo-acoustic}
\bibinfo{author}{\bibfnamefont{A.}~\bibnamefont{Schubnel}},
  \bibinfo{author}{\bibfnamefont{S.}~\bibnamefont{Nielsen}},
  \bibinfo{author}{\bibfnamefont{J.}~\bibnamefont{Taddeucci}},
  \bibinfo{author}{\bibfnamefont{S.}~\bibnamefont{Vinciguerra}},
  \bibnamefont{and} \bibinfo{author}{\bibfnamefont{S.}~\bibnamefont{Rao}},
  \bibinfo{journal}{Earth Planet. Sci. Lett.} \textbf{\bibinfo{volume}{308}},
  \bibinfo{pages}{424 } (\bibinfo{year}{2011}), ISSN \bibinfo{issn}{0012-821X},
  \urlprefix\url{http://www.sciencedirect.com/science/article/pii/S0012821X11003724}.

\bibitem[{\citenamefont{Latour et~al.}(2011)\citenamefont{Latour, Gallot,
  Catheline, Voisin, Renard, Larose, and Campillo}}]{Latour2011ultrafast}
\bibinfo{author}{\bibfnamefont{S.}~\bibnamefont{Latour}},
  \bibinfo{author}{\bibfnamefont{T.}~\bibnamefont{Gallot}},
  \bibinfo{author}{\bibfnamefont{S.}~\bibnamefont{Catheline}},
  \bibinfo{author}{\bibfnamefont{C.}~\bibnamefont{Voisin}},
  \bibinfo{author}{\bibfnamefont{F.}~\bibnamefont{Renard}},
  \bibinfo{author}{\bibfnamefont{E.}~\bibnamefont{Larose}}, \bibnamefont{and}
  \bibinfo{author}{\bibfnamefont{M.}~\bibnamefont{Campillo}},
  \bibinfo{journal}{Europhys. Lett.} \textbf{\bibinfo{volume}{96}},
  \bibinfo{pages}{59003} (\bibinfo{year}{2011}),
  \urlprefix\url{http://stacks.iop.org/0295-5075/96/i=5/a=59003}.

\bibitem[{\citenamefont{Audry et~al.}(2012)\citenamefont{Audry, Fretigny,
  Chateauminois, Teissere, and Barthel}}]{Audry2012slip}
\bibinfo{author}{\bibfnamefont{M.~C.} \bibnamefont{Audry}},
  \bibinfo{author}{\bibfnamefont{C.}~\bibnamefont{Fretigny}},
  \bibinfo{author}{\bibfnamefont{A.}~\bibnamefont{Chateauminois}},
  \bibinfo{author}{\bibfnamefont{J.}~\bibnamefont{Teissere}}, \bibnamefont{and}
  \bibinfo{author}{\bibfnamefont{E.}~\bibnamefont{Barthel}},
  \bibinfo{journal}{Eur. Phys. J. E} \textbf{\bibinfo{volume}{35}}
  (\bibinfo{year}{2012}), ISSN \bibinfo{issn}{1292-8941},
  \urlprefix\url{http://dx.doi.org/10.1140/epje/i2012-12083-0}.

\bibitem[{\citenamefont{Br{\"{o}}rmann
  et~al.}(2013)\citenamefont{Br{\"{o}}rmann, Barel, Urbakh, and
  Bennewitz}}]{Broermann2013friction}
\bibinfo{author}{\bibfnamefont{K.}~\bibnamefont{Br{\"{o}}rmann}},
  \bibinfo{author}{\bibfnamefont{I.}~\bibnamefont{Barel}},
  \bibinfo{author}{\bibfnamefont{M.}~\bibnamefont{Urbakh}}, \bibnamefont{and}
  \bibinfo{author}{\bibfnamefont{R.}~\bibnamefont{Bennewitz}},
  \bibinfo{journal}{Tribol. Lett.} \textbf{\bibinfo{volume}{50}},
  \bibinfo{pages}{3} (\bibinfo{year}{2013}), ISSN \bibinfo{issn}{1023-8883},
  \urlprefix\url{http://dx.doi.org/10.1007/s11249-012-0044-3}.

\bibitem[{\citenamefont{McLaskey et~al.}(2012)\citenamefont{McLaskey, Thomas,
  Glaser, and Nadeau}}]{McLaskey2012fault}
\bibinfo{author}{\bibfnamefont{G.~C.} \bibnamefont{McLaskey}},
  \bibinfo{author}{\bibfnamefont{A.~M.} \bibnamefont{Thomas}},
  \bibinfo{author}{\bibfnamefont{S.~D.} \bibnamefont{Glaser}},
  \bibnamefont{and} \bibinfo{author}{\bibfnamefont{R.~M.}
  \bibnamefont{Nadeau}}, \bibinfo{journal}{Nature}
  \textbf{\bibinfo{volume}{491}}, \bibinfo{pages}{101} (\bibinfo{year}{2012}),
  \urlprefix\url{http://dx.doi.org/10.1038/nature11512}.

\bibitem[{\citenamefont{Svetlizky and Fineberg}(2014)}]{Svetlizky2014classical}
\bibinfo{author}{\bibfnamefont{I.}~\bibnamefont{Svetlizky}} \bibnamefont{and}
  \bibinfo{author}{\bibfnamefont{J.}~\bibnamefont{Fineberg}},
  \bibinfo{journal}{Nature} \textbf{\bibinfo{volume}{509}},
  \bibinfo{pages}{205} (\bibinfo{year}{2014}), ISSN \bibinfo{issn}{0028-0836},
  \urlprefix\url{http://dx.doi.org/10.1038/nature13202}.

\bibitem[{\citenamefont{Freund}({1990})}]{Freund1990dynamic}
\bibinfo{author}{\bibfnamefont{L.~B.} \bibnamefont{Freund}},
  \emph{\bibinfo{title}{{Dynamic Fracture Mechanics}}}
  (\bibinfo{publisher}{{Cambridge University Press}}, \bibinfo{year}{{1990}}).

\bibitem[{\citenamefont{Degrandi-Contraires
  et~al.}(2012)\citenamefont{Degrandi-Contraires, Poulard, Restagno, and
  Leger}}]{Degrandi2012sliding}
\bibinfo{author}{\bibfnamefont{E.}~\bibnamefont{Degrandi-Contraires}},
  \bibinfo{author}{\bibfnamefont{C.}~\bibnamefont{Poulard}},
  \bibinfo{author}{\bibfnamefont{F.}~\bibnamefont{Restagno}}, \bibnamefont{and}
  \bibinfo{author}{\bibfnamefont{L.}~\bibnamefont{Leger}},
  \bibinfo{journal}{Faraday Discuss.} \textbf{\bibinfo{volume}{156}},
  \bibinfo{pages}{255} (\bibinfo{year}{2012}),
  \urlprefix\url{http://dx.doi.org/10.1039/C2FD00121G}.

\bibitem[{\citenamefont{Braun et~al.}(2009)\citenamefont{Braun, Barel, and
  Urbakh}}]{Braun2009dynamics}
\bibinfo{author}{\bibfnamefont{O.~M.} \bibnamefont{Braun}},
  \bibinfo{author}{\bibfnamefont{I.}~\bibnamefont{Barel}}, \bibnamefont{and}
  \bibinfo{author}{\bibfnamefont{M.}~\bibnamefont{Urbakh}},
  \bibinfo{journal}{Phys. Rev. Lett.} \textbf{\bibinfo{volume}{103}},
  \bibinfo{pages}{194301} (\bibinfo{year}{2009}),
  \urlprefix\url{http://prl.aps.org/abstract/PRL/v103/i19/e194301}.

\bibitem[{\citenamefont{Capozza and Urbakh}(2012)}]{Capozza2012static}
\bibinfo{author}{\bibfnamefont{R.}~\bibnamefont{Capozza}} \bibnamefont{and}
  \bibinfo{author}{\bibfnamefont{M.}~\bibnamefont{Urbakh}},
  \bibinfo{journal}{Phys. Rev. B} \textbf{\bibinfo{volume}{86}},
  \bibinfo{pages}{085430} (\bibinfo{year}{2012}),
  \urlprefix\url{http://link.aps.org/doi/10.1103/PhysRevB.86.085430}.

\bibitem[{\citenamefont{Tr{\o}mborg et~al.}(2014)\citenamefont{Tr{\o}mborg,
  Sveinsson, Scheibert, Th{\o}gersen, Amundsen, and
  Malthe-S{\o}renssen}}]{Tromborg2014slow}
\bibinfo{author}{\bibfnamefont{J.~K.} \bibnamefont{Tr{\o}mborg}},
  \bibinfo{author}{\bibfnamefont{H.~A.} \bibnamefont{Sveinsson}},
  \bibinfo{author}{\bibfnamefont{J.}~\bibnamefont{Scheibert}},
  \bibinfo{author}{\bibfnamefont{K.}~\bibnamefont{Th{\o}gersen}},
  \bibinfo{author}{\bibfnamefont{D.~S.} \bibnamefont{Amundsen}},
  \bibnamefont{and}
  \bibinfo{author}{\bibfnamefont{A.}~\bibnamefont{Malthe-S{\o}renssen}},
  \bibinfo{journal}{Proc. Natl. Acad. Sci. U. S. A.}
  \textbf{\bibinfo{volume}{111}}, \bibinfo{pages}{8764} (\bibinfo{year}{2014}),
  \urlprefix\url{http://www.pnas.org/content/111/24/8764.abstract}.

\bibitem[{\citenamefont{Bouchbinder et~al.}(2011)\citenamefont{Bouchbinder,
  Brener, Barel, and Urbakh}}]{Bouchbinder2011slow}
\bibinfo{author}{\bibfnamefont{E.}~\bibnamefont{Bouchbinder}},
  \bibinfo{author}{\bibfnamefont{E.~A.} \bibnamefont{Brener}},
  \bibinfo{author}{\bibfnamefont{I.}~\bibnamefont{Barel}}, \bibnamefont{and}
  \bibinfo{author}{\bibfnamefont{M.}~\bibnamefont{Urbakh}},
  \bibinfo{journal}{Phys. Rev. Lett.} \textbf{\bibinfo{volume}{107}},
  \bibinfo{pages}{235501} (\bibinfo{year}{2011}),
  \urlprefix\url{http://dx.doi.org/10.1103/PhysRevLett.107.235501}.

\bibitem[{\citenamefont{Bar~Sinai et~al.}(2012)\citenamefont{Bar~Sinai, Brener,
  and Bouchbinder}}]{BarSinai2012slow}
\bibinfo{author}{\bibfnamefont{Y.}~\bibnamefont{Bar~Sinai}},
  \bibinfo{author}{\bibfnamefont{E.~A.} \bibnamefont{Brener}},
  \bibnamefont{and}
  \bibinfo{author}{\bibfnamefont{E.}~\bibnamefont{Bouchbinder}},
  \bibinfo{journal}{Geophys. Res. Lett.} \textbf{\bibinfo{volume}{39}}
  (\bibinfo{year}{2012}), ISSN \bibinfo{issn}{0094-8276},
  \urlprefix\url{http://dx.doi.org/10.1029/2011GL050554}.

\bibitem[{\citenamefont{Braun and Peyrard}(2012)}]{Braun2012crack}
\bibinfo{author}{\bibfnamefont{O.~M.} \bibnamefont{Braun}} \bibnamefont{and}
  \bibinfo{author}{\bibfnamefont{M.}~\bibnamefont{Peyrard}},
  \bibinfo{journal}{Phys. Rev. E} \textbf{\bibinfo{volume}{85}},
  \bibinfo{pages}{026111} (\bibinfo{year}{2012}), ISSN
  \bibinfo{issn}{1539-3755},
  \urlprefix\url{http://link.aps.org/doi/10.1103/PhysRevE.85.026111}.

\bibitem[{\citenamefont{Bartolomeo et~al.}(2012)\citenamefont{Bartolomeo,
  Massi, Baillet, Culla, Fregolent, and Berthier}}]{DiBartolomeo2012wave}
\bibinfo{author}{\bibfnamefont{M.~D.} \bibnamefont{Bartolomeo}},
  \bibinfo{author}{\bibfnamefont{F.}~\bibnamefont{Massi}},
  \bibinfo{author}{\bibfnamefont{L.}~\bibnamefont{Baillet}},
  \bibinfo{author}{\bibfnamefont{A.}~\bibnamefont{Culla}},
  \bibinfo{author}{\bibfnamefont{A.}~\bibnamefont{Fregolent}},
  \bibnamefont{and} \bibinfo{author}{\bibfnamefont{Y.}~\bibnamefont{Berthier}},
  \bibinfo{journal}{Tribol. Int.} \textbf{\bibinfo{volume}{52}},
  \bibinfo{pages}{117} (\bibinfo{year}{2012}), ISSN \bibinfo{issn}{0301-679X},
  \urlprefix\url{http://www.sciencedirect.com/science/article/pii/S0301679X12001053}.

\bibitem[{\citenamefont{Kammer et~al.}(2012)\citenamefont{Kammer, Yastrebov,
  Spijker, and Molinari}}]{Kammer2012propagation}
\bibinfo{author}{\bibfnamefont{D.~S.} \bibnamefont{Kammer}},
  \bibinfo{author}{\bibfnamefont{V.~A.} \bibnamefont{Yastrebov}},
  \bibinfo{author}{\bibfnamefont{P.}~\bibnamefont{Spijker}}, \bibnamefont{and}
  \bibinfo{author}{\bibfnamefont{J.-F.} \bibnamefont{Molinari}},
  \bibinfo{journal}{Tribol. Lett.} \textbf{\bibinfo{volume}{48}},
  \bibinfo{pages}{27} (\bibinfo{year}{2012}), ISSN \bibinfo{issn}{1023-8883},
  \urlprefix\url{http://dx.doi.org/10.1007/s11249-012-9920-0}.

\bibitem[{\citenamefont{Otsuki and Matsukawa}(2013)}]{Otsuki2013systematic}
\bibinfo{author}{\bibfnamefont{M.}~\bibnamefont{Otsuki}} \bibnamefont{and}
  \bibinfo{author}{\bibfnamefont{H.}~\bibnamefont{Matsukawa}},
  \bibinfo{journal}{Sci. Rep.} \textbf{\bibinfo{volume}{3}}
  (\bibinfo{year}{2013}), ISSN \bibinfo{issn}{2045-2322},
  \urlprefix\url{http://dx.doi.org/10.1038/srep01586}.

\bibitem[{\citenamefont{Bar-Sinai et~al.}(2013)\citenamefont{Bar-Sinai,
  Spatschek, Brener, and Bouchbinder}}]{BarSinai2013instabilities}
\bibinfo{author}{\bibfnamefont{Y.}~\bibnamefont{Bar-Sinai}},
  \bibinfo{author}{\bibfnamefont{R.}~\bibnamefont{Spatschek}},
  \bibinfo{author}{\bibfnamefont{E.~A.} \bibnamefont{Brener}},
  \bibnamefont{and}
  \bibinfo{author}{\bibfnamefont{E.}~\bibnamefont{Bouchbinder}},
  \bibinfo{journal}{Phys. Rev. E} \textbf{\bibinfo{volume}{88}},
  \bibinfo{pages}{060403} (\bibinfo{year}{2013}),
  \urlprefix\url{http://link.aps.org/doi/10.1103/PhysRevE.88.060403}.

\bibitem[{\citenamefont{Papangelo et~al.}(2014)\citenamefont{Papangelo, Stingl,
  Hoffmann, and Ciavarella}}]{Papangelo2014simple}
\bibinfo{author}{\bibfnamefont{A.}~\bibnamefont{Papangelo}},
  \bibinfo{author}{\bibfnamefont{B.}~\bibnamefont{Stingl}},
  \bibinfo{author}{\bibfnamefont{N.~P.} \bibnamefont{Hoffmann}},
  \bibnamefont{and}
  \bibinfo{author}{\bibfnamefont{M.}~\bibnamefont{Ciavarella}},
  \bibinfo{journal}{Physical Mesomechanics} \textbf{\bibinfo{volume}{17}},
  \bibinfo{pages}{311} (\bibinfo{year}{2014}), ISSN \bibinfo{issn}{1029-9599},
  \urlprefix\url{http://dx.doi.org/10.1134/S1029959914040080}.

\bibitem[{\citenamefont{Dieterich}(1979)}]{Dieterich1979modeling1}
\bibinfo{author}{\bibfnamefont{J.~H.} \bibnamefont{Dieterich}},
  \bibinfo{journal}{J. Geophys. Res.} \textbf{\bibinfo{volume}{84}},
  \bibinfo{pages}{2161} (\bibinfo{year}{1979}),
  \urlprefix\url{http://dx.doi.org/10.1029/JB084iB05p02161}.

\bibitem[{\citenamefont{Ruina}(1983)}]{Ruina1983slip}
\bibinfo{author}{\bibfnamefont{A.}~\bibnamefont{Ruina}}, \bibinfo{journal}{J.
  Geophys. Res.} \textbf{\bibinfo{volume}{88}}, \bibinfo{pages}{10359}
  (\bibinfo{year}{1983}), ISSN \bibinfo{issn}{0148-0227},
  \urlprefix\url{http://doi.wiley.com/10.1029/JB088iB12p10359}.

\bibitem[{\citenamefont{Marone}(1998)}]{Marone1998laboratory-derived}
\bibinfo{author}{\bibfnamefont{C.}~\bibnamefont{Marone}},
  \bibinfo{journal}{Annu. Rev. Earth Planet. Sci.}
  \textbf{\bibinfo{volume}{26}}, \bibinfo{pages}{643} (\bibinfo{year}{1998}),
  ISSN \bibinfo{issn}{0084-6597},
  \urlprefix\url{http://www.annualreviews.org/doi/abs/10.1146/annurev.earth.26.1.643}.

\bibitem[{\citenamefont{Baumberger and Caroli}(2006)}]{Baumberger2006solid}
\bibinfo{author}{\bibfnamefont{T.}~\bibnamefont{Baumberger}} \bibnamefont{and}
  \bibinfo{author}{\bibfnamefont{C.}~\bibnamefont{Caroli}},
  \bibinfo{journal}{Adv. Phys.} \textbf{\bibinfo{volume}{55}},
  \bibinfo{pages}{279} (\bibinfo{year}{2006}), ISSN \bibinfo{issn}{0001-8732},
  \urlprefix\url{http://www.tandfonline.com/doi/abs/10.1080/00018730600732186}.

\bibitem[{\citenamefont{Li et~al.}(2011)\citenamefont{Li, Tullis, Goldsby, and
  Carpick}}]{Li2011ageing}
\bibinfo{author}{\bibfnamefont{Q.}~\bibnamefont{Li}},
  \bibinfo{author}{\bibfnamefont{T.~E.} \bibnamefont{Tullis}},
  \bibinfo{author}{\bibfnamefont{D.}~\bibnamefont{Goldsby}}, \bibnamefont{and}
  \bibinfo{author}{\bibfnamefont{R.~W.} \bibnamefont{Carpick}},
  \bibinfo{journal}{Nature} \textbf{\bibinfo{volume}{480}},
  \bibinfo{pages}{233} (\bibinfo{year}{2011}), ISSN \bibinfo{issn}{0028-0836},
  \urlprefix\url{http://dx.doi.org/10.1038/nature10589}.

\bibitem[{\citenamefont{Braun and R{\"o}der}(2002)}]{Braun2002transition}
\bibinfo{author}{\bibfnamefont{O.~M.} \bibnamefont{Braun}} \bibnamefont{and}
  \bibinfo{author}{\bibfnamefont{J.}~\bibnamefont{R{\"o}der}},
  \bibinfo{journal}{Phys. Rev. Lett.} \textbf{\bibinfo{volume}{88}},
  \bibinfo{pages}{096102} (\bibinfo{year}{2002}),
  \urlprefix\url{http://dx.doi.org/10.1103/PhysRevLett.88.096102}.

\bibitem[{\citenamefont{Farkas et~al.}(2005)\citenamefont{Farkas, Dahmen, and
  Wolf}}]{Farkas2005static}
\bibinfo{author}{\bibfnamefont{Z.}~\bibnamefont{Farkas}},
  \bibinfo{author}{\bibfnamefont{S.~R.} \bibnamefont{Dahmen}},
  \bibnamefont{and} \bibinfo{author}{\bibfnamefont{D.~E.} \bibnamefont{Wolf}},
  \bibinfo{journal}{J. Stat. Mech.: Theory Exp.}
  \textbf{\bibinfo{volume}{2005}}, \bibinfo{pages}{P06015}
  (\bibinfo{year}{2005}),
  \urlprefix\url{http://iopscience.iop.org/1742-5468/2005/06/P06015}.

\bibitem[{\citenamefont{Persson}(1995)}]{Persson1995theory}
\bibinfo{author}{\bibfnamefont{B.~N.~J.} \bibnamefont{Persson}},
  \bibinfo{journal}{Phys. Rev. B} \textbf{\bibinfo{volume}{51}},
  \bibinfo{pages}{13568} (\bibinfo{year}{1995}),
  \urlprefix\url{http://link.aps.org/doi/10.1103/PhysRevB.51.13568}.

\bibitem[{\citenamefont{Persson}(2000)}]{Persson2000sliding}
\bibinfo{author}{\bibfnamefont{B.~N.~J.} \bibnamefont{Persson}},
  \emph{\bibinfo{title}{Sliding friction: {P}hysical principles and
  applications}} (\bibinfo{publisher}{Springer}, \bibinfo{year}{2000}),
  \bibinfo{edition}{2nd} ed.

\bibitem[{\citenamefont{Braun and Peyrard}(2008)}]{Braun2008modeling}
\bibinfo{author}{\bibfnamefont{O.~M.} \bibnamefont{Braun}} \bibnamefont{and}
  \bibinfo{author}{\bibfnamefont{M.}~\bibnamefont{Peyrard}},
  \bibinfo{journal}{Phys. Rev. Lett.} \textbf{\bibinfo{volume}{100}},
  \bibinfo{pages}{125501} (\bibinfo{year}{2008}),
  \urlprefix\url{http://dx.doi.org/10.1103/PhysRevLett.100.125501}.

\bibitem[{\citenamefont{Braun and Peyrard}(2010)}]{Braun2010master}
\bibinfo{author}{\bibfnamefont{O.~M.} \bibnamefont{Braun}} \bibnamefont{and}
  \bibinfo{author}{\bibfnamefont{M.}~\bibnamefont{Peyrard}},
  \bibinfo{journal}{Phys. Rev. E} \textbf{\bibinfo{volume}{82}},
  \bibinfo{pages}{036117} (\bibinfo{year}{2010}),
  \urlprefix\url{http://dx.doi.org/10.1103/PhysRevE.82.036117}.

\bibitem[{\citenamefont{Th\o{}gersen et~al.}(2014)\citenamefont{Th\o{}gersen,
  Tr\o{}mborg, Sveinsson, Malthe-S\o{}renssen, and
  Scheibert}}]{Thogersen2014history-dependent}
\bibinfo{author}{\bibfnamefont{K.}~\bibnamefont{Th\o{}gersen}},
  \bibinfo{author}{\bibfnamefont{J.~K.} \bibnamefont{Tr\o{}mborg}},
  \bibinfo{author}{\bibfnamefont{H.~A.} \bibnamefont{Sveinsson}},
  \bibinfo{author}{\bibfnamefont{A.}~\bibnamefont{Malthe-S\o{}renssen}},
  \bibnamefont{and}
  \bibinfo{author}{\bibfnamefont{J.}~\bibnamefont{Scheibert}},
  \bibinfo{journal}{Phys. Rev. E} \textbf{\bibinfo{volume}{89}},
  \bibinfo{pages}{052401} (\bibinfo{year}{2014}),
  \urlprefix\url{http://link.aps.org/doi/10.1103/PhysRevE.89.052401}.

\bibitem[{\citenamefont{Reguzzoni et~al.}(2010)\citenamefont{Reguzzoni,
  Ferrario, Zapperi, and Righi}}]{Reguzzoni2010onset}
\bibinfo{author}{\bibfnamefont{M.}~\bibnamefont{Reguzzoni}},
  \bibinfo{author}{\bibfnamefont{M.}~\bibnamefont{Ferrario}},
  \bibinfo{author}{\bibfnamefont{S.}~\bibnamefont{Zapperi}}, \bibnamefont{and}
  \bibinfo{author}{\bibfnamefont{M.}~\bibnamefont{Righi}},
  \bibinfo{journal}{Proc. Natl. Acad. Sci. U. S. A.}
  \textbf{\bibinfo{volume}{107}}, \bibinfo{pages}{1311} (\bibinfo{year}{2010}),
  \urlprefix\url{http://www.pnas.org/content/107/4/1311.abstract}.

\bibitem[{\citenamefont{Filippov et~al.}(2004)\citenamefont{Filippov, Klafter,
  and Urbakh}}]{Filippov2004friction}
\bibinfo{author}{\bibfnamefont{A.~E.} \bibnamefont{Filippov}},
  \bibinfo{author}{\bibfnamefont{J.}~\bibnamefont{Klafter}}, \bibnamefont{and}
  \bibinfo{author}{\bibfnamefont{M.}~\bibnamefont{Urbakh}},
  \bibinfo{journal}{Phys. Rev. Lett.} \textbf{\bibinfo{volume}{92}},
  \bibinfo{pages}{135503} (\bibinfo{year}{2004}),
  \urlprefix\url{http://link.aps.org/doi/10.1103/PhysRevLett.92.135503}.

\bibitem[{\citenamefont{Srinivasan and Walcott}(2009)}]{Srinivasan2009binding}
\bibinfo{author}{\bibfnamefont{M.}~\bibnamefont{Srinivasan}} \bibnamefont{and}
  \bibinfo{author}{\bibfnamefont{S.}~\bibnamefont{Walcott}},
  \bibinfo{journal}{Phys. Rev. E} \textbf{\bibinfo{volume}{80}},
  \bibinfo{pages}{046124} (\bibinfo{year}{2009}), ISSN
  \bibinfo{issn}{1539-3755},
  \urlprefix\url{http://link.aps.org/doi/10.1103/PhysRevE.80.046124}.

\bibitem[{\citenamefont{Brace and Byerlee}(1966)}]{Brace1966stick-slip}
\bibinfo{author}{\bibfnamefont{W.~F.} \bibnamefont{Brace}} \bibnamefont{and}
  \bibinfo{author}{\bibfnamefont{J.~D.} \bibnamefont{Byerlee}},
  \bibinfo{journal}{Science} \textbf{\bibinfo{volume}{153}},
  \bibinfo{pages}{990} (\bibinfo{year}{1966}),
  \urlprefix\url{http://www.sciencemag.org/content/153/3739/990.abstract}.

\bibitem[{\citenamefont{Burridge and Knopoff}(1967)}]{Burridge1967model}
\bibinfo{author}{\bibfnamefont{R.}~\bibnamefont{Burridge}} \bibnamefont{and}
  \bibinfo{author}{\bibfnamefont{L.}~\bibnamefont{Knopoff}},
  \bibinfo{journal}{Bull. Seismol. Soc. Am.} \textbf{\bibinfo{volume}{57}},
  \bibinfo{pages}{341} (\bibinfo{year}{1967}),
  \urlprefix\url{http://www.bssaonline.org/content/57/3/341.abstract}.

\bibitem[{\citenamefont{Carlson et~al.}(1991)\citenamefont{Carlson, Langer,
  Shaw, and Tang}}]{Carlson1991intrinsic}
\bibinfo{author}{\bibfnamefont{J.~M.} \bibnamefont{Carlson}},
  \bibinfo{author}{\bibfnamefont{J.~S.} \bibnamefont{Langer}},
  \bibinfo{author}{\bibfnamefont{B.~E.} \bibnamefont{Shaw}}, \bibnamefont{and}
  \bibinfo{author}{\bibfnamefont{C.}~\bibnamefont{Tang}},
  \bibinfo{journal}{Phys. Rev. A} \textbf{\bibinfo{volume}{44}},
  \bibinfo{pages}{884} (\bibinfo{year}{1991}),
  \urlprefix\url{http://dx.doi.org/10.1103/PhysRevA.44.884}.

\bibitem[{\citenamefont{Carlson et~al.}(1994)\citenamefont{Carlson, Langer, and
  Shaw}}]{Carlson1994dynamics}
\bibinfo{author}{\bibfnamefont{J.~M.} \bibnamefont{Carlson}},
  \bibinfo{author}{\bibfnamefont{J.~S.} \bibnamefont{Langer}},
  \bibnamefont{and} \bibinfo{author}{\bibfnamefont{B.~E.} \bibnamefont{Shaw}},
  \bibinfo{journal}{Rev. Mod. Phys.} \textbf{\bibinfo{volume}{66}},
  \bibinfo{pages}{657} (\bibinfo{year}{1994}), ISSN \bibinfo{issn}{0034-6861},
  \urlprefix\url{http://link.aps.org/doi/10.1103/RevModPhys.66.657}.

\bibitem[{\citenamefont{Capozza et~al.}(2011)\citenamefont{Capozza, Rubinstein,
  Barel, Urbakh, and Fineberg}}]{Capozza2011stabilizing}
\bibinfo{author}{\bibfnamefont{R.}~\bibnamefont{Capozza}},
  \bibinfo{author}{\bibfnamefont{S.~M.} \bibnamefont{Rubinstein}},
  \bibinfo{author}{\bibfnamefont{I.}~\bibnamefont{Barel}},
  \bibinfo{author}{\bibfnamefont{M.}~\bibnamefont{Urbakh}}, \bibnamefont{and}
  \bibinfo{author}{\bibfnamefont{J.}~\bibnamefont{Fineberg}},
  \bibinfo{journal}{Phys. Rev. Lett.} \textbf{\bibinfo{volume}{107}},
  \bibinfo{pages}{024301} (\bibinfo{year}{2011}),
  \urlprefix\url{http://link.aps.org/doi/10.1103/PhysRevLett.107.024301}.

\bibitem[{\citenamefont{Myers and Langer}(1993)}]{Myers1993rupture}
\bibinfo{author}{\bibfnamefont{C.~R.} \bibnamefont{Myers}} \bibnamefont{and}
  \bibinfo{author}{\bibfnamefont{J.~S.} \bibnamefont{Langer}},
  \bibinfo{journal}{Phys. Rev. E} \textbf{\bibinfo{volume}{47}},
  \bibinfo{pages}{3048} (\bibinfo{year}{1993}),
  \urlprefix\url{http://link.aps.org/doi/10.1103/PhysRevE.47.3048}.

\bibitem[{\citenamefont{Ben-Zion}(2001)}]{Ben-Zion2001dynamic}
\bibinfo{author}{\bibfnamefont{Y.}~\bibnamefont{Ben-Zion}},
  \bibinfo{journal}{J. Mech. Phys. Solids} \textbf{\bibinfo{volume}{49}},
  \bibinfo{pages}{2209} (\bibinfo{year}{2001}),
  \urlprefix\url{http://dx.doi.org/10.1016/S0022-5096(01)00036-9}.

\bibitem[{\citenamefont{Yim and Sohn}(2000)}]{Yim2000numerical}
\bibinfo{author}{\bibfnamefont{H.}~\bibnamefont{Yim}} \bibnamefont{and}
  \bibinfo{author}{\bibfnamefont{Y.}~\bibnamefont{Sohn}},
  \bibinfo{journal}{{IEEE} T. Ultrason. Ferr.} \textbf{\bibinfo{volume}{47}},
  \bibinfo{pages}{549} (\bibinfo{year}{2000}), ISSN \bibinfo{issn}{0885-3010},
  \urlprefix\url{http://dx.doi.org/10.1109/58.842041}.

\bibitem[{\citenamefont{Ben-David
  et~al.}(2010{\natexlab{b}})\citenamefont{Ben-David, Rubinstein, and
  Fineberg}}]{Ben-David2010slip-stick}
\bibinfo{author}{\bibfnamefont{O.}~\bibnamefont{Ben-David}},
  \bibinfo{author}{\bibfnamefont{S.~M.} \bibnamefont{Rubinstein}},
  \bibnamefont{and} \bibinfo{author}{\bibfnamefont{J.}~\bibnamefont{Fineberg}},
  \bibinfo{journal}{Nature} \textbf{\bibinfo{volume}{463}}, \bibinfo{pages}{76}
  (\bibinfo{year}{2010}{\natexlab{b}}),
  \urlprefix\url{http://dx.doi.org/10.1038/nature08676}.

\bibitem[{\citenamefont{Giorgi}(1999)}]{Giorgi1999income}
\bibinfo{author}{\bibfnamefont{G.~M.} \bibnamefont{Giorgi}}, in
  \emph{\bibinfo{booktitle}{Handbook on Income Inequality Measurement}}, edited
  by \bibinfo{editor}{\bibfnamefont{J.}~\bibnamefont{Silber}}
  (\bibinfo{publisher}{Kluwer Academic Publishers}, \bibinfo{address}{New
  York}, \bibinfo{year}{1999}), chap.~\bibinfo{chapter}{8}, pp.
  \bibinfo{pages}{245--267}, ISBN \bibinfo{isbn}{978-94-011-4413-1},
  \urlprefix\url{http://dx.doi.org/10.1007/978-94-011-4413-1}.

\bibitem[{\citenamefont{Giorgi}(1990)}]{Giorgi1990bibliographic}
\bibinfo{author}{\bibfnamefont{G.~M.} \bibnamefont{Giorgi}},
  \bibinfo{journal}{METRON: International Journal of Statistics}
  \textbf{\bibinfo{volume}{XLVIII}}, \bibinfo{pages}{183}
  (\bibinfo{year}{1990}),
  \urlprefix\url{http://econpapers.repec.org/RePEc:wpa:wuwpem:0511004}.

\bibitem[{\citenamefont{{Di Toro} et~al.}(2011)\citenamefont{{Di Toro}, Han,
  Hirose, {De Paola}, Nielsen, Mizoguchi, Ferri, Cocco, and
  Shimamoto}}]{DiToro2011fault}
\bibinfo{author}{\bibfnamefont{G.}~\bibnamefont{{Di Toro}}},
  \bibinfo{author}{\bibfnamefont{R.}~\bibnamefont{Han}},
  \bibinfo{author}{\bibfnamefont{T.}~\bibnamefont{Hirose}},
  \bibinfo{author}{\bibfnamefont{N.}~\bibnamefont{{De Paola}}},
  \bibinfo{author}{\bibfnamefont{S.}~\bibnamefont{Nielsen}},
  \bibinfo{author}{\bibfnamefont{K.}~\bibnamefont{Mizoguchi}},
  \bibinfo{author}{\bibfnamefont{F.}~\bibnamefont{Ferri}},
  \bibinfo{author}{\bibfnamefont{M.}~\bibnamefont{Cocco}}, \bibnamefont{and}
  \bibinfo{author}{\bibfnamefont{T.}~\bibnamefont{Shimamoto}},
  \bibinfo{journal}{Nature} \textbf{\bibinfo{volume}{471}},
  \bibinfo{pages}{494} (\bibinfo{year}{2011}), ISSN \bibinfo{issn}{1476-4687},
  \urlprefix\url{http://dx.doi.org/10.1038/nature09838}.

\bibitem[{\citenamefont{Caroli and Nozi\`{e}res}(1998)}]{Caroli1998hysteresis}
\bibinfo{author}{\bibfnamefont{C.}~\bibnamefont{Caroli}} \bibnamefont{and}
  \bibinfo{author}{\bibfnamefont{P.}~\bibnamefont{Nozi\`{e}res}},
  \bibinfo{journal}{Eur. Phys. J. B} \textbf{\bibinfo{volume}{4}},
  \bibinfo{pages}{233} (\bibinfo{year}{1998}),
  \urlprefix\url{http://link.springer.com/article/10.1007/s100510050374}.

\bibitem[{\citenamefont{Braun et~al.}(2012)\citenamefont{Braun, Peyrard,
  Stryzheus, and Tosatti}}]{Braun2012collective}
\bibinfo{author}{\bibfnamefont{O.~M.} \bibnamefont{Braun}},
  \bibinfo{author}{\bibfnamefont{M.}~\bibnamefont{Peyrard}},
  \bibinfo{author}{\bibfnamefont{D.~V.} \bibnamefont{Stryzheus}},
  \bibnamefont{and} \bibinfo{author}{\bibfnamefont{E.}~\bibnamefont{Tosatti}},
  \bibinfo{journal}{Tribol. Lett.} \textbf{\bibinfo{volume}{48}},
  \bibinfo{pages}{11} (\bibinfo{year}{2012}), ISSN \bibinfo{issn}{1023-8883},
  \urlprefix\url{http://dx.doi.org/10.1007/s11249-012-9913-z}.

\bibitem[{\citenamefont{Ben-David
  et~al.}(2010{\natexlab{c}})\citenamefont{Ben-David, Cohen, and
  Fineberg}}]{Ben-David2010short-time}
\bibinfo{author}{\bibfnamefont{O.}~\bibnamefont{Ben-David}},
  \bibinfo{author}{\bibfnamefont{G.}~\bibnamefont{Cohen}}, \bibnamefont{and}
  \bibinfo{author}{\bibfnamefont{J.}~\bibnamefont{Fineberg}},
  \bibinfo{journal}{Tribol. Lett.} \textbf{\bibinfo{volume}{39}},
  \bibinfo{pages}{235} (\bibinfo{year}{2010}{\natexlab{c}}), ISSN
  \bibinfo{issn}{1023-8883},
  \urlprefix\url{http://link.springer.com/10.1007/s11249-010-9601-9}.

\bibitem[{\citenamefont{Braun and Peyrard}(2011)}]{Braun2011dependence}
\bibinfo{author}{\bibfnamefont{O.~M.} \bibnamefont{Braun}} \bibnamefont{and}
  \bibinfo{author}{\bibfnamefont{M.}~\bibnamefont{Peyrard}},
  \bibinfo{journal}{Phys. Rev. E} \textbf{\bibinfo{volume}{83}},
  \bibinfo{pages}{046129} (\bibinfo{year}{2011}),
  \urlprefix\url{http://dx.doi.org/10.1103/PhysRevE.83.046129}.

\bibitem[{\citenamefont{Katano et~al.}(2014)\citenamefont{Katano, Nakano,
  Otsuki, and Matsukawa}}]{Katano2014novel}
\bibinfo{author}{\bibfnamefont{Y.}~\bibnamefont{Katano}},
  \bibinfo{author}{\bibfnamefont{K.}~\bibnamefont{Nakano}},
  \bibinfo{author}{\bibfnamefont{M.}~\bibnamefont{Otsuki}}, \bibnamefont{and}
  \bibinfo{author}{\bibfnamefont{H.}~\bibnamefont{Matsukawa}},
  \bibinfo{journal}{Sci. Rep.} \textbf{\bibinfo{volume}{4}},
  \bibinfo{pages}{6324} (\bibinfo{year}{2014}), ISSN \bibinfo{issn}{2045-2322},
  \urlprefix\url{http://dx.doi.org/10.1038/srep06324}.

\bibitem[{\citenamefont{Ben-David and Fineberg}(2011)}]{Ben-David2011static}
\bibinfo{author}{\bibfnamefont{O.}~\bibnamefont{Ben-David}} \bibnamefont{and}
  \bibinfo{author}{\bibfnamefont{J.}~\bibnamefont{Fineberg}},
  \bibinfo{journal}{Phys. Rev. Lett.} \textbf{\bibinfo{volume}{106}},
  \bibinfo{pages}{254301} (\bibinfo{year}{2011}), ISSN
  \bibinfo{issn}{0031-9007},
  \urlprefix\url{http://link.aps.org/doi/10.1103/PhysRevLett.106.254301}.

\bibitem[{\citenamefont{Bonamy and Bouchaud}(2011)}]{Bonamy2011failure}
\bibinfo{author}{\bibfnamefont{D.}~\bibnamefont{Bonamy}} \bibnamefont{and}
  \bibinfo{author}{\bibfnamefont{E.}~\bibnamefont{Bouchaud}},
  \bibinfo{journal}{Phys. Rep.} \textbf{\bibinfo{volume}{498}},
  \bibinfo{pages}{1} (\bibinfo{year}{2011}), ISSN \bibinfo{issn}{03701573},
  \urlprefix\url{http://www.sciencedirect.com/science/article/pii/S0370157310002115}.

\bibitem[{\citenamefont{Pradhan et~al.}(2010)\citenamefont{Pradhan, Hansen, and
  Chakrabarti}}]{Pradhan2010failure}
\bibinfo{author}{\bibfnamefont{S.}~\bibnamefont{Pradhan}},
  \bibinfo{author}{\bibfnamefont{A.}~\bibnamefont{Hansen}}, \bibnamefont{and}
  \bibinfo{author}{\bibfnamefont{B.}~\bibnamefont{Chakrabarti}},
  \bibinfo{journal}{Rev. Mod. Phys.} \textbf{\bibinfo{volume}{82}},
  \bibinfo{pages}{499} (\bibinfo{year}{2010}),
  \urlprefix\url{http://link.aps.org/doi/10.1103/RevModPhys.82.499}.

\bibitem[{\citenamefont{Coker et~al.}(2003)\citenamefont{Coker, Rosakis, and
  Needleman}}]{Coker2003dynamic}
\bibinfo{author}{\bibfnamefont{D.}~\bibnamefont{Coker}},
  \bibinfo{author}{\bibfnamefont{A.~J.} \bibnamefont{Rosakis}},
  \bibnamefont{and}
  \bibinfo{author}{\bibfnamefont{A.}~\bibnamefont{Needleman}},
  \bibinfo{journal}{J. Mech. Phys. Solids} \textbf{\bibinfo{volume}{51}},
  \bibinfo{pages}{425} (\bibinfo{year}{2003}),
  \urlprefix\url{http://www.sciencedirect.com/science/article/pii/S0022509602000820}.

\bibitem[{\citenamefont{Komarov}(2010)}]{Komarov2010gini}
\bibinfo{author}{\bibfnamefont{O.}~\bibnamefont{Komarov}},
  \emph{\bibinfo{title}{{GINI coefficient}}} (\bibinfo{year}{2010}),
  \urlprefix\url{http://www.mathworks.com/matlabcentral/fileexchange/26452-gini-coefficient}.

\bibitem[{\citenamefont{Latour et~al.}(2013)\citenamefont{Latour, Voisin,
  Renard, Larose, Catheline, and Campillo}}]{Latour2013effect}
\bibinfo{author}{\bibfnamefont{S.}~\bibnamefont{Latour}},
  \bibinfo{author}{\bibfnamefont{C.}~\bibnamefont{Voisin}},
  \bibinfo{author}{\bibfnamefont{F.}~\bibnamefont{Renard}},
  \bibinfo{author}{\bibfnamefont{E.}~\bibnamefont{Larose}},
  \bibinfo{author}{\bibfnamefont{S.}~\bibnamefont{Catheline}},
  \bibnamefont{and} \bibinfo{author}{\bibfnamefont{M.}~\bibnamefont{Campillo}},
  \bibinfo{journal}{J. Geophys. Res.: Solid Earth}
  \textbf{\bibinfo{volume}{118}}, \bibinfo{pages}{5888} (\bibinfo{year}{2013}),
  ISSN \bibinfo{issn}{21699313},
  \urlprefix\url{http://doi.wiley.com/10.1002/2013JB010231}.

\bibitem[{\citenamefont{Muratov}(1999)}]{Muratov1999traveling}
\bibinfo{author}{\bibfnamefont{C.~B.} \bibnamefont{Muratov}},
  \bibinfo{journal}{Phys. Rev. E} \textbf{\bibinfo{volume}{59}},
  \bibinfo{pages}{3847} (\bibinfo{year}{1999}),
  \urlprefix\url{http://dx.doi.org/10.1103/PhysRevE.59.3847}.

\bibitem[{\citenamefont{Scholz}(2002)}]{Scholz2002mechanics}
\bibinfo{author}{\bibfnamefont{C.~H.} \bibnamefont{Scholz}},
  \emph{\bibinfo{title}{The Mechanics of Earthquakes and Faulting}}
  (\bibinfo{publisher}{Cambridge University Press},
  \bibinfo{address}{Cambridge}, \bibinfo{year}{2002}), \bibinfo{edition}{2nd}
  ed.

\end{thebibliography}

\end{document}